\documentclass[aps,twocolumn,groupaddress,usenatbib,showkeys,showpacs]{revtex4-1}
\usepackage{graphicx}
\usepackage{dcolumn}
\usepackage{amssymb}
\usepackage{amsfonts}
\usepackage{amsbsy}
\usepackage{color}
\usepackage{rotating}
\usepackage[english]{babel}
\usepackage{epsfig}

\begin{document}
\title{Testing  feasibility of scalar-tensor gravity by  scale dependent mass and coupling to matter}
\author{D. F. Mota,}
\affiliation{Institute of Theoretical Astrophysics, University of Oslo, 0315 Oslo, Norway}
\author{V. Salzano,}
\affiliation{Fisika Teorikoaren eta Zientziaren Historia Saila, Zientzia eta Teknologia Fakultatea, \\ Euskal
Herriko Unibertsitatea, 644 Posta Kutxatila, 48080 Bilbao, Spain}
\author{and S. Capozziello}
\affiliation{Dipartimento di Scienze Fisiche, Universita' degli Studi di
Napoli ``Federico II" and INFN, Sezione di Napoli, Complesso
Universitario di Monte S. Angelo, Via Cinthia, Edificio N, 80126
Napoli, Italy}

\begin{abstract}
We investigate whether there are any cosmological evidences for a
scalar field with a mass and coupling to matter which change accordingly to the
properties of the astrophysical system it ``lives in'', without directly
focusing on the  underlying mechanism that drives the scalar field scale-dependent-properties.  We assume a  Yukawa type of coupling between the field and matter and also that the scalar field mass grows with density,
in order to overcome all gravity constraints within the solar system.
We analyse three
different gravitational systems assumed as ``cosmological
indicators'': supernovae type Ia, low surface brightness spiral galaxies
and clusters of galaxies. Results show that:  \textit{a.} a quite good fit to the
rotation curves of low surface brightness galaxies only using
visible stellar and gas mass components is obtained; \textit{b.} a scalar field can
fairly well reproduce the matter profile in clusters of galaxies,
estimated by X-ray observations and without the need of any
additional dark matter; \textit{c.} there is
an intrinsic difficulty in extracting information about the possibility
of a scale-dependent massive scalar field (or more generally about a varying gravitational
constant) from supernovae type Ia.
\end{abstract}


\maketitle

\section{Introduction}
\label{sec:introduction}

Present cosmological and astrophysical observations clearly depict
a universe dominated by ``dark'' components, being it made by dark
matter for $25 \%$ and by dark energy for $71 \%$, while ordinary
baryonic matter contributes for only the remaining $4 \%$
\citep{Komatsu10}.

Dark matter has a long history, being ``introduced'' by
\citep{Zwi33}: for solving the problems of high mass-to-light
ratios of galaxy clusters and of the rotation curves of spiral
galaxies.
Later on, several versions of the the so-called cold dark matter
model (CDM) have been built starting from the assumption that a
large amount of non-baryonic matter,  matter non-interacting with
the electromagnetic radiation but only detectable by its
gravitational interaction with visible matter, could account for
the observations in the framework of the standard Newtonian
dynamics. But even if its clustering and distribution properties
are fairly well known at every scale (see \citep{NFW96} for the most used model),
its nature is unknown, up to
now, at a fundamental level.

Dark energy has a more recent history: it was ``introduced'' about
ten years ago while reconstructing the Hubble diagram of Type Ia
Supernovae (SNeIa) observations, from which it was deduced that
the universe is now accelerating
\citep{Perlmutter99,Riess04,ast05,clo05}. While growing the
quantity of available cosmological data (measurements of cluster
properties as the mass, the correlation function and the evolution
with redshift of their abundance \citep{eke98,vnl02,bach03,bb03};
the already mentioned Hubble diagram of SNeIa; the optical surveys
of large scale structure \citep{pope04,cole05,eis05}; the
anisotropies in the cosmic microwave background
\citep{Boom,WMAP,Komatsu10}; the cosmic shear measured from weak
lensing surveys \citep{vW01,refr03} and the Lyman\,-\,$\alpha$
forest absorption \citep{chd99,mcd04}), more evidences towards a
spatially flat universe with a subcritical matter content and
undergoing a phase of accelerated
expansion have been collected.\\
Despite of the existence of universe acceleration has been clearly
undisclosed, the nature and the fundamental properties of the
underneath physical mechanism remain essentially unknown
notwithstanding the great theoretical efforts made up to now. By
\textit{simply} adding a constant to the dynamical equations of
universe, the cosmological constant $\Lambda$
\citep{CarLam,Sahni}, in the context of CDM model, it was defined
a new model, the $\Lambda$CDM, which quickly became the {\it
consensus model} because it provides a good fit to most of the
data \citep{Teg03,Sel04,sanch05} giving a reliable snapshot of the
today observed universe. Nevertheless, it is affected by serious
theoretical shortcomings that have motivated the search for more
general and alternative candidates generically referred to as dark
energy. Such models range from  scalar fields rolling down self
interaction potentials to phantom fields, from phenomenological
unified models of dark energy and dark matter to alternative
gravity theories \citep{Cap02,PB03,Pad03,Copeland06,CapFra}.

In the last three decades, scalar fields have played an important
role in both cosmology and particle physics \cite{Linde,Binetruy}.
Scalar fields have been postulated as means to explain the early
and late time acceleration of the universe. However, it is almost
always the case that such fields interact with standard matter:
either due to a direct Lagrangian coupling or indirectly through a
coupling to the Ricci Scalar or as the result of quantum loop
corrections. Both for inflation in the early universe and for dark
energy, such couplings can lead to problems. In inflation, for
example, couplings might destroy the flatness of the potential
needed to drive a period of inflation. If there are scalar fields
which permeate the universe today and have non-zero couplings to
matter, then they would induce an additional force in nature. If
the scalar field self-interactions are negligible, then the
experimental bounds on such a field are very strong: either the
couplings to matter are much smaller than gravity, or the scalar
fields are very heavy, so that they mediate a  short-ranged
interaction.

A certain class of theories have been proposed, in which the
scalar field properties depend on the environment: its mass
depends on the local environmental density. These are the so
called {\it Chameleon Field Theories}, proposed by \cite{Khoury04}, that employs a combination of
self-interaction and couplings to matter of the scalar field to
avoid the most restrictive of the current gravity bounds. In these
models a scalar field couples to matter with gravitational
strength, in harmony with general expectations from string theory,
whilst, at the same time, remaining relatively light on
cosmological scales. It was found that local gravity constraints
are (roughly) satisfied as long as the mass-scale of the potential
satisfies $M\lesssim (1mm)^{-1}$. This coincides with the scale
associated with the late time acceleration of the universe, and it
is surprising that it should come from local experiments.
Chameleon models have been subject to many studies, from
laboratory experiments up to cosmological probes
\citep{c1,c2,c3,c4,c5,c6,c7,c8,c9,c10,c11,c12,c13,c14,Mota:2008ne,c15}.
Lately, chameleons models where also studied in the context of
Modified Gravity, in particular the so called $f(R)$-gravity
\citep{fr1,fr2,fr3}. In such a case, good results of general
relativity,   according to the standard probes, are reproduced at
local scales while dark energy (accelerating behavior) and dark
matter (dynamical clustering) effects are reproduced at larger
scales. The major issue is to select suitable $f(R)$-models
capable of matching the density profiles at the various
gravitational scales, as discussed e.g. in \citep{fr3}.

Chameleon fields evade tight gravity constraints via the so called
chameleon mechanism: it consists on a field-generated  force, a
sort of {\it fifth force}, to become short-ranged in highly dense
regions, and long-ranged in low density regions. Such feature
would imply that at different astrophysical scales the fifth force
felt by matter would be suppressed or enhanced according to the
local astrophysical density.

In this work we investigate whether there is evidence for
a coupling between baryonic matter and a massive scalar field  which could mimic and replace
the contributions of a possible dark matter component.
In particular, motivated by chameleon models, we aim to verify if it is
possible to observationally detect a scalar field whose mass (or, equivalently,
interaction length) and coupling may change with scale, matching different astrophysical
observations. In this case, by astrophysical observations, we mean data from SNeIa, low surface
brightness (LSB) dwarf galaxies and, finally, clusters of galaxies. The range of scales
is very wide and we are adopting photometric and spectroscopic data
to probe the mechanism at different redshifts.\\

We want to stress here that we are just interested to investigate whether the data
would favor a model where baryons may be coupled to a scalar field whose mass
and coupling change with scale. If such a change is due to a chameleon
mechanism (associated to a local-density variation) or due to some other mechanism,
that is not the main concern in this article and so we will not compute specific predictions for any
particular model.

Chameleon models are highly non linear so that, in principle, no superposition principle for such
a non-linear field would be possible. Due to that, it is extremely difficult to
build the gravitational potential of an extended astrophysical system as the ones we analyse in this work.
Fortunately, we are not specifically investigating chameleon
models; we are here studying a general coupled scalar field model and asking
the data what are the preferred values of its mass and coupling  at different
scales. In order to compute an extended gravitational potential (in a
non-linear case as the chameleon model would be), we would need to properly
study the non-linear regime of structure formation within this type of models.
But this is well beyond the scope of this article. In spite of being of
utmost importance to set up N-body simulations with heavy scalar field models,
such task is extremely complex. In fact up to nowadays most of the N-body
simulations assume light scalar fields which do not cluster at small scales.
That is obviously not the case of chameleon fields and such investigation is
not considered in the present paper.


The article is organized as follows: In
\S~(\ref{sec:scalar_theory}) we give a brief but exhaustive
summary of all the main properties of the scalar field theory. In
\S~(\ref{sec:obs_data}) we accurately describe the used
astrophysical and cosmological data and the theoretical model
defined for any of them. In \S~(\ref{sec:results}) we expose our
results with a discussion on their implications for a more general
and comprehensive theory of gravity. Conclusions are drawn in
\S~(\ref{sec:conclusions})

\section{The Scalar field theory}
\label{sec:scalar_theory}

\begin{figure*}
\centering
  \includegraphics[width=80mm]{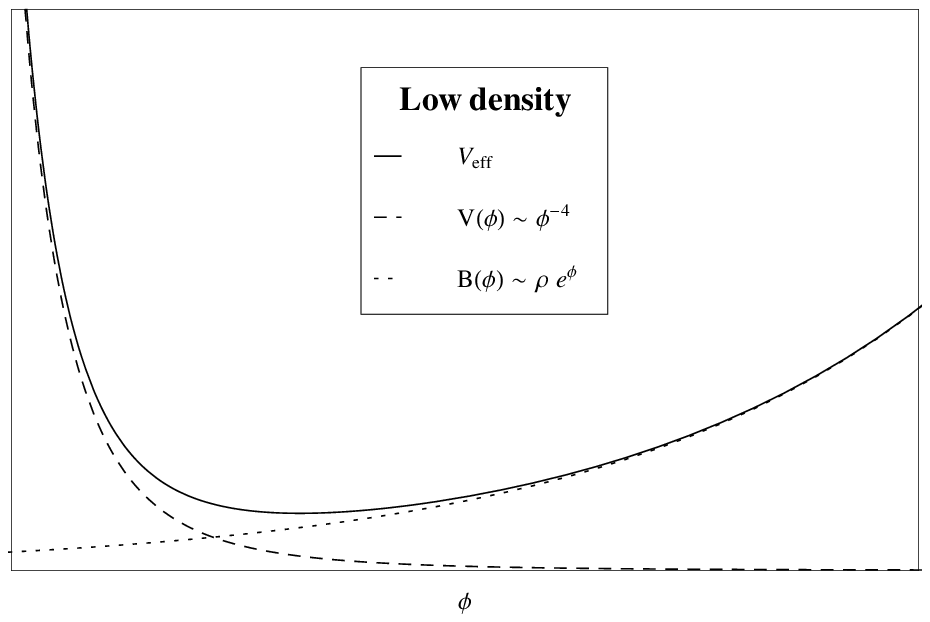}
  \includegraphics[width=80mm]{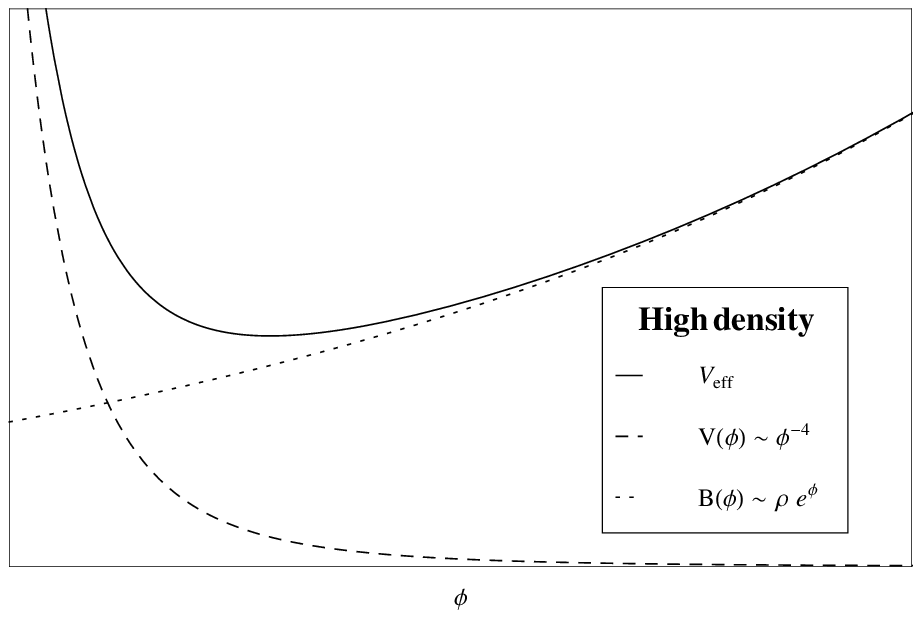}
  \caption{Chameleon potential in low and high local density environment.\label{fig:cham_pot}}
\end{figure*}

A general action governing the dynamics of the chameleon (scalar) field
$\phi$ can be of the form:
\begin{eqnarray}\label{eq:action_cham}
S &=& \int {\mathrm{d}}^{4} x \sqrt{-g} \left\{ \frac{M_{Pl}^{2}}{2} {\mathcal{R}}
- \frac{1}{2} (\partial \phi)^{2} - V(\phi) \right\} \nonumber \\
&-& \int {\mathrm{d}}^{4}x \, {\mathcal{L}}_{m}(\psi_{m}^{(i)},
g_{\mu \nu}^{i}) \; ,
\end{eqnarray}
where $g$ is the determinant of the metric $g_{\mu\nu}$,
$\mathcal{R}$ is the Ricci scalar, $\psi_{m}^{(i)}$ are the
various matter fields and ${\mathcal{L}}_{m}$ is the Lagrangian
density of ordinary matter.\\
In the expression for the reduced Planck Mass, $M_{Pl} \equiv (8
\pi G_{\ast})^{-1/2}$, $G_{\ast}$ is the bare gravitational
constant and differs from the usually measured one.\\
This can be better understood if we consider the most general case
of a scalar-tensor model, whose action (Eq.~\ref{eq:action_cham} is a particular case
of this) is \citep{Esposito01}:
\begin{eqnarray}\label{eq:scal-tens-action}
S &=& \frac{1}{16 \pi G_{\ast}} \int {\mathrm{d}}^{4}x \sqrt{-g} \left\{ F(\phi) {\mathcal{R}}
- Z(\phi) (\partial \phi)^{2} + \right. \nonumber \\
&-& \left. V(\phi)\right\} - \int d^{4}x \,
{\mathcal{L}}_{m}(\psi_{m}^{(i)}, g_{\mu \nu}^{i}) \; .
\end{eqnarray}
$f(R)$-gravity models too can  be enclosed  in  this general case
since it is straightforward to show that, by suitable
manipulations, they are a subclass of Eq.~(\ref{eq:scal-tens-action});
see for example \citep{CapFra}. From this action one can naively
define Newton's gravitational constant as:
\begin{equation}\label{eq:Geff_GF}
G_{N} \doteq \frac{G_{\ast}}{F} \, ,
\end{equation}
but $G_{N}$   has not  the same physical meaning as Newton's
gravitational constant in general relativity. The actual Newtonian
force measured in a Cavendish-type experiment between two test
masses will experience an effective coupling constant
\begin{equation}\label{eq:G_scal}
G_{eff} = \frac{G_{\ast}}{F} \left\{1+ \alpha(\phi)\right\} \doteq G_{N} \left\{1+ \alpha(\phi)\right\} \, ,
\end{equation}
where the term $G_{\ast}/F$ is due to the (average) exchange of
gravitons between the two bodies, while $G_{\ast}/F \cdot
\alpha(\phi)$ comes from the exchange of a scalar particle between
them, with the analytical expression for $\alpha$ depending on the
particular scalar theory one considers. It is also clear from this
expression, that what in general relativity is a \textit{true}
constant, now becomes a possible function of time and radius, so
the use of the term \textit{``constant"} is quite inappropriate.
When we will speak about a \textit{varying gravitational constant}
we always refer to the effective scalar field gravitational constant
expression.

The key ingredient in studying the cosmological dynamics of a
field, when it supports a chameleon mechanism, is that the scalar
field feels both a potential, $V(\phi)$, and a coupling to matter
depending on $\rho$, the local density of matter, and on the
coupling constant, $\beta$. At the end the field dynamics is
governed by an effective potential,
\begin{equation}\label{eq:V_cham}
V_{eff}(\phi) = V(\phi) + \sum_{i} \rho_{i} e^{\beta_{i} \phi /
M_{Pl}} \; .
\end{equation}
If $V(\phi)$ is a run-away potential and $\beta_{i}>0$, the
effective potential has a minimum at $\phi_{min}$ satisfying the
condition:
\begin{equation}
V_{,\phi}(\phi_{min}) + \sum_{i} \frac{\beta}{M_{Pl}} \rho_{i}
e^{\beta_{i} \phi / M_{Pl}} = 0 \; ,
\end{equation}
and the effective mass of the field of small perturbations about
the minimum is
\begin{equation}
m^2 = V_{,\phi\phi}^{eff} = V_{,\phi\phi}(\phi_{min}) + \sum_{i}
\frac{\beta^2}{M_{Pl}^2} \rho_{i} e^{\beta_{i} \phi / M_{Pl}} \; .
\end{equation}
So the effective potential depends on the matter density and both
the minimum in potential and the mass of the scalar field are
function of the local density, as shown in Fig.~(\ref{fig:cham_pot}).
As density increases, the minimum
in potential shifts to smaller values of the field and the mass of
small fluctuations increases. This last property, in particular,
makes the chameleon field able to satisfy the constraints from
laboratory tests of the principle of equivalence, because in high
densities environment, such as terrestrial laboratories, the field
can be heavy enough so to evade them.

It must be stressed, however, that even with such a mechanism, it
is very difficult to build a theory with a late time cosmology
observationally indistinguishable from the standard $\Lambda$CDM
model.

Another interesting consequence of this model comes out when
considering linear  perturbations of matter and their related
equation. In the most general case of a scalar-tensor theory \citep{Brax04}
we have:
\begin{equation}\label{eq:pert_eq}
\delta_{c}'' + a H \delta_{c}' = \frac{3}{2} a^2 H^2 \left[ 1 +
\frac{2 \beta^2}{1+a^2 V_{,\phi\phi}/k^2}\right] \; ,
\end{equation}
where $a$ is the scale factor, $H$ is the Hubble function,
$\delta_{c} = \delta(\rho_{m} e^{\beta \phi/M_{Pl}})$ is the
matter density contrast and, if the field is at the minimum, its
mass is $m^2 = V_{,\phi\phi}$. The quantity in brackets can be
interpreted as the expression for the effective gravitational
constant within the context of massive scalar field models \citep{Gannouji09}:
\begin{equation}\label{eq:G_cham}
G_{eff}(a; \beta, m; k) = G_{N} \left( 1 + 2 \beta^{2}
\frac{\frac{k^{2}}{a^{2} m^{2}}}{1 + \frac{k^{2}}{a^{2} m^{2}}}
\right) \; .
\end{equation}
In particular the term proportional to $m^2$ results from the
scalar field-mediated force, which is negligible if the physical
length scale of the perturbation is much larger than the range of
the chameleon-mediated force, namely, if $a/k \gg m^{-1}$. In this
case the left hand side of Eq.~(\ref{eq:pert_eq}) is well
approximated by $3 a^{2} H^{2} \delta_{c}/2$ and the matter
fluctuations grow as in general relativity.

Anyway chameleon theories (i.e. based on Eq.~\ref{eq:V_cham}) do not behave like linear theories (as anticipated
in the introductory section) of massive scalar fields when massive bodies are involved.
Varying the action Eq.~(\ref{eq:action_cham}) with respect to the chameleon (scalar)
field $\phi$ in a spherically symmetric spacetime gives:
\begin{equation}
\frac{{\mathrm{d}}^2 \phi}{{\mathrm{d}} r^2} + \frac{2}{r}
\frac{{\mathrm{d}} \phi}{{\mathrm{d}} r} = \frac{{\mathrm{d}} V_{eff}}{{\mathrm{d}} \phi} \; ,
\end{equation}
because of the effective potential for $\phi$ changes in different
density environments, this differential equation is highly non-linear.
These non-linear features have been studied in \citep{Khoury04},
where it was found that for a spherically symmetric object of mass
$M_{c}$ and radius $R_{c}$ surrounded by a gas of asymptotic density
$\rho_{\infty}$, the profile of the field is governed by the so-called
``thin-shell'' parameter,
\begin{equation}
\Delta = \frac{\mid \phi_{min}^{\infty} - \phi_{min}^{c} \mid}{6 \beta M_{Pl} \Phi_{c}}
\end{equation}
where $\phi_{min}^{\infty}$ and $\phi_{min}^{c}$ are the minima of the
effective potential outside and inside the object respectively and
$\Phi_{c} = M_{c}/8 \pi M_{Pl}^2 R_{c}$ is the Newtonian potential
at the surface of the object. Thus $\Delta$ is the ratio of the
difference in $\phi$ potential to the Newtonian potential and
quantifies how perturbing the object is for the $\phi$ field.
If $\Delta$ is large, which happens for small objects, then the
external (to the object) profile of $\phi$ is the usual Yukawa profile,
\begin{equation}
\phi(r) = - \left( \frac{\beta}{4 \pi M_{Pl}} \right) \frac{M_{c} \exp^{-m_{\infty}}}{r} + \phi_{min}^{\infty} \; ,
\end{equation}
where $m_{\infty}$ is the mass of the scalar field. For large and
compact objects, $\Delta$ is small and the Yukawa profile is
suppressed by a factor of $\Delta$. Thus the term ``thin-shell''
comes from the fact that only a portion of such a ``thin-shell''
contributes to the external Yukawa profile. A discussion of this
issue in the framework of $f(R)$-gravity is in \citep{fr3}.\\

In summary the  main ingredients  of our model are:
\begin{itemize}
 \item a massive scalar field coupled with ordinary observable
       baryonic mass;
 \item its mass $m$, or its interaction length $L \propto m^{-1}$;
 \item its coupling constant with baryonic mass $\beta$
\end{itemize}

\subsection{Modified gravitational potential}
\label{sec:modified_potential}

Taking the inverse Fourier transform of Eq.~(\ref{eq:G_cham}) it
is straightforward to obtain the corresponding expression of the
gravitational potential for a point mass distribution, $\psi(r)$.
Remembering that a potential $\propto \frac{1}{r}$ in real space
yields a $k^{-2}$ term in Fourier space, we can recognize in
Eq.~(\ref{eq:G_cham}) the point-like gravitational potential per
unit mass:
\begin{eqnarray}\label{eq:grav_pot_point}
\psi (r) &=& -\frac{G}{r} \left( 1 + 2 \beta^2 e^{-m r}\right) = \nonumber \\
&=& -\frac{G}{r} \left( 1 + 2 \beta^2 e^{-r/L}\right) \; ,
\end{eqnarray}
where $m$ is the mass of the scalar field, $L \propto m^{-1}$
is the interaction range of the modified gravitational potential,
i.e. the length where the scalar field is effective, and
$\beta$ still being the coupling constant between matter and the
scalar field. The gravitational potential given in
Eq.~(\ref{eq:grav_pot_point}) is a point-like one, so that we have
to generalize it to extended systems if we want to use it for
clusters of galaxies and LSB galaxies. As we will discuss later,
we are going to model galaxy clusters as spherically symmetric
systems: we simply consider the system composed by many
infinitesimal mass elements each one contributing with a
point-like gravitational potential. Then, summing up all terms,
namely integrating them on a spherical volume, we obtain a
suitable potential. Specifically, we have to solve the integral:
\begin{equation}
\Psi(r) = \int_{0}^{\infty} r'^{2} dr' \int_{0}^{\pi} \sin \theta'
d\theta' \int_{0}^{2\pi} d\omega' \psi(r') \; .
\end{equation}
We make explicit that with an abuse of notation we are writing
inside the point-like potential $r'$, while it should be replaced
by $|\vec{\mathbf{x}}-\vec{\mathbf{x}}'| = (r^{2}+r'^{2}-2rr' \cos
\theta')^{1/2}$.

The point-like potential can be split in two terms. The
\textit{Newtonian} component is:
\begin{equation}
\psi_{N}(r) = -\frac{G M}{r} \; ,
\end{equation}
and its extended integral is the well-known expression:
\begin{equation}
\Psi_{N}(r) = -\frac{G M(<r)}{r} \; ,
\end{equation}
where $M(<r)$ is the mass enclosed in a sphere with radius $r$.
The \textit{correction} term coming from the scalar field is:
\begin{equation}
\psi_{C}(r) = -\frac{G M}{r} \left(2 \beta^2 e^{-\frac{r}{L}}
\right) \; ;
\end{equation}
from the integration of the angular part, we have:
\begin{eqnarray}
\Psi_{C}(r) &=& - 2 \pi G \, (2 \beta^2 L) \int_{0}^{\infty}
{\mathrm{d}}r' r' \rho(r') \cdot \nonumber \\
&\cdot& \frac{e^{-\frac{|r-r'|}{L}} -
e^{-\frac{|r+r'|}{L}}}{r}
\end{eqnarray}
The radial integral is  numerically estimated once the mass
density is given. A fundamental difference between such a term and
the Newtonian one is that while in the latter the matter outside
the spherical shell of radius $r$ does not contribute to the
potential, in the former external matter takes part to the
integration procedure, even if its contribution is really
negligible in most cases. \\
At the end, the total potential of the spherical mass distribution
will be:
\begin{equation}\label{eq:total corrected potential}
\Psi(r) = \Psi_{N}(r) + \Psi_{C}(r) \; .
\end{equation}
As we will show below, for our purpose we need the gravitational
potential derivative with respect to the variable $r$; this may
not be evaluated analytically so we estimate it
numerically, once we have given an expression for the mass density
$\rho(r)$. While the Newtonian term gives the simple expression:
\begin{equation}
\frac{{\mathrm{d}}\Phi_{N}}{{\mathrm{d}}r}(r) = \frac{G
M(<r)}{r^{2}} \; ,
\end{equation}
the derivative of the corrective potential term is more involved.
We do not give it explicitly for sake of brevity, but only
remind that it is an integral-function of the form
\begin{equation}
{{\mathcal{F}}}(r, r') = \int_{\alpha(r)}^{\beta(r)} dr' \ f(r,r')
\; ;
\end{equation}
from it one has: {\setlength\arraycolsep{0.2pt}
\begin{eqnarray}
\frac{{\mathrm{d}}{\mathcal{F}}(r, r')}{\mathrm{d}r} &=&
\int_{\alpha(r)}^{\beta(r)} dr'
\frac{{\mathrm{d}}f(r,r')}{{\mathrm{d}}r} - f(r,\alpha(r))
\frac{{\mathrm{d}}\alpha}{{\mathrm{d}}r}(r)+ \nonumber \\
&+& f(r,\beta(r)) \frac{{\mathrm{d}}\beta}{{\mathrm{d}}r}(r) \; .
\end{eqnarray}}
Such an expression is numerically derived once the integration
extremes are given.\\
For spiral galaxies, we have the same theoretical apparatus but a
different geometric configuration. Matter in spiral galaxies is
generally modeled as distributed in a thin axis-symmetric disk,
so that the extended gravitational potential is given by:
\begin{equation}
\Psi(r, z) = \int_{0}^{\infty} R' {\mathrm{d}}R' \int_{0}^{2\pi}
{\mathrm{d}}\omega' \hspace{0.1cm} \psi(R',z')\,.
\end{equation}
Even in this case, for being more exact, the couple of variables
$(R',z')$ inside the point-like potential should be replaced by
$|\vec{\mathbf{x}}-\vec{\mathbf{x}}'| = (R^{2}+R'^{2}-2RR' \cos
\omega'+(z-z')^2)^{1/2}$. Once the gravitational potential is
given, the rotation curve for the disk can be easily computed
starting from the relation \citep{Binney}:
\begin{equation}\label{eq:rot_vel}
v_{c}^2 = R \frac{{\mathrm{d}} \Psi(R,z)}{{\mathrm{d}}
R}\bigg{|}_{z=0} \; .
\end{equation}

It is important to underline that both in the clusters of
galaxies case and in the LSB galaxies one, we have to perform
derivatives with respect of the distance from the center
of any system, $r$, of the numerically derived gravitational
potential. To be completely rigorous we should add a term to all the previously
written relations, coming from the derivative of the \textit{function}
$\beta(\rho) \sim \beta(r)$. In our scalar-field approach we have treated the coupling parameter as a
\textit{constant} at all, while in the definition of the chameleon mechanism it is
also possible and not trivial that it may depend on the local density of the gravitational
system one is going to consider. At the same time, we do not know what is
the possible analytical behavior of this quantity or, better: it is
one of our purposes trying to reconstruct it. It is also evident that
if we want to use previous relations in the form we have shown before,
we are implicitly assuming that $\beta$ can satisfy two different
scenarios: \textit{a.} it is really constant, and in this case one expects
not to detect any change in it when comparing different gravitational
scales; or (as we will shown to be our case) \textit{b.} it is a function
of the gravitational scale, but its derivative is supposed to be
negligible, i.e. ${\mathrm{d}}\beta/{\mathrm{d}}r \approx 0$.

\subsection{Modified distance modulus}
\label{sec:modified_distance}

In \citep{Brax04} the Friedmann equation is derived from the action governing
the dynamics of the chameleon field $\phi$ in the Jordan frame:
\begin{equation}\label{eq:fried_cham}
3 H^{2} M_{Pl}^{2} = \rho_{m} e^{\beta \phi / M_{Pl}} +
\frac{1}{2} \dot{\phi}^{2} + V(\phi) + \rho_{r} \, ,
\end{equation}
with contribution from matter, radiation and the scalar field.
Making explicit the expression for the Planck mass, Eq.~(\ref{eq:fried_cham}) becomes:
{\setlength\arraycolsep{0.2pt}
\begin{equation}\label{eq:fried_gannouji}
3 H^{2} = 8 \pi G_{\ast} \left[ \rho_{m} e^{\beta \phi / M_{Pl}} +
\left( \frac{1}{2} \dot{\phi}^{2} + V(\phi) \right) + \rho_{r} \right] \; .
\end{equation}}
We can show that this equation in the chameleon case easily converts in
the most general expression for a given scalar field  $\phi$.\\
If we assume the chameleon field $\phi$ is in the minimum of the
effective potential from the early stages of the universe on, then
we have $\frac{\phi}{M_{PL}} \ll 1$ \citep{Gannouji09} during the
subsequent evolution until today. This also means that $e^{\beta
\phi / M_{PL}} = 1$ to very high accuracy so it will disappear
from equations and does not have to be considered here. Then,
considering that the function $F$ appearing in Eq.~(\ref{eq:Geff_GF})
is equal to unity in the scalar field case, we also have $G_{N} =
G_{\ast}$. At the end the Friedmann equation is completely equal
to the usual expression: {\setlength\arraycolsep{0.2pt}
\begin{equation}
3 H^{2} = 8 \pi G_{N} \left[\rho_{m} + \left( \frac{1}{2}
\dot{\phi}^{2} + V(\phi) \right) + \rho_{r} \right],
\end{equation}}
and finally, avoiding radiation contribution while suffix $Sc$ corresponds
to the scalar field:
\begin{equation}
h^{2}(z) = \Omega_{m,0} (1 + z)^{3} + \Omega_{Sc,0} \epsilon(z) \; ,
\end{equation}
where:
{\setlength\arraycolsep{0.2pt}
\begin{eqnarray}
h(z)            & \doteq & \frac{H(z)}{H_{0}} \nonumber \\
\Omega_{m,0}    & \doteq & \frac{8 \pi G_{N}}{3 H_{0}^{2}} \rho_{m,0} \\
\Omega_{Sc,0}    & \doteq & \frac{8 \pi G_{N}}{3 H_{0}^{2}}
\left(\frac{1}{2} \dot{\phi}^{2} + V{\phi} \right)\Bigg{|}_{z=0} =
\frac{8 \pi G_{N}}{3 H_{0}^{2}} \rho_{Sc,0} \nonumber
\end{eqnarray}}
The function $\epsilon(z)$ is unknown; but one knows (assumes) that the
scalar field works like a cosmological constant on cosmological scales, so we may choose it
to be constant in redshift or one can use more general and
extended models as the Chevallier-Polarski-Linder (CPL)
parametrization \citep{Chevallier01,Linder03} usually used to
\textit{phenomenologically} describe dark energy fluids.

Even if we do not have any possibility to discriminate between
$\Lambda$CDM and a scalar field scenario only by using $h(z)$, we
have a discriminating tool in the distance modulus, the main
observable quantity derivable from SNeIa observations, modified
from the usual expression by assuming that the gravitational constant can
vary with time:
\begin{eqnarray}\label{eq:mod_dist_cham}
\mu(z; \beta, m; k) &=& 5 \log \left( (1+z) \int_{0}^{z}
\frac{{\mathrm{d}}z}{h(z)}\right) + \mu_{0} \nonumber \\
&+& \frac{15}{4} \log \frac{G_{eff}(z; \beta, m; k)}{G_{eff}(0; \beta, m; k)} \, .
\end{eqnarray}
In this expression there is an additional term made of with the
ratio between the value of effective gravitational constant at any
redshift and the same quantity evaluated at the present ($z=0$). As
accurately described in \citep{Riazuelo02}, a time-varying
gravitational constant can affect light curves from SNeIa by
changing both the thermonuclear energy release, since the
luminosity at the maximum in the light curve is proportional to
the mass of synthesized nickel, and the time scale of stellar
explosion. This means that by using Eq.~(\ref{eq:mod_dist_cham})
we are going to test the scalar field mechanism on cosmological scales,
in particular considering the consequent role of a possible variation of the effective gravitational
constant in the Universe acceleration rate history.

In the scalar field theory one has an analytical expression for the
effective gravitational constant, i.e. Eq.~(\ref{eq:G_cham}),
which we have modified in the
following one for uniforming all the results:
\begin{equation}\label{eq:G_sn}
G_{eff}(z; \beta, L; \lambda) = G_{N} \left( 1 + 2 \beta^{2}
\frac{(1+z)^2 L^2}{(1+z)^2 L^2 + \lambda^2} \right) \; ,
\end{equation}
which depends on the following variables:
\begin{itemize}
   \item the redshift $z$ ($a = 1/(1+z)$);
   \item the wavelength $k$ (or the length $\lambda \propto k^{-1}$), which one could fix or vary on a grid;
   \item the intrinsic model parameters, i.e., the coupling constant $\beta$ and the interaction length $L \propto
         m^{-1}$, which can be constrained with a fitting procedure.
\end{itemize}

\section{Observational data}
\label{sec:obs_data}


We are going to test the scalar field mechanism on three
different scale ranges:
\begin{itemize}
 \item on cosmological scales, by means of supernovae luminosity distance;
 \item on a Mpc-astrophysical scale, using mass profiles of clusters of galaxies;
 \item on a kpc-astrophysical scale, analyzing rotation curves from spiral galaxies.
\end{itemize}
For any of them, we have found out the necessary sample data in literature.

\subsection{Cosmological scale: Supernovae}
\label{sec:SNdata}

SNeIa are useful because of the possibility to easily modify the expression of distance modulus
for more general theories with a varying gravitational constant
(as in the scalar field case). Moreover, the distance modulus is the
main observable quantity derivable from this kind of astrophysical
objects. Adding to that, the possibility of using data ranging up to
redshift values much larger than those ones from galaxies or clusters
of galaxies ($z \approx 2$) makes us possible to test and verify a
possible temporal variation of the gravitational constant (if
there is any) and so a possible alternative gravity scenario. It
is interesting to underline that with the modified expression of
the distance modulus we can also verify the coexistence, at the
same time, of both \textit{dark energy} and \textit{dark matter},
both explained as different consequences on different scales of
the same unified context, namely, the scalar field. In
fact, in the expression for the distance modulus we are going to
describe in next sections, we employ both a term with a
\textit{dark energy-modeled fluid}, coming out nothing else that
from the effective behavior of the scalar field on cosmological
scales (which does not differ much from the cosmological constant
behavior) and a term acting as a \textit{dark matter-modeled
component}, coming out from the scalar field working on
gravitational scales smaller than the cosmological one.
In this sense, we have also explored how the scalar field process can
mimic dark matter profiles in clusters and spiral galaxies.

We use the \textit{Constitution} sample described in \citep{Hicken09}, which is a data set obtained by combining
the Union data set by \citep{Kowalski08} with new $90$ nearby
objects from the CfA3 release described in \citep{Hicken09A}.

The Union SNeIa compilation is a data set of low-redshift
nearby-Hubble-flow SNeIa and is built with new analysis procedures
for working with several heterogeneous SNeIa compilations. It
includes $13$ independent sets with SNe from the SCP, High-z
Supernovae Search (HZSNS) team, Supernovae Legacy Survey and
ESSENCE Survey, the older data sets, as well as the recently
extended data set of distant supernovae observed with HST. After
various selection cuts were applied in order to create a
homogeneous and high-signal-to-noise data set, we have final $307$
SNeIa events distributed over the redshift interval $0.15 \leq z
\leq 1.55$.

The CfA3 data set is originally made of 185 multi-band optical
SNeIa light curves taken at the F.L. Whipple Observatory of the
Harvard-Smithsonian Center for Astrophysics (CfA); 90 of the
original 185 objects pass the quality cuts of \citep{Kowalski08}
and are added to the Union data set to form the Constitution
one.

The statistical analysis of Constitution SNeIa sample rests on the
definition of the distance modulus given in
Eq.~(\ref{eq:mod_dist_cham}). The best fits were obtained by
minimizing the quantity
\begin{equation}\label{eq: sn_chi}
\chi^{2}_{\mathrm{SN}}(\mu_{0}, \lambda, \{\theta_{i}\}) = \sum^{{\mathcal{N}}}_{j =
1} \frac{(\mu(z_{j}; \mu_{0}, \lambda, \{\theta_{i})\} -
\mu_{obs}(z_{j}))^{2}}{\sigma^{2}_{\mathrm{\mu}}(z_{j})}
\end{equation}
where ${\mathcal{N}}=397$ is the number of observed SNeIa, $\mu$
is the distance modulus (the observed, $\mu_{obs}$, and the
theoretical one, $\mu(z_{j}; \mu_{0}, \lambda, \{\theta_{i})$),
$\sigma^{2}_{\mathrm{\mu}}$ are the measurement variances and
$\{\theta_{i}\} = \{\beta, L\}$ is the parameters theory vector.
The nuisance parameter $\mu_{0}$ encodes the Hubble parameter and
the absolute magnitude $M$, and has to be marginalized over.
Giving the heterogeneous origin of the Constitution data set, and
the procedures described in \citep{Kowalski08} and \citep{Hicken09}
for reducing data, we have worked with an alternative version
Eq.~(\ref{eq: sn_chi}), which consists in minimizing the quantity
\begin{equation}\label{eq: sn_chi_mod}
\tilde{\chi}^{2}_{\mathrm{SN}}(\{\theta_{i}\}) = c_{1} -
\frac{c^{2}_{2}}{c_{3}}
\end{equation}
with respect to the other parameters. Here
\begin{equation}
c_{1} = \sum^{{\mathcal{N}}}_{j = 1} \frac{(\mu(z_{j}; \mu_{0}=0,
\{\theta_{i})\} -
\mu_{obs}(z_{j}))^{2}}{\sigma^{2}_{\mathrm{\mu}}(z_{j})}\, ,
\end{equation}
\begin{equation}
c_{2} = \sum^{{\mathcal{N}}}_{j = 1} \frac{(\mu(z_{j}; \mu_{0}=0,
\{\theta_{i})\} -
\mu_{obs}(z_{j}))}{\sigma^{2}_{\mathrm{\mu}}(z_{j})}\, ,
\end{equation}
\begin{equation}
c_{3} = \sum^{{\mathcal{N}}}_{j = 1}
\frac{1}{\sigma^{2}_{\mathrm{\mu}}(z_{j})}\,.
\end{equation}
It is trivial to see that $\tilde{\chi}^{2}_{SN}$ is just a
version of $\chi^{2}_{SN}$, minimized with respect to $\mu_{0}$.
To that end it suffices to notice that
\begin{equation}
\chi^{2}_{\mathrm{SN}}(\mu_{0}, \lambda, \{\theta_{i}\}) = c_{1} - 2 c_{2}
\mu_{0} + c_{3} \mu^{2}_{0} \,
\end{equation}
which clearly becomes minimum for $\mu_{0} = c_{2}/c_{3}$, and so
we can see $\tilde{\chi}^{2}_{\mathrm{SN}} \equiv
\chi^{2}_{\mathrm{SN}}(\mu_{0} = 0, \lambda, \{\theta_{i}\})$. Furthermore,
one can check that the difference between $\chi^{2}_{SN}$ and
$\tilde{\chi}^{2}_{SN}$ is negligible.

We minimize the $\chi$-square using the Markov Chains Monte Carlo
Method (MCMC) and testing their convergence with the method
described by \citep{Dunkley05}. The $i\sigma$ confidence levels are
easily estimated deriving them from the final samples, using the
$15.87$-th and $84.13$-th quartiles (which define the $68\%$
confidence interval) for $i=1$, the $2.28$-th and $97.72$-th
quartiles (which define the $95\%$ confidence interval) for $i=2$
and the $0.13$-th and $99.87$-th quantiles (which define the
$99\%$ confidence interval) for $i=3$.

\subsection{Galaxy Cluster Sample}
\label{sec:cluster}

Clusters of galaxies are uniquely useful tracer of cosmological
evolution and so ineludible tests in the field of alternative
gravities other than general relativity \citep{Voit05}. They are
fundamental tracers for two main features. First of all, they are
the largest gravitational objects whose masses can be adequately
measured, and the largest objects to have undergone gravitational
relaxation and entered into virial equilibrium. Second, clusters
are essentially ``closed boxes'' that retain all their gaseous
mass content because their gravitational wells are much deep. The
most accepted paradigm is that clusters of galaxies are mostly
made of collisionless cold dark matter particles (CDM model) and
are virialized systems from scale-free Gaussian initial density
perturbations. The CDM paradigm and the numerical simulations make
clear predictions for the structure of clusters of galaxies;
comparisons of these predictions with the results of high-quality
observations is a necessary consistent check and any significant
deviation can place important constraints on their theoretical
model but also on cosmological models and, as in our case, on the
exploration of different gravity theories.

The formalism described in \S~\ref{sec:modified_potential} can be
applied to a sample of $12$ galaxy clusters. We shall use the
cluster sample studied in \citep{Vik05,Vik06} which consists of
$13$ low-redshift clusters spanning a temperature range $0.7\div
9.0\ {\rm keV}$ derived from high quality {\it Chandra} archival
data. In all these clusters, the surface brightness and the gas
temperature profiles are measured out to large radii, so that mass
estimates can be extended up to $r_{500}$ or beyond.

Clusters of galaxies are generally considered self-bound
gravitational systems with spherical symmetry and in hydrostatic
equilibrium if virialized. The last two hypotheses are still
widely used, despite of the fact that it has been widely proved
that most clusters show more complex morphologies and/or signs of
strong interactions or dynamical
activity, especially in their innermost regions \citep{Chak08,DeFil05}. \\
Under the hypothesis of  spherical symmetry in hydrostatic
equilibrium, the structure equation can be derived from the
collisionless Boltzmann equation
\begin{eqnarray}\label{Boltzmann equation}
\frac{{\mathrm{d}}}{{\mathrm{d}}r}(\rho_{gas}(r)
\sigma^{2}_{r}) &+&
\frac{2\rho_{gas}(r)}{r}(\sigma^{2}_{r}-\sigma^{2}_{\theta,\omega})
= \nonumber \\
&=& -\rho_{gas}(r) \cdot \frac{{\mathrm{d}}\Psi(r)}{{\mathrm{d}}r}
\end{eqnarray}
where $\Psi$ is the gravitational potential of the cluster,
$\sigma_{r}$ and $\sigma_{\theta,\omega}$ are the mass-weighted
velocity dispersions in the radial and tangential directions
respectively, and $\rho$ is the gas-mass density. For an isotropic
system, it is
\begin{equation}\label{velocity dispersion}
\sigma_{r} = \sigma_{\theta,\omega} \; ;
\end{equation}
the pressure profile can be related to these quantities by
\begin{equation}\label{pressure}
P(r) = \sigma^{2}_{r} \cdot \rho_{gas}(r) \; .
\end{equation}
Substituting Eqs.~(\ref{velocity dispersion})~-~(\ref{pressure})
into Eq.~(\ref{Boltzmann equation}), we have, for an isotropic
sphere,
\begin{equation}\label{isotropic sphere}
\frac{{\mathrm{d}} P(r)}{{\mathrm{d}}r} = - \rho_{gas}(r)
\frac{{\mathrm{d}}\Psi(r)}{{\mathrm{d}}r} \; .
\end{equation}
For a gas sphere with temperature profile $T(r)$, the velocity
dispersion becomes
\begin{equation}\label{temperature}
\sigma^{2}_{r} = \frac{k T(r)}{\mu m_{p}} \; ,
\end{equation}
where $k$ is the Boltzmann constant, $\mu \approx 0.609$ is the
mean mass particle and $m_{p}$ is the proton mass. Substituting
Eqs.~(\ref{pressure})~-~(\ref{temperature}) into
Eq.~(\ref{isotropic sphere}), we obtain
\[
\frac{{\mathrm{d}}}{{\mathrm{d}}r} \left( \frac{k T(r)}{\mu m_{p}}
\rho_{gas}(r) \right) = -\rho_{gas}(r) \frac{{\mathrm{d}}
\Psi(r)}{{\mathrm{d}}r} \; ,
\]
or, equivalently,
\begin{equation}\label{eq:Boltzmann potential}
-\frac{{\mathrm{d}}\Psi(r)}{{\mathrm{d}}r} = \frac{k T(r)}{\mu m_{p}
r}\left[\frac{{\mathrm{d}}\ln\rho_{gas}(r)}{{\mathrm{d}}\ln r} +
\frac{{\mathrm{d}}\ln T(r)}{{\mathrm{d}}\ln r}\right] \; .
\end{equation}
Now the total gravitational potential of the cluster is:
\begin{equation}\label{eq:total corrected potential1}
\Psi(r) = \Psi_{N}(r) + \Psi_{C}(r) \; .
\end{equation}
It is worth underlining that if we consider \textit{only} the
standard Newtonian  potential and its derivative in
Eq.~(\ref{eq:Boltzmann potential}), the \textit{total} cluster
mass $M_{cl,N}(r)$ (the standard estimation of clusters mass in a
CDM scenario) is composed by gas mass $+$ galaxies mass $+$ dark
matter and it is given by the expression:
{\setlength\arraycolsep{0.2pt}
\begin{eqnarray}
\label{eq:M_tot} M_{cl,N}(r) &=& M_{gas}(r) + M_{gal}(r) +
M_{DM}(r) = \nonumber
\\
&=& - \frac{k T(r)}{\mu m_{p} G} r
\left[\frac{{\mathrm{d}}\ln\rho_{gas}(r)}{{\mathrm{d}}\ln
r}+\frac{{\mathrm{d}}\ln T(r)}{{\mathrm{d}}\ln r}\right] \; .
\end{eqnarray}}
Generally the galaxy contribution is considered negligible with
respect to the other two components so we have:
\[
M_{cl,N}(r) \approx M_{gas}(r) + M_{DM}(r) \approx
\]
\[
\hspace{1.35cm} \approx - \frac{k T(r)}{\mu m_{p}} r
\left[\frac{{\mathrm{d}}\ln\rho_{gas}(r)}{{\mathrm{d}}\ln
r}+\frac{{\mathrm{d}}\ln T(r)}{{\mathrm{d}}\ln r}\right] \; .
\]
Since the gas-mass estimates are provided by X-ray observations, the
equilibrium equation can be used to derive the amount of dark
matter and to reconstruct its spatial distribution in a cluster of galaxies.

Inserting the previously defined \textit{extended-corrected}
potential of Eq.~(\ref{eq:total corrected potential1}) into
Eq.~(\ref{eq:Boltzmann potential}), we obtain:
\begin{equation}
\label{eq:corrected_mass} -\frac{\mathrm{d}\Psi_{N}}{\mathrm{d}r}
-\frac{\mathrm{d}\Psi_{C}}{\mathrm{d}r} =\frac{k T(r)}{\mu m_{p}
r}\left[\frac{\mathrm{d}\ln\rho_{gas}(r)}{\mathrm{d}\ln r} +
\frac{\mathrm{d}\ln T(r)}{\mathrm{d}\ln r}\right] \; ,
\end{equation}
from which the \textit{extended-corrected} mass estimate follows:
{\setlength\arraycolsep{0.2pt}
\begin{eqnarray}\label{eq:fit relation}
M_{cl,EC}(r) &+& \frac{r^{2}}{G}
\frac{\mathrm{d}\Psi_{C}(r)}{\mathrm{d}r} = \\ &=&
 - \frac{k T(r)}{\mu m_{p}G} r
\left(\frac{{\mathrm{d}}\ln\rho_{gas}(r)}{{\mathrm{d}}\ln
r}+\frac{{\mathrm{d}}\ln T(r)}{{\mathrm{d}}\ln r}\right) \nonumber \; .
\end{eqnarray}}
Since the use of a corrected potential avoids, in principle, the additional
requirement of dark matter, the total cluster mass, in this case,
is given by only the baryonic matter counterparts:
\begin{equation}
M_{cl,EC}(r) = M_{gas}(r) + M_{gal}(r) \; ,
\end{equation}
that can be entirely evaluated by observational data. The mass density in the $\Psi_{C}$ term is
\begin{equation}
\rho_{cl,EC}(r) = \rho_{gas}(r) + \rho_{gal}(r) \; ,
\end{equation}
with the  density components derived from observations.
Considering that the right term in Eq.~(\ref{eq:fit relation}) is
the total Newtonian mass estimation for a cluster of galaxies, we
easily derive that the corrective term in the gravitational
potential works in mimicking an \textit{effective} dark matter
contribution:
\begin{equation}\label{eq:mass_contr}
\frac{r^{2}}{G} \frac{\mathrm{d}\Psi_{C}}{\mathrm{d}r}(r) =
M_{cl,N}(r) - M_{cl,EC}(r) \; .
\end{equation}
But in our approach, instead of requiring new kind of particles,
it arises by the interaction of baryonic matter with the scalar field
scalar field.

We have hence performed a best-fit analysis of the theoretical
estimation of dark matter, Eq.~(\ref{eq:mass_contr})
{\setlength\arraycolsep{0.2pt}
\begin{equation}\label{eq:theo_dark}
M_{dm,th}(r; \beta, L) \doteq M_{eff}(r; \beta, L) =
\frac{r^{2}}{G} \frac{\mathrm{d}\Psi_{C}(r)}{\mathrm{d}r}(r)
\end{equation}}
which depends on scalar field parameters through the potential
$\Psi_{C}$, versus the same but observationally-derived quantity,
\begin{equation}\label{eq:obs_dark}
M_{dm,obs}(r) = M_{cl,N}(r) - M_{cl,EC}(r) \; .
\end{equation}
We underline here that we could not fit directly the observed
baryonic mass because of the great difference in order of
magnitude between $M_{cl,N}(r)$ and $M_{cl,EC}(r)$, working this
last one like a \textit{small perturbation} to the total mass
estimation. It is clear that the term corresponding to our
theoretical derived dark matter quantity is much bigger than the
baryonic contribution, and even a small and acceptable deviation
of only $1 \%$ in it would have been translated in a larger
deviation of $10 \%$ in the baryonic one.\\
Since not all the data involved in the above estimations have
measurable errors, we cannot perform an \textit{exact}
$\chi$-square minimization. Actually, we can minimize the
quantity:
\begin{equation}
\chi_{Cl}^{2}(\{\theta_{i}\}) = \frac{1}{{\mathcal{N}}-n_{p}-1} \cdot \sum_{i=1}^{{\mathcal{N}}}
\frac{(M_{dm,obs}^{i}-M_{dm,th}^{i}(\{\theta_{i}\})^{2}}{M_{dm,th}^{i}(\{\theta_{i}\})}
\end{equation}
where ${\mathcal{N}}$ is the number of data and $n_{p} = 2$ is the
free parameters number of the model and $\{\theta_{i}\} = \{\beta,
L\}$. As usual we find the minimum in $\chi$-square running MCMCs.
Even if the convergence is achieved after few thousand steps of
the chain, we have decided to run longer chains of $10^{5}$ points
to reduce the noise from the histograms and avoid under- or over-
estimations of errors on the parameters.

\subsubsection{Gas Density Model}
\label{sec:gas_model}

The gas density distribution of the clusters in the
sample is described by the analytic model proposed in~\citep{Vik06}. Such a model
modifies the classical $\beta-$model to represent the characteristic
properties of the observed X-ray surface brightness profiles, i.e.
the power-law-type cusps of gas density in the cluster center,
instead of a flat core and the steepening of the
brightness profiles at large radii. Eventually, a second $\beta-$model,
with a small core radius, is added to improve  the model
close to the cluster cores. The  analytical
form for the particle emission  is given by:
{\setlength\arraycolsep{0.2pt}
\begin{eqnarray}
\label{gas density vik} n_{p}n_{e} &=& n_{0}^{2} \cdot
\frac{(r/r_{c})^{-\alpha}}{(1+r^{2}/r_{c}^{2})^{3\beta-\alpha/2}}
\cdot \frac{1}{(1+r^{\gamma}/r_{s}^{\gamma})^{\epsilon/\gamma}}+
\nonumber \\
&+& \frac{n_{02}^{2}}{(1+r^{2}/r_{c2}^{2})^{3\beta_{2}}} \; ,
\end{eqnarray}}
which can be easily converted to a mass density using the relation:
\begin{equation}
\label{eq:gas_density} \rho_{gas} = n_T \cdot \mu m_p =
\frac{1.4}{1.2} n_e m_p \; ,
\end{equation}
where $n_T$ is the total number density of particles in the gas.
The resulting model has a large number of parameters, some of
which do not have a direct physical interpretation. While this can often
be inappropriate and computationally inconvenient, it suits well
our case, where the main requirement is a detailed qualitative
description of the cluster profiles.\\
In \citep{Vik06}, Eq.~(\ref{gas density vik}) is applied to a
restricted range of distance from the cluster center, i.e. between
an inner cutoff $r_{min}$, chosen to exclude the central
temperature bin ($\approx 10\div 20\ {\rm kpc}$) where the ICM is
likely to be multi-phase, and $r_{det}$, where the X-ray surface
brightness is at least $3 \sigma$ significant. We have
extrapolated the above function to values outside this restricted
range using the following criteria:
\begin{itemize}
  \item for $r < r_{min}$, we have performed a linear extrapolation
  of the first three terms out to $r = 0$ kpc;
  \item for $r > r_{det}$, we have performed a linear extrapolation
  of the last three terms out to a distance $\bar{r}$ for which
  $\rho_{gas}(\bar{r})=\rho_{c}$, $\rho_{c}$ being the critical
  density of the Universe at the cluster redshift:
  $\rho_{c} = \rho_{c,0} \cdot (1 + z)^{3}$. For radii larger than $\bar{r}$,
  the gas density is assumed constant at $\rho_{gas}(\bar{r})$.
\end{itemize}
We point out that, in Table~\ref{tabcluster}, the radius limit $r_{min}$
is almost the same as given in the previous definition. When the
value given by \citep{Vik06} is less than the cD-galaxy radius, which is
defined in the next section, we  choose  this last one as the lower
limit. On the contrary, $r_{max}$ is quite different from
$r_{det}$: it is fixed by considering the higher value of temperature profile
and not by imaging methods. \\
We then compute the gas mass $M_{gas}(r)$ and the total mass
$M_{cl,N}(r)$, respectively, for all clusters in our sample,
substituting Eq.~(\ref{gas density vik}) into
Eqs.~(\ref{eq:gas_density}) and (\ref{eq:M_tot}), respectively;
the gas temperature profile is described in details in
\S~\ref{sec:T_prof}. The resulting mass values, estimated at
$r=r_{max}$, are listed in Table~\ref{tabcluster}.

\subsubsection{Temperature Profiles}
\label{sec:T_prof}

As  stressed in \S~\ref{sec:gas_model}, for the purpose of
this work, we need an accurate qualitative description of the
radial behavior of the gas properties. Standard isothermal or
polytropic models, or even the more complex one proposed in
\citep{Vik06}, do not provide a good description of the data at all
radii and for all clusters in the present sample. We hence describe the
gas temperature profiles using the straightforward X-ray spectral analysis
results, without the introduction of any analytic model.\\
X-ray spectral values have been provided by A. Vikhlinin (private
communication). A detailed description of the relative spectral
analysis can be found in \citep{Vik05}.


\subsubsection{Galaxy Distribution Model}
\label{sec:gal_model}

The galaxy density can be modelled as proposed by \citep{Bah96}.
Even if the galaxy distribution is a \textit{point}-distribution
instead of a continuous function, assuming that galaxies are in
equilibrium with gas, we can use a $\beta-$model, $\propto
r^{-3}$, for $r < R_{c}$ from the  cluster center, and a steeper
one, $\propto r^{-2.6}$, for $r > R_{c}$, where $R_{c}$ is the
cluster core radius (its value is taken from \citep{Vik06}). Its
final expression is:
\begin{equation}\label{gal density bahcall}
\rho_{gal}(r) = \left\{%
\begin{array}{ll}
    \rho_{gal,1} \cdot \left[1+
\left(\frac{r}{R_{c}}\right)^{2} \right]^{-\frac{3}{2}} & \hbox{$r < R_{c}$} \\
    \rho_{gal,2} \cdot \left[1+
\left(\frac{r}{R_{c}}\right)^{2} \right]^{-\frac{2.6}{2}} & \hbox{$r > R_{c}$} \\
\end{array}%
\right.
\end{equation}
where the constants $\rho_{gal,1}$ and $\rho_{gal,2}$ are chosen
in the following way:
\begin{itemize}
 \item \citep{Bah96} provides the central number density of galaxies in
 rich compact clusters for  galaxies located within a
 $1.5$ h$^{-1}$Mpc radius from the cluster center and brighter than $m_3+2^m$
 (where $m_3$ is the magnitude of the third brightest galaxy):
 $n_{gal,0} \sim 10^{3} h^{3}$ galaxies Mpc$^{-3}$. Then we  fix
 $\rho_{gal,1}$ in the range $\sim 10^{34}\div 10^{36}$ kg/kpc$^{3}$.
 \item the constant $\rho_{gal,2}$ has been fixed with the only
 requirement that the galaxy density function has to be continuous at
 $R_{c}$.
\end{itemize}
For any cluster we assume that the galaxy population also consists
in a cD galaxy, a giant elliptical galaxy with a diffuse envelope
which is generally located at the center of clusters and whose
typical mass is in the range $10^{12}\div 10^{13} M_{\odot}$. The
cD galaxy density has been modeled as described in \citep{SA06};
they use a Jaffe model of the form:
\begin{equation}\label{jaffe cd galaxy}
\rho_{CDgal} = \frac{\rho_{0,J}}{\left(\frac{r}{r_{c}}\right)^{2}
\left(1+\frac{r}{r_{c}}\right)^{2}} \; ,
\end{equation}
where $r_{c}$ is the core radius while the central density is
obtained from ${\displaystyle M_{J} = \frac{4}{3} \pi R_{c}^{3}
\rho_{0,J}}$. The mass of the cD galaxy has been fixed at $1.14
\times 10^{12}$ $M_{\odot}$, with $r_{c} = R_{e}/0.76$, with
$R_{e} = 25$ kpc being the effective radius of the galaxy. The
central galaxy for each cluster in the sample is assumed to have
approximately this stellar mass.

We have tested the effect of varying galaxy density with the
central density parameter $\rho_{gal,1}$ in the above range $\sim
10^{34}\div 10^{36}$ kg/kpc$^{3}$ on the cluster with the lowest
mass, namely A262. In this case, we would expect greater
variations with respect to other clusters; the result is that the
contribution due to galaxies and cD-galaxy gives a variation $\leq
1\%$ to the final estimate of fit parameters.

Finally, we have assumed that the total galaxy-component mass
(galaxies plus cD-galaxy masses) is $\approx 20\div25\%$ of the
gas mass: in \citep{Schindler02}, the mean fraction of gas versus
the total mass (with dark matter) for a cluster is estimated to be
$15\div 20\%$, while the same quantity for galaxies is $3\div
5\%$. This means that the relative mean mass ratio gal-to-gas in a
cluster is $\approx 20\div 25\%$. We have varied the parameters
$\rho_{gal,1}$, $\rho_{gal,2}$ and $M_{J}$ in their previous
defined ranges to obtain a mass ratio between total galaxy mass
and total gas mass which lies in this range. At the end the
cD-galaxy is dominant with respect to the other galaxies only in
the inner region (below $100$ kpc). As already stated in
\S~\ref{sec:gas_model}, cluster innermost regions have been
excluded from our analysis and so the contribution due to the
cD-galaxy is practically negligible in our analysis. The gas is,
as a consequence, the dominant visible component, starting from
innermost regions out to large radii, being galaxy mass only
$20\div 25\%$ of gas mass.

\subsubsection{Uncertainties on mass profiles}
\label{sec:uncertainties}

Uncertainties on the cluster total mass profiles have been
estimated performing Monte-Carlo simulations \citep{NeuBoh95}. We
proceed to simulate temperature profiles and choose random
radius-temperature values couples for each bin which we have in
our temperature data given by \cite
{Vik05}. Random temperature
values have been extracted from a Gaussian distribution centered
on the spectral values, and with a dispersion fixed to its $68\%$
confidence level. For the radius, we choose a random value inside
each bin. We have performed 2000 simulations for each cluster and
perform two cuts on the simulated profile. First, we exclude those
profiles that give an unphysical negative estimate of the mass:
this is possible when our simulated couples of quantities give
rise to too high temperature-gradient. After this cut, we have
$\approx1500$ simulations for any cluster. Then we have ordered
the resulting mass values for increasing radius values. Extreme
mass estimates (outside the $10\div90\%$ range) are excluded from
the obtained distribution, in order to avoid other high mass
gradients which give rise to masses too different from real data.
The resulting limits provide the errors on the total mass.
Uncertainties on the electron-density profiles  have not been
included in the simulations, being them negligible with respect to
those of the gas-temperature profiles.

\subsection{Low surface brightness galaxies}
\label{sec:LSB}

For the analysis of galactic scales we have used a sample of the
so-called low surface brightness (LSB) galaxies and dwarf
galaxies. It is still unclear to what extent rotation curves in
bright spiral galaxies may give clues about the profile of both
visual and dark matter, mainly because they are poor in gas
content, so that rotation curves can be hardly detected out to
sufficiently large radii where they are supposed to be dark matter
dominated. Moreover they also show some typical complex features,
such as extended spiral arms or barred structure that can lead to
consistent non-circular motions and thus making difficult the
interpretation of data. On the other side, LSB galaxies are
supposed to be dark matter dominated at all radii, and therefore
the analysis of their rotation curves can yield important clues
about it. Effectively, LSB galaxies exhibit a large discrepancy
between the detectable and the Newtonian dynamics mass within the
optical disk \citep{deBlok97,McGaugh98,Swaters00,bour}. They are
also a little challenge in the framework of CDM model because
predictions from CDM based simulations have revealed disagreement
with observational profiles of several dwarf galaxies
\citep{Moore94}. In particular, data indicate much less cuspy
distributions of matter than the simulations, and possible
solutions such as feedback effect due to star formation have been
excluded \citep{vandenHoek00} by the observed low star formation
ratio, at present and in the past, from which it can be deduced
that star formation rate has never been important enough in these
galaxies to modify their structure. Then, alternative solutions to
Newtonian dynamics are also possible.

The chosen galaxies for our analysis come from a sample of $15$
elements with high resolution H$\alpha$/HI rotation curves
extracted from the larger sample described in \citep{deBlok02}.
This sample was selected by \citep{Capozziello07} using the
criteria of the contemporary availability of data on rotation
curves, on surface photometry in the R band and on surface
gas-mass density. Rotation curves were derived from spectrographic
observations performed by the authors themselves of
\citep{deBlok02}, while the photometry of the stellar disk and
H$\alpha$/HI surface densities are collected from literature. For
a more complete and detailed discussion one can mainly refers to
\citep{SwatersPhD}. \\
The final sample does not constitute a complete sample of dwarf
and LSB galaxies, but contains galaxies in a wide range in
luminosity and surface brightness for which high-quality rotation
curves are available. Therefore it is well suited for testing
scalar field mechanism in late-type dwarf and LSB galaxies.

Moving to the way of modeling our spiral galaxies, we have to
specify the properties of stellar and gas distribution. Stars are
generally supposed being distributed in a thin and circularly
symmetric disk, with the surface density $\Sigma(R)$ derived from
the observed surface brightness distribution through the relation:
\begin{equation}
\Sigma(R) = 10^{-0.4 (\mu(R) - \mu_{R,\odot} - C)} \; ,
\end{equation}
where $\mu_{R,\odot}=4.46$ is the solar magnitude in the R band
and $C = 21.572$ is the needed constant for converting surface
brightness from magnitude units, $\mathrm{mag}/\mathrm{arcsec}^2$,
to linear units, $L_{\odot}/ \, \mathrm{pc}^2$. The luminosity
surface density is commonly fitted with the exponential disk model
\cite
{Freeman70}:
\begin{equation}
\Sigma(R) = \Sigma_{0} \exp (-R/R_{d}) \; ,
\end{equation}
where $\Sigma_{0}$ is the central surface brightness and $R_{d}$
is the disk scale length. Of course, in determining our
theoretical rotation curve, we need the stellar mass density, that
can be obtained from the luminosity one by simply multiplying it
for the stellar mass-to-luminosity ratio, $Y_{\ast}$, which is,
together with our scalar field parameters, the third free parameter
having to be constrained with the fitting procedure.

Modeling the gas density is more complicated because we do not
have an analytical function able to describe its behavior at all
radii and because the profile is very disturbed. Using
\cite
{SwatersPhD} plots and images, we fit the outer radii profile
with a linear relation (in magnitude unities), while the inner one
is generally fitted by simply interpolating data, with any
analytical expression (generally a polynomial one) which well fits
data points. We then check if the model works well by comparing
the obtained total gas mass with the same quantity but evaluated by extrapolation
from observational data. We verified that we only need very small
normalization constants, in the range $(0.95,1.05)$, to fully match
data values. Finally, we multiply the gas density with the factor $1.4$ for including
helium contribution.

With these model components, the general expression for the
rotation velocity
\begin{equation}
v_{c}^2(R) = R \frac{{\mathrm{d}} \Psi(R,z)}{{\mathrm{d}}
R}\bigg{|}_{z=0} \; ,
\end{equation}
can be decomposed in the following terms:
\begin{equation}
v_{c}^2(R) = v_{c,N}^2(R) + v_{c,C}^2(R) \; ,
\end{equation}
with
\begin{equation}
v_{c,N}^2(R) = R \frac{{\mathrm{d}} \Psi_{N}(R,z)}{{\mathrm{d}}
R}\bigg{|}_{z=0} \; ,
\end{equation}
and
\begin{equation}
v_{c,C}^2(R) = R \frac{{\mathrm{d}} \Psi_{C}(R,z)}{{\mathrm{d}}
R}\bigg{|}_{z=0} \, ,
\end{equation}
the Newtonian and the corrective contributions to the rotational
velocity from the respective terms in the total gravitational
potential. Anyone of the previous terms can then be decomposed in
two different component elements, concerning the two mass
components, namely stars and gas. For stars we have:
\begin{equation}
v_{c,N}^{star}\,^2(R; Y_{\ast}) = R \, \frac{{\mathrm{d}}
\Psi_{N}^{star}(R,z; Y_{\ast})}{{\mathrm{d}} R}\bigg{|}_{z=0} \; ,
\end{equation}
where we make explicit that this term depends on the free parameter $Y_{\ast}$
which appears in the stellar density inside the potential expression. When this term
is not present in the available data, we use the expression given in \citep{Binney} for the case
of an exponential disk and with $Y_{\ast} = 1$, i.e.
\begin{eqnarray}
v_{c,N}^{star}\,^2(R) &=& 4 \pi \, {\mathrm{G}} \Sigma_{0} R_{d}(y)^2 \cdot \\
&\cdot& \left[ I_{0}(y) K_{0}(y) - I_{1}(y) K_{1}(y)\right] \; , \nonumber
\end{eqnarray}
with $y = R/2 R_{d}$. For gas we have
\begin{equation}
v_{c,N}^{gas}\,^2(R) = R \, \frac{{\mathrm{d}}
\Psi_{N}^{gas}(R,z)}{{\mathrm{d}} R}\bigg{|}_{z=0} \; ,
\end{equation}
and, as in the stellar case, when this kind of data is not
available, we derive it in a numerical way by using the modeled gas
density. From the corrective term to the potential we have
\begin{equation}
v_{c,C}^{star}\,^2(R; Y_{\ast}, \beta, L) = R \, \frac{{\mathrm{d}} \Psi_{C}^{star}(R,z; Y_{\ast}, \beta, L)}{{\mathrm{d}} R}\bigg{|}_{z=0}
\end{equation}
and
\begin{equation}
v_{c,C}^{gas}\,^2(R; \beta, L) = R \, \frac{{\mathrm{d}}
\Psi_{C}^{gas}(R,z; \beta, L)}{{\mathrm{d}} R}\bigg{|}_{z=0} \; ;
\end{equation}
both these two quantities are derived numerically. Finally, the
total rotation velocity is the sum in quadrature of all these
elements, i.e.
\begin{eqnarray}\label{eq:rot_vel_final}
v_{c}^2(R) &=& v_{c,N}^{star}\,^2(R; Y_{\ast}) + v_{c,C}^{star}\,^2(R; Y_{\ast}, \beta, L) + \nonumber \\
&+& v_{c,N}^{gas}\,^2(R) + v_{c,C}^{gas}\,^2(R; \beta, L) \; ;
\end{eqnarray}
and the chi-square function is:
\begin{equation}
\chi^{2}_{\mathrm{LSB}}(\{\theta_{i}\}) = \sum^{\mathcal{N}}_{j =
1} \frac{(v_{c,th}(R_{i}, \{\theta_{i})\} -
v_{c,obs}(R_{i}))^2}{\sigma^{2}_{i}}
\end{equation}
where $\mathcal{N}$ is the number of data, $\sigma^{2}_{i}$ are the measurement
variances and $\{\theta_{i}\} = \{\beta, L, Y_{\ast}\}$ is the parameters theory vector. Even in this case,
like for supernovae and clusters, we use MCMC method for minimizing the
chi-square function and deriving errors on the fit parameters.

\section{Results and discussion}
\label{sec:results}

In this section we are going to discuss the obtained results with
the goal  to achieve a  comprehensive  background where scalar field
works at various scales.  In particular, we are going to
search for  possible trends and correlations between observable
quantities and theoretical scalar field parameters in the various
class of the considered astrophysical objects. SNeIa  will be
discussed alone because of the  difficulties in interpreting their
results in this  general context. On the other hand, clusters and
galaxies seem to be tied in a  common  picture by a rescaling
process.

\subsection{Supernovae: results}
\label{sec:SN_results}

Difficulties  come from the SNeIa analysis. In this case we have
performed MCMC analysis leaving the scalar field parameters free,
with the only minimal requirement that $L>0$, being it a length,
and that $\beta>0$, because of only the term $\beta^{2}$ appears
to be involved in the Eq.~(\ref{eq:G_sn}) and no possibility to
discriminate between positive and negative values is given.

The parameter $\lambda$ may be considered as a length related to
stellar scale, to typical supernovae explosion, and/or to stellar
formation. Since its exact value is unknown, we have varied it on
a grid ranging from $\lambda = \, 10^{-3} {\mathrm{h}}^{-1}$ as
the minimum, and corresponding to a length of  $\approx 1 \,
{\mathrm{kpc}}$ (assuming $H_{0}= 74.2$ as in
\citep{Komatsu10}), up to $\lambda = \, 1 \mathrm{h}^{-1}$ as the
maximum, and corresponding to a length $\approx 1 \,
\mathrm{Mpc}$.

{\renewcommand{\arraystretch}{1.5}
\begin{table*}
\begin{center}
  \caption{\textit{Supernovae.} Column 1: $\lambda$ value. Column 2: chi square value evaluated at the best fit values for fitting parameters.
  Column 3: present matter content, $\Omega_{m,0}$. Column 4: coupling parameter $\beta$ from scalar field
  ($1\sigma$ confidence interval). Column 5: gravitational length $L$ from scalar field ($1\sigma$ confidence
  interval).\label{tabsn}}
 \begin{tabular}{ccccc}
  \tableline
  $\lambda$  & $\chi^{2}_{best}$ &$\Omega_{m,0}$   & $\beta$ & $L$      \\
  $(h^{-1})$ &                   &                 &         & (Mpc) \\
  \tableline
  \tableline
  $10^{-3} \, (\approx 1 \, kpc)$   & $465.610$ &$0.292^{+0.023}_{-0.023}$ & $0.135^{+0.459}_{-0.099}$ & $0.145^{+0.677}_{-0.110}$ \\
  $10^{-2} \, (\approx 10 \, kpc)$  & $465.659$ &$0.295^{+0.027}_{-0.024}$ & $0.120^{+0.326}_{-0.088}$ & $0.158^{+0.743}_{-0.119}$ \\
  $10^{-1} \, (\approx 100 \, kpc)$ & $465.633$ &$0.301^{+0.040}_{-0.026}$ & $0.114^{+0.305}_{-0.083}$ & $0.174^{+0.758}_{-0.136}$ \\
  $10^0    \, (\approx 1 \, Mpc)$   & $465.707$ &$0.298^{+0.031}_{-0.024}$ & $0.129^{+0.347}_{-0.094}$ & $0.159^{+0.991}_{-0.123}$ \\
  \tableline
 \end{tabular}
\end{center}
\end{table*}}

In Table~\ref{tabsn} we have also reproduced the values of
chi-square function evaluated at the best fit points: they are
completely equal to the same quantity evaluated for a CPL model
with $\Omega_{m,0}, w_{0}$ and $w_{a}$ as fit parameters and a
gravitational constant fixed at its usual Newtonian value
(\textit{constant in time and scale}). For CPL model we have
obtained the following results:
\begin{eqnarray}
\Omega_{m,0} &=& 0.273^{+0.091}_{-0.133} \nonumber \\
w_{0} &=& -0.954^{+0.210}_{-0.275} \nonumber \\
w_{a} &=& 0.003^{+0.201}_{-0.150} \nonumber \\
\chi_{best} &=& 465.665
\end{eqnarray}
If we want to compare the  reliability of our model against the
CPL one, we can use a Bayesian type test as BIC. It is defined as
${\mathrm{BIC}} = -2 \ln {\mathcal{L}} + k \ln N$
\citep{Schwarz78}, with $-2 \ln \mathcal{L}$ being the chi-square
value and $\mathcal{L}$ the likelihood function, $k$ the number of
parameters of the model and $N$ the number of points in the
dataset. The best model has the lowest BIC; in particular, when
comparing two different models, if the difference between the two
BIC values is $\Delta \mathrm{BIC}<2$, than there is no
significant difference between the models; if $2<\Delta
\mathrm{BIC}<5$, this difference is substantial; if $5<\Delta
\mathrm{BIC}<10$ there is a ``strong'' evidence in favor of the
model with the lowest BIC value; while for $\Delta
\mathrm{BIC}>10$ this evidence is ``decisive''(following
the most used value scale, the ``Jeffreys' scale'', defined and
reported in \cite{Jeffreys}).\\
If we consider that in the CPL case we have three fit parameters
as in our scalar field model, we can easily deduce that there are no
significative differences between the CPL and the scalar field model.\\
Things change slightly if we consider a $\Lambda$CDM model with a
Newtonian gravitational constant and only one free parameter,
$\Omega_{m,0}$. In this case, even if the chi-square has the same
value of our scalar field model, we have only one parameter, so that
$\Lambda$CDM model with a Newtonian gravitational constant is
strongly favored with respect to a scalar field model with a varying
gravitational constant (we may underline that this is an obvious
consequence of the BIC definition, where the $k \ln N$ tends to
support models with a smaller number of parameters). We will
return later on the chi-square values for a further discussion.

Now we can take a look to the best fit values of model parameters.
For what concerning the matter content parameter, we can observe
that for $\Omega_{m,0}$ they are slightly higher than the latest
estimation \citep{Komatsu10}, $\Omega_{m,0} \approx 0.24$, from combining
WMAP results with BAO and and the Hubble constant measurements,
but they are aligned with usual values derived from only-SN
analysis, which generally lead to higher values for this parameter
\citep{BuenoSanchez09}.\\
We have also to stress that such a high value for $\Omega_{m,0}$,
although comparable with the usual contribution generally attributed
to dark matter, is not contradictory at all with our intent of explaining dark matter
on smaller than cosmological scales and dark energy on cosmological
scales with the same source, namely, the scalar field effective scalar field.\\
In fact we can easily consider that there are two elements contributing
to the $\Omega_{m,0}$ value: one coming from ordinary baryonic matter,
$\Omega_{b,0} \approx 0.04$, and one coming from the scalar field,
thus acting as a dark matter-type component and scaling as $(1+z)^3$.
As the discussion in next pages will show, this is a feasible possibility.
Then, we have a contribution from scalar field to the acceleration
of universe through an effective dark energy component, for an amount
equal to the usually derived one ($\Omega_{ch,0} \approx 0.70$) and
acting as a cosmological constant.\\
Discussions about the scalar field parameters are more tricky: they
do not show significative statistical changes while varying
$\lambda$. 
By applying Eq.~(\ref{eq:mod_dist_cham}) to the Constitution data set we are
implicitly assuming that the $\lambda$ length is the same for all
supernovae; if we assume that these objects constitute an
homogeneous astrophysical family, our hypothesis looks general
and quite good.\\ 
The parameter $\beta$ quantifies the coupling of the
scalar field  with ordinary matter and mainly measures how much
the effective gravitational constant deviates from the usual
Newtonian one. From the cosmological analysis it is in the interval
$[0.114, 0.135]$. 
 On the other side, the length L may be considered in this case
as the \textit{``minimal''} length for which the variation of the gravitational
constant has ``cosmological'' effects ( i.e. detectable with the Hubble SNeIa diagram)
and with the scalar-field mimicking a dark energy-type component. Our analysis
shows that such a scale is $\sim 100$ kpc.

\begin{figure*}
\centering
  \includegraphics[width=100mm]{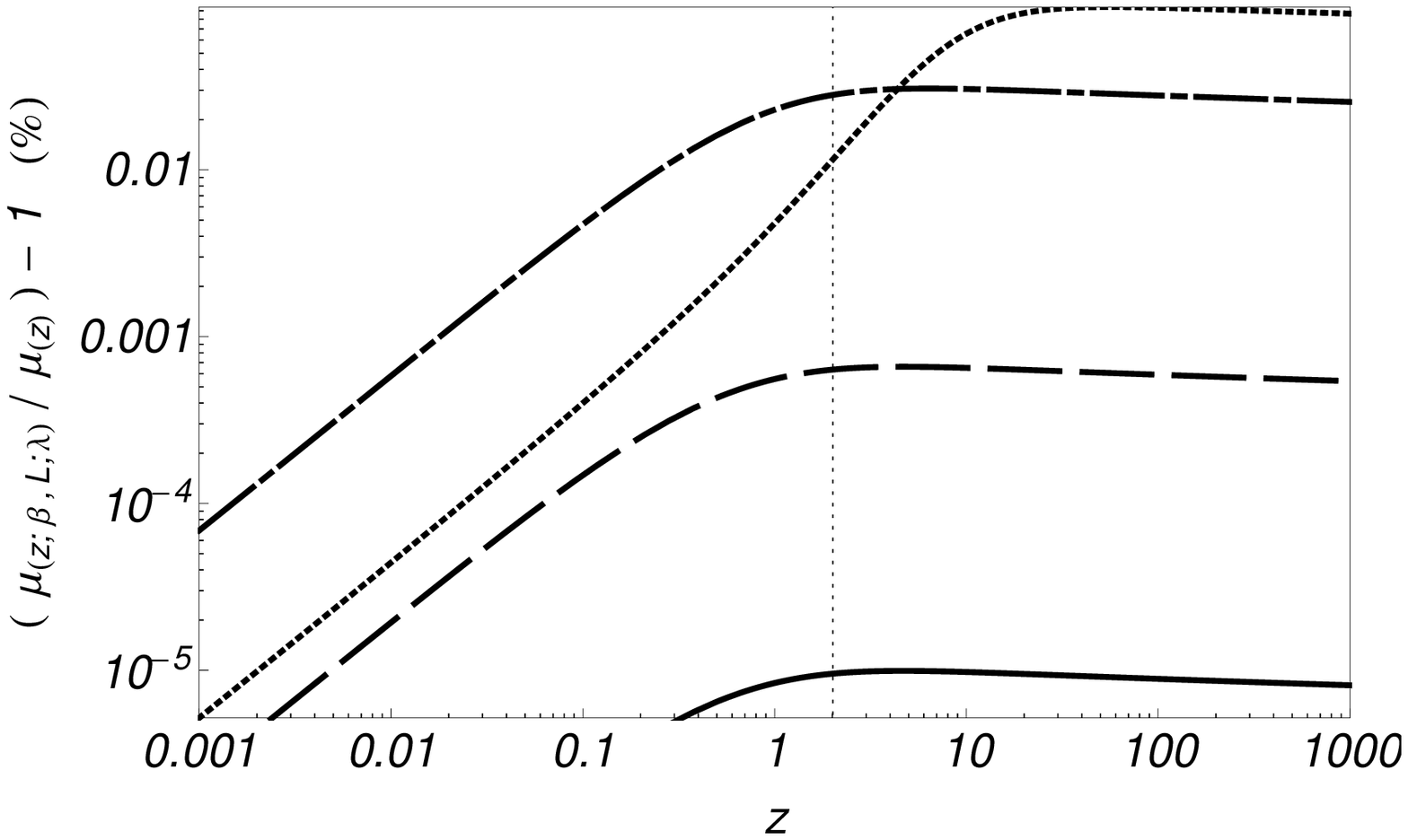}
  \includegraphics[width=93mm]{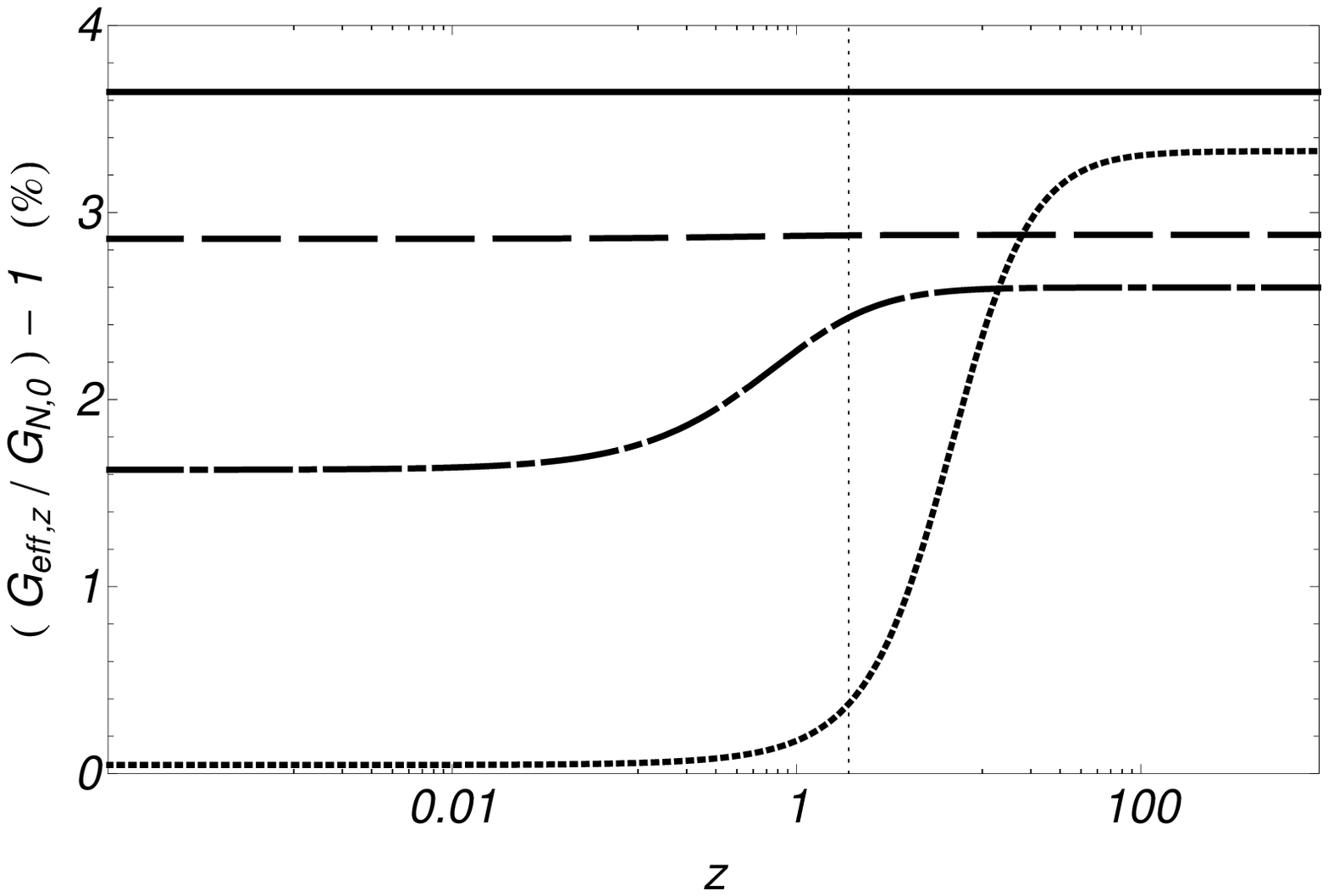}
  \includegraphics[width=100mm]{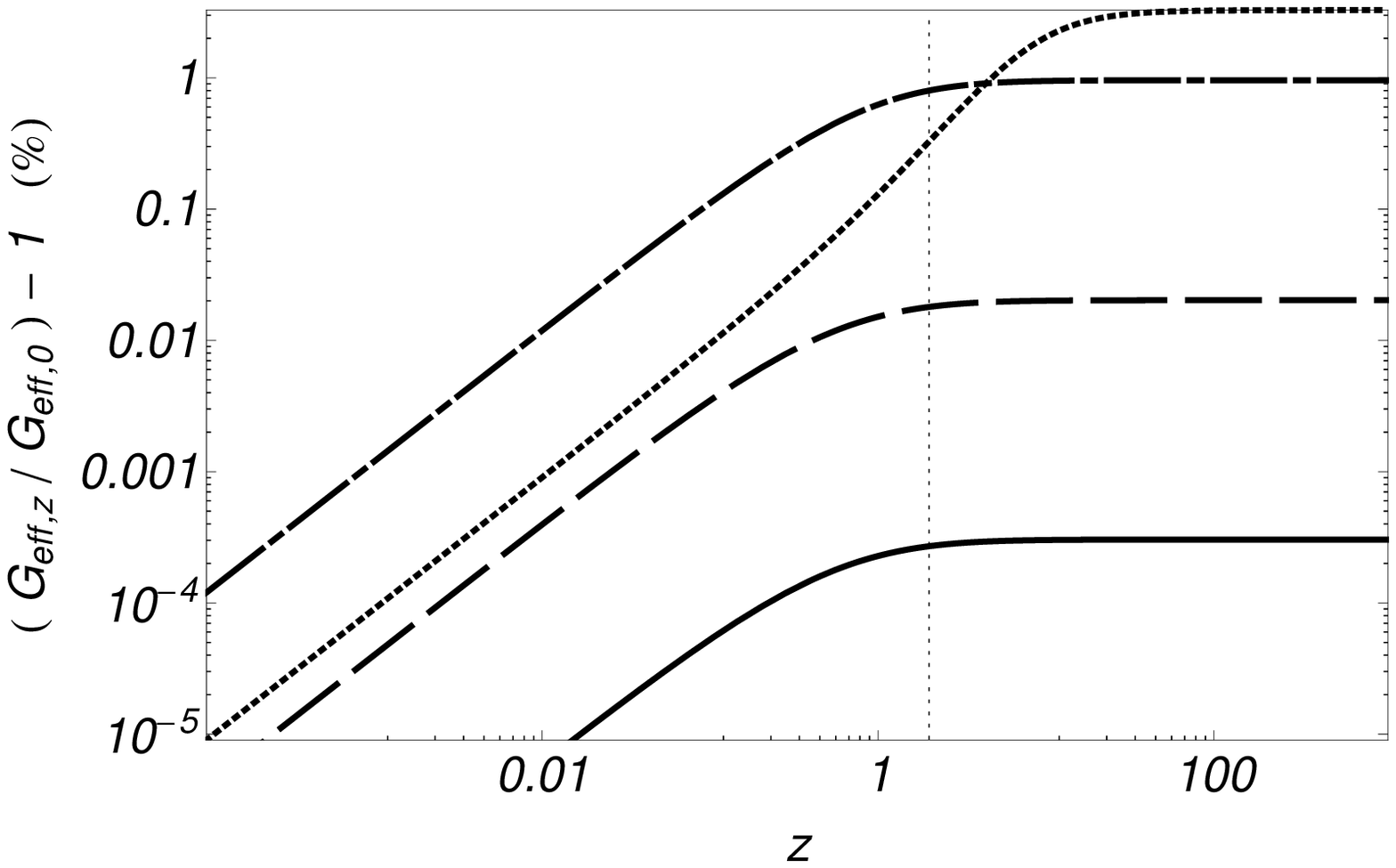}
  \caption{Continuous line
  is for $\lambda = 10^{-3}$ ($\beta = 0.135$; $L=0.145$ Mpc);  dashed line is for $\lambda = 10^{-2}$ ($\beta = 0.120$; $L=0.158$ Mpc); dot-dashed line is for $\lambda = 10^{-1}$ ($\beta = 0.114$; $L=0.174$ Mpc);
  dotted line is for $\lambda = 10^{0}$ ($\beta = 0.129$; $L=0.159$ Mpc). The vertical dotted line shows the
  maximum redshift from SNeIa sample. \textit{Top panel.} Deviations of the corrected
  version of distance modulus from the usual one. \textit{Middle panel.} Ratio between the effective gravitational constant, Eq.~(\ref{eq:G_cham}),
  and the Newton gravitational constant value. This ratio quantifies the temporal evolution of the deviation of scalar field mechanism
  from Newtonian gravity. \textit{Bottom panel.} Ratio between the effective
  gravitational constant, Eq.~(\ref{eq:G_cham}), and the present value of the same quantity. This ratio quantifies the
  temporal evolution of the effective gravitational constant.\label{fig:sn_results}}
\end{figure*}

Anyway, we have to combine all these results with the evidence that chi-square
values are completely equal for any $\lambda$, then we have also
to consider the possibility that our analysis is not well-based.
In the top panel of Fig.~(\ref{fig:sn_results}) we have plotted how
much the modified distance modulus in Eq.~(\ref{eq:mod_dist_cham})
differs from the usual expression,
\begin{equation}\label{eq:normmod}
\mu(z) = 5 \log \left( (1+z) \int_{0}^{z} \frac{{\mathrm{d}}z}{h(z)}\right) + \mu_{0} \, .
\end{equation}
If we consider that the modified distance modulus can be written
as:
\begin{equation}
\mu(z; \beta, L; \lambda) = \mu(z) + \frac{15}{4} \log \frac{G_{eff}(z; \beta, L; \lambda)}{G_{eff}(0; \beta, L; \lambda)}
\end{equation}
the correction coming from the effective scalar field gravitational
constant is negligible when compared to the usual distance modulus
expression, being at most $\approx 0.08
\%$. Facing with this question, we have two possibilities:
\begin{enumerate}
  \item our results do not come from an effective test of a scalar field mechanism, but are got stuck by the
        impossibility to detect changes in the gravitational constant, being the distance modulus mainly
        driven by the $\Omega_{m,0}$ parameter in Eq.~(\ref{eq:normmod});
  \item we are effectively testing a scalar field mechanism, but then we have the problem of including the obtained
        values for scalar field length $L$ in a consistent theoretical background.
\end{enumerate}

As we can see in the middle and bottom panels of Fig.~(\ref{fig:sn_results}),
we have different scenarios depending on the value of $\lambda$:
\begin{itemize}
 \item for $\lambda = 10^{-3}$ we have an effective gravitational constant which is practically constant,
       exhibiting a change of only $\sim 0.003 \%$ in the redshift range $[0.005, 1]$ and assessing at a value which
       differs from the Newtonian gravitational constant value for $\sim 3.6 \%$. In this case the relative contribution
       of the corrective term with respect of the usual one is $\approx 10^{-5} \%$;
 \item for $\lambda = 10^{-2}$, we have again a practically constant effective gravitational constant,
       with a change of $\sim 0.7 \%$ in the SNeIa redshift range, and assessing at a value which differs from the Newtonian gravitational constant value for $\sim 2.85 \%$. In this case the relative contribution of the corrective term with respect of the usual
       one is really small, $\approx 0.0006 \%$;
 \item for $\lambda = 10^{-1}$ we have a more sensible rise of effective gravitational constant in the supernovae redshift range, reaching an
       asymptotic value bigger than Newtonian one for $\sim 2.6 \%$ at $z > 50$. Following the large change in the redsfhift
       dependence of the effective gravitational constant, which can vary for $1 \%$ even in the supernovae redshift interval,
       the ratio between the corrective and the usual expression of distance modulus in this case is $\approx 0.028 \%$;
 \item for $\lambda = 1$ the rising is less pronounced in the supernovae redshift range with respect of previous case,
       but goes over at $z \approx 15$ and reaches its asymptotic value at $z \sim 500$,
       being it $\sim 3.32 \%$ bigger than Newtonian gravitational constant. In this case the corrective distance modulus term
       reaches a maximum deviation of $\approx 0.08 \%$ around $z \sim 50$; in the same range the effective gravitational constant
       can vary for $\approx 3 \%$.
\end{itemize}

\subsection{Clusters of galaxies: results}
\label{sec:Cl results}

When considering clusters of galaxies we remind that in this case
we left free the scalar field parameters, $\beta$ and $L$, with only
the minimum requirement of their positiveness.

\begin{figure*}
\centering
  \includegraphics[width=80mm]{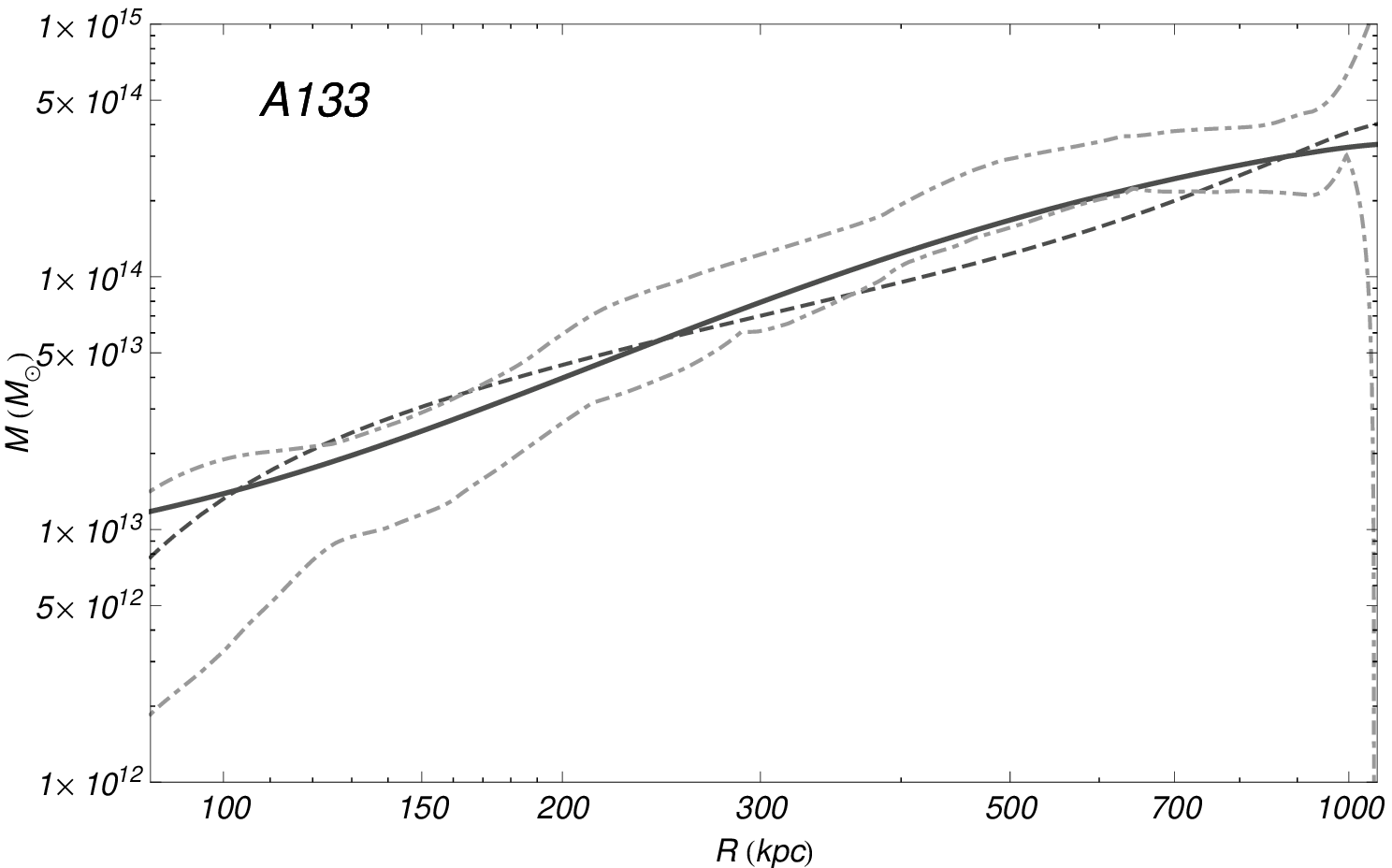}
  \includegraphics[width=80mm]{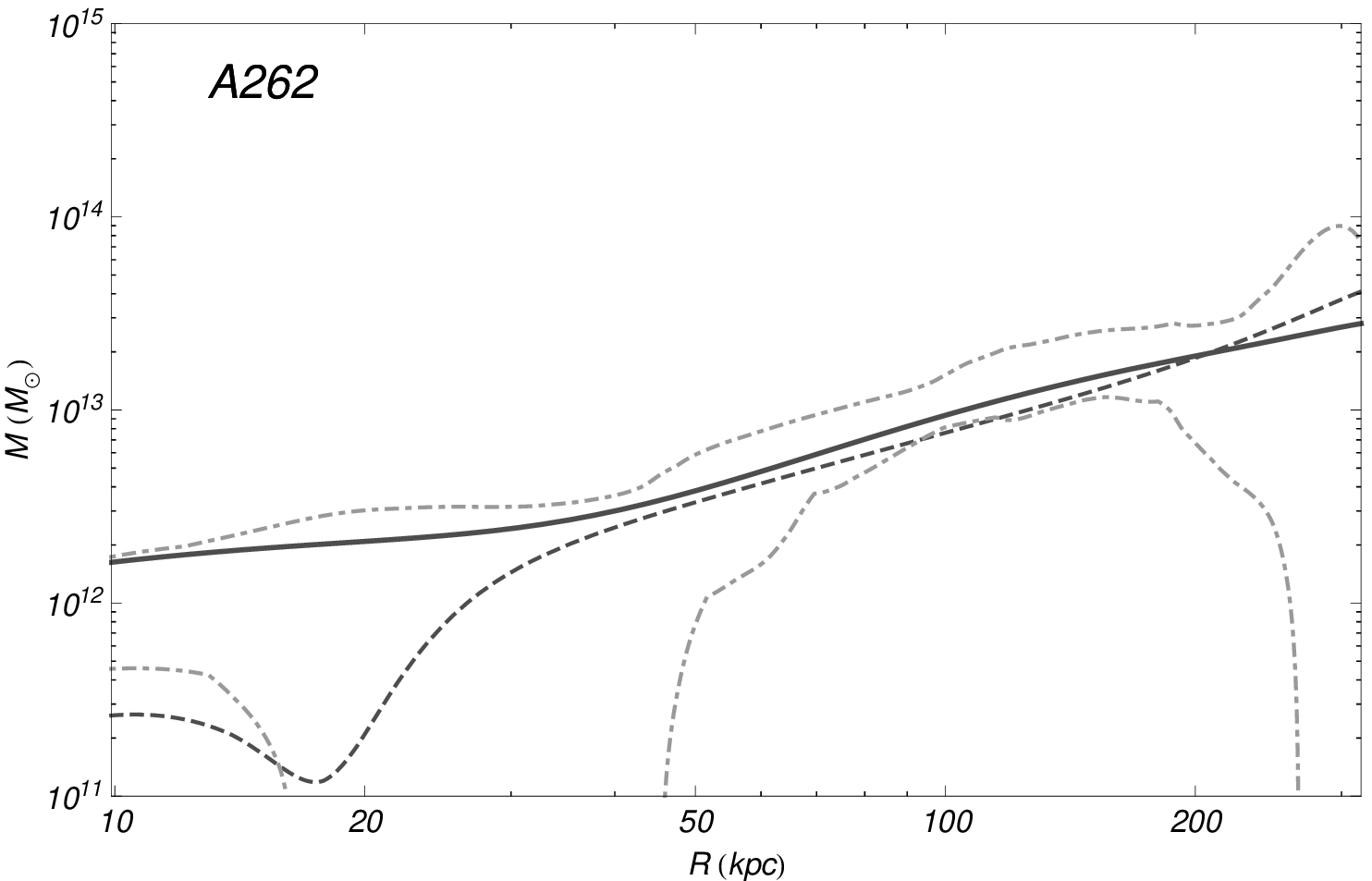}
  \includegraphics[width=80mm]{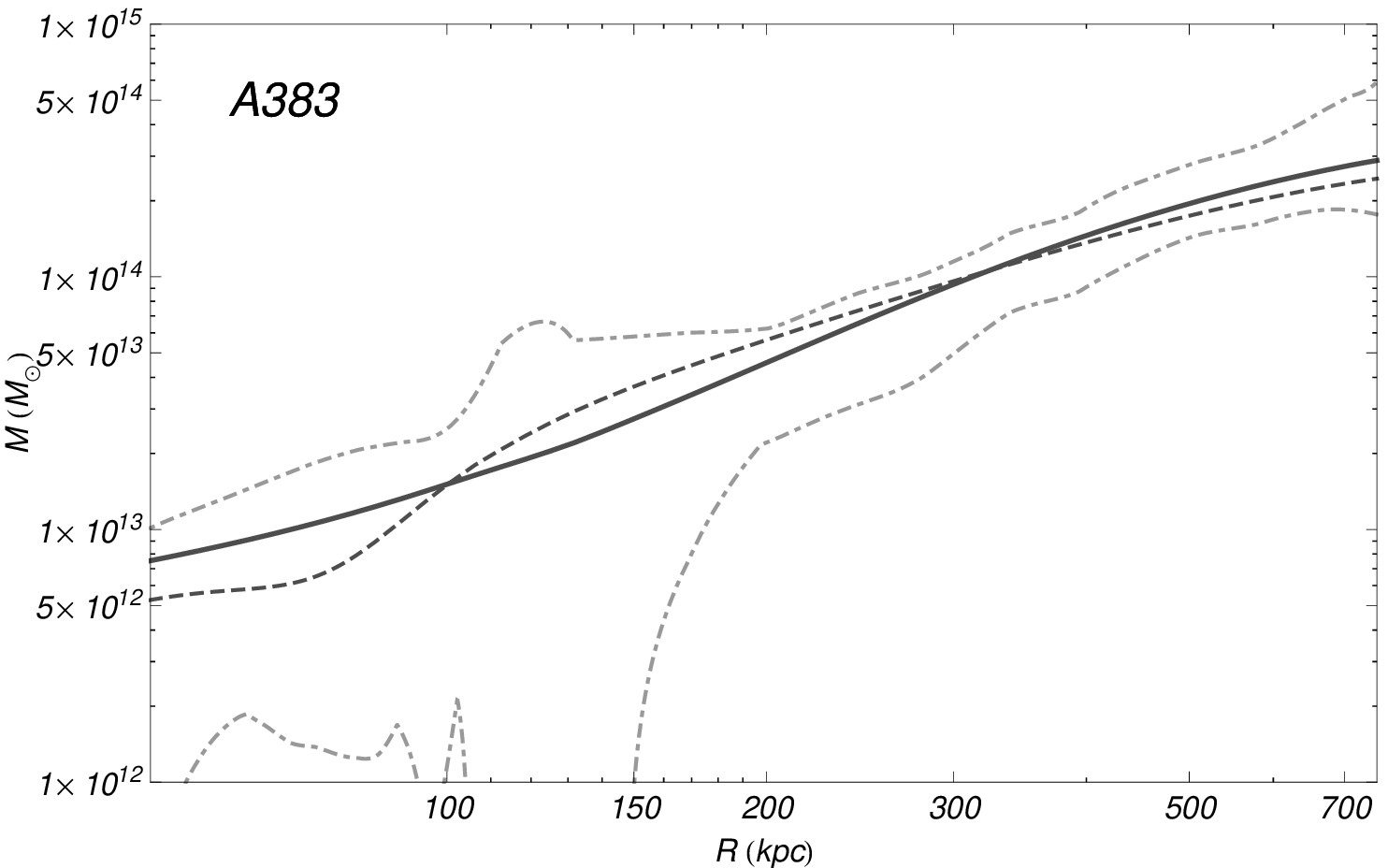}
  \includegraphics[width=80mm]{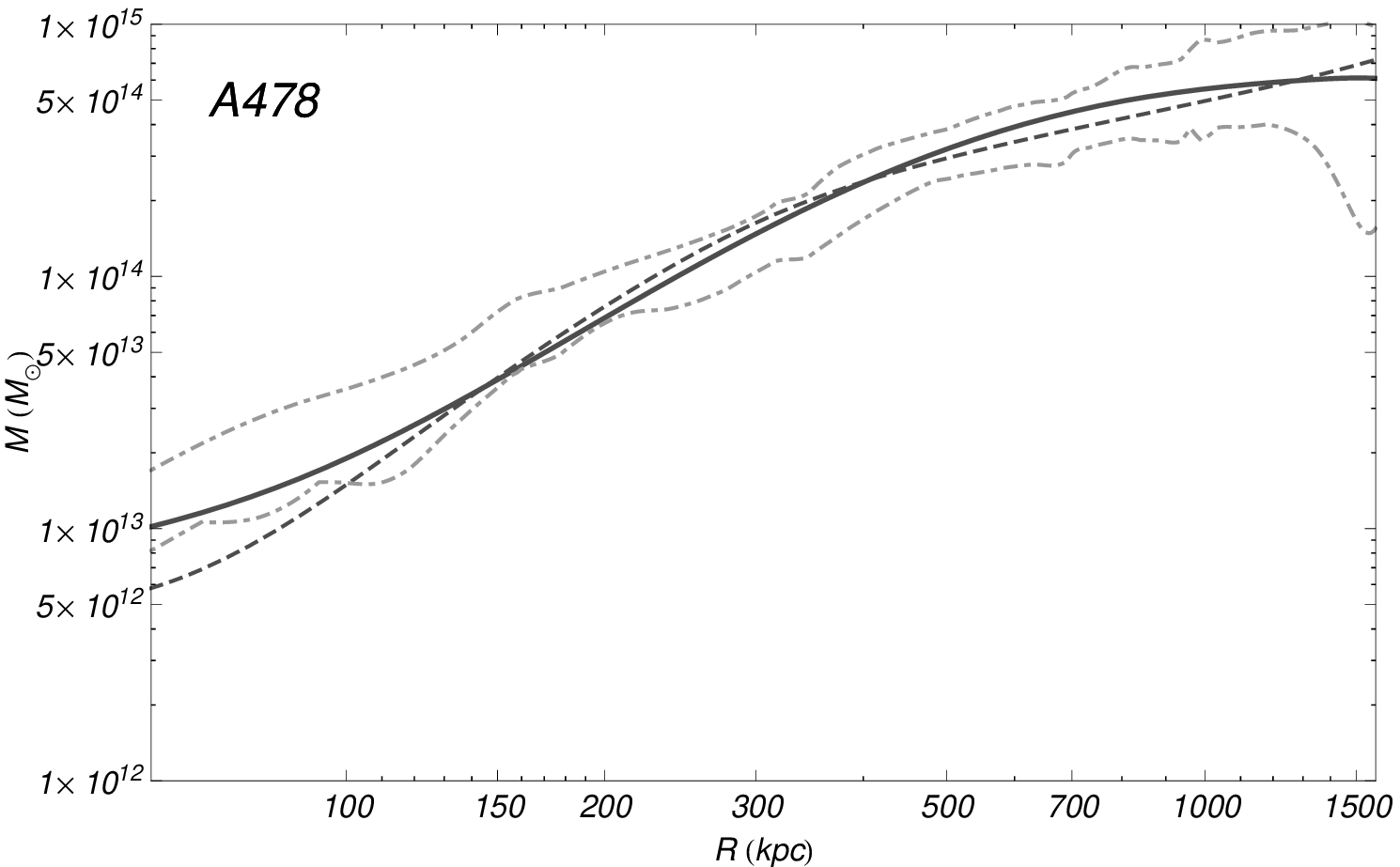}
  \includegraphics[width=80mm]{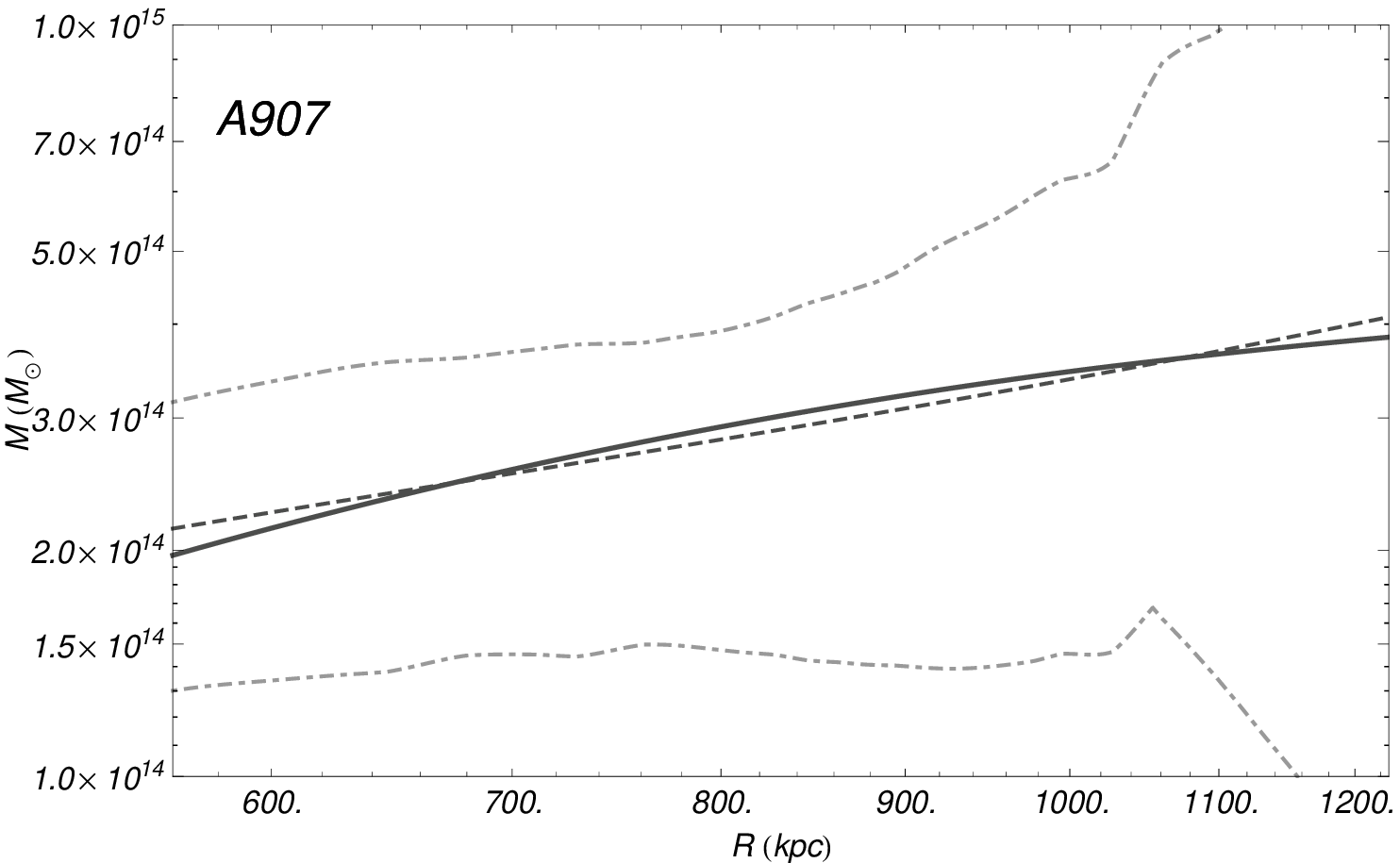}
  \includegraphics[width=80mm]{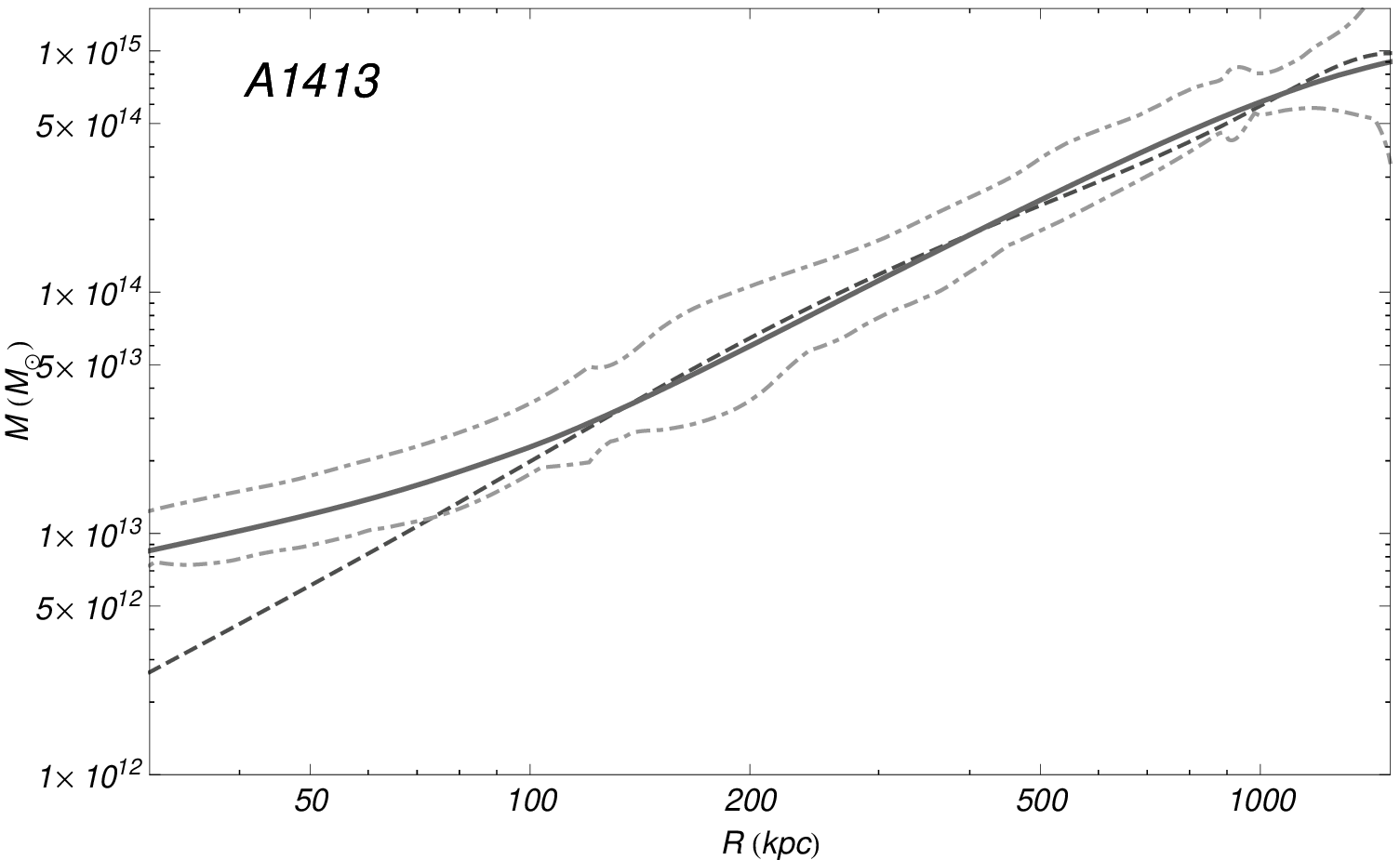}
  \caption{Dark matter profile vs radii for clusters of galaxies. Dashed line is the observationally derived estimation of dark matter, Eq.~(\ref{eq:obs_dark});
  solid line is the theoretical estimation for the effective dark matter component, Eq.~(\ref{eq:theo_dark});
  dot-dashed lines are the 1-$\sigma$ confidence levels given by errors on fitting parameters plus statistical errors on mass profiles as
  discussed in \S~\ref{sec:uncertainties}.\label{fig:cham_cl1}}
\end{figure*}

\begin{figure*}
\centering
  \includegraphics[width=80mm]{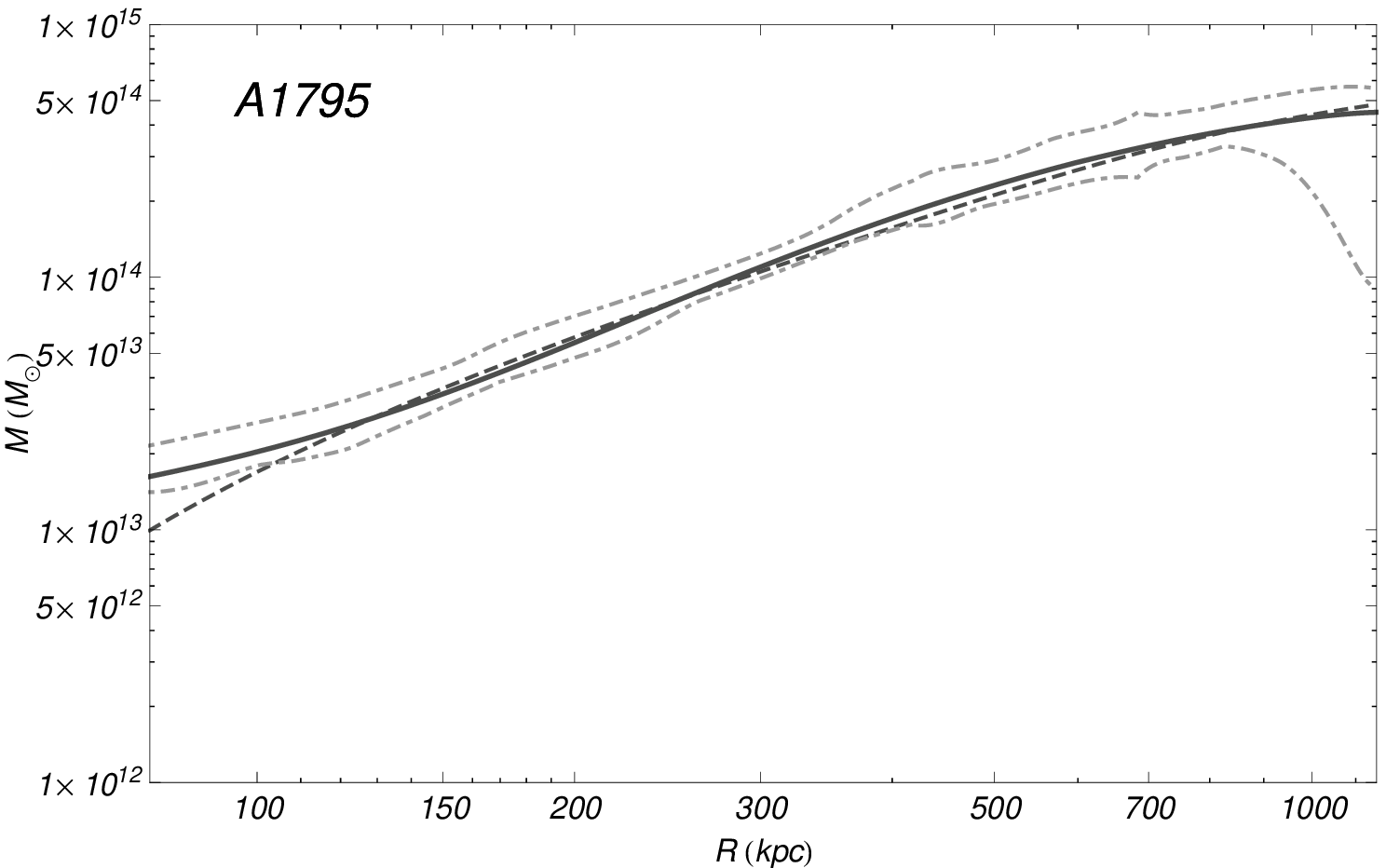}
  \includegraphics[width=80mm]{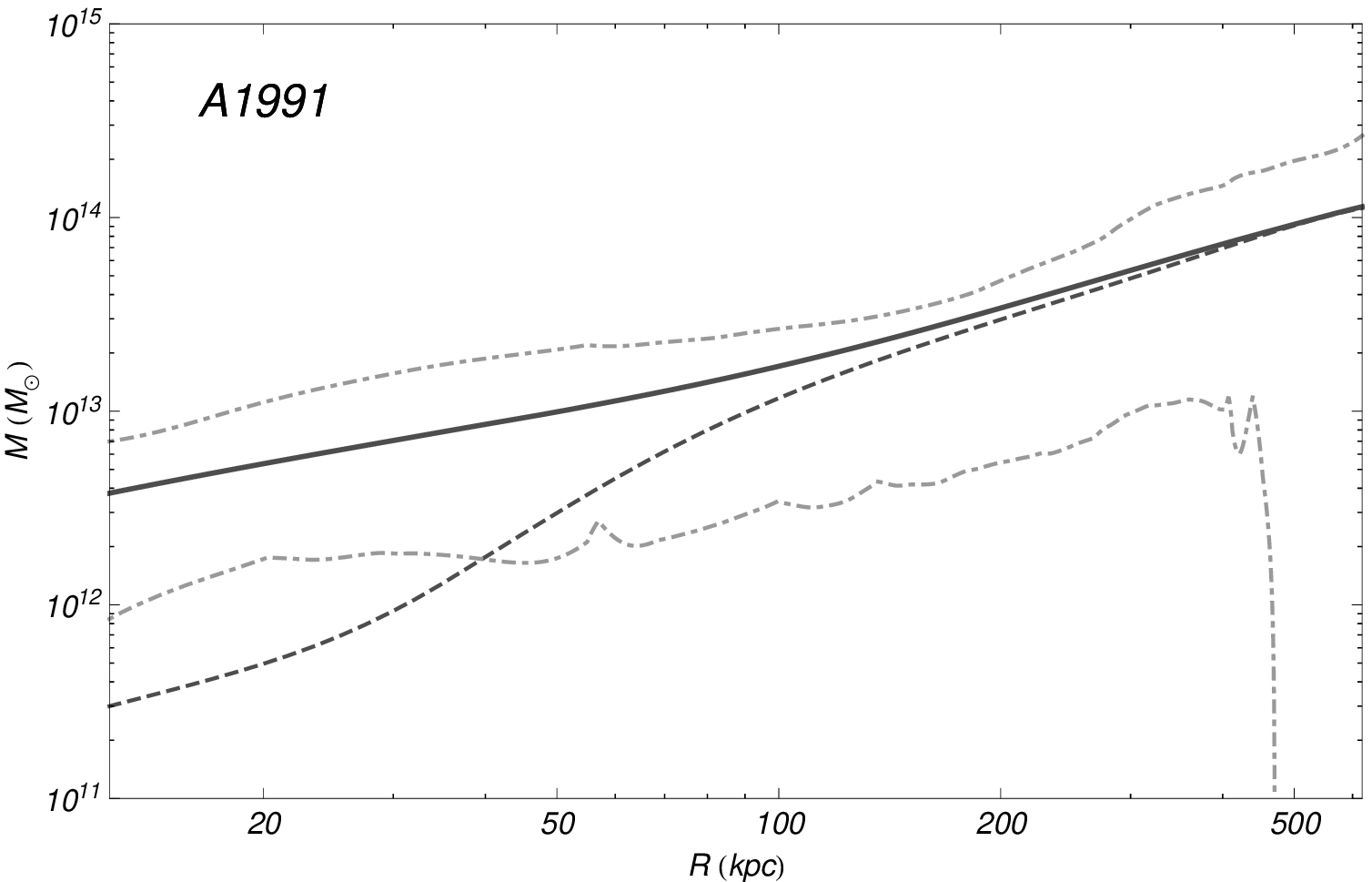}
  \includegraphics[width=80mm]{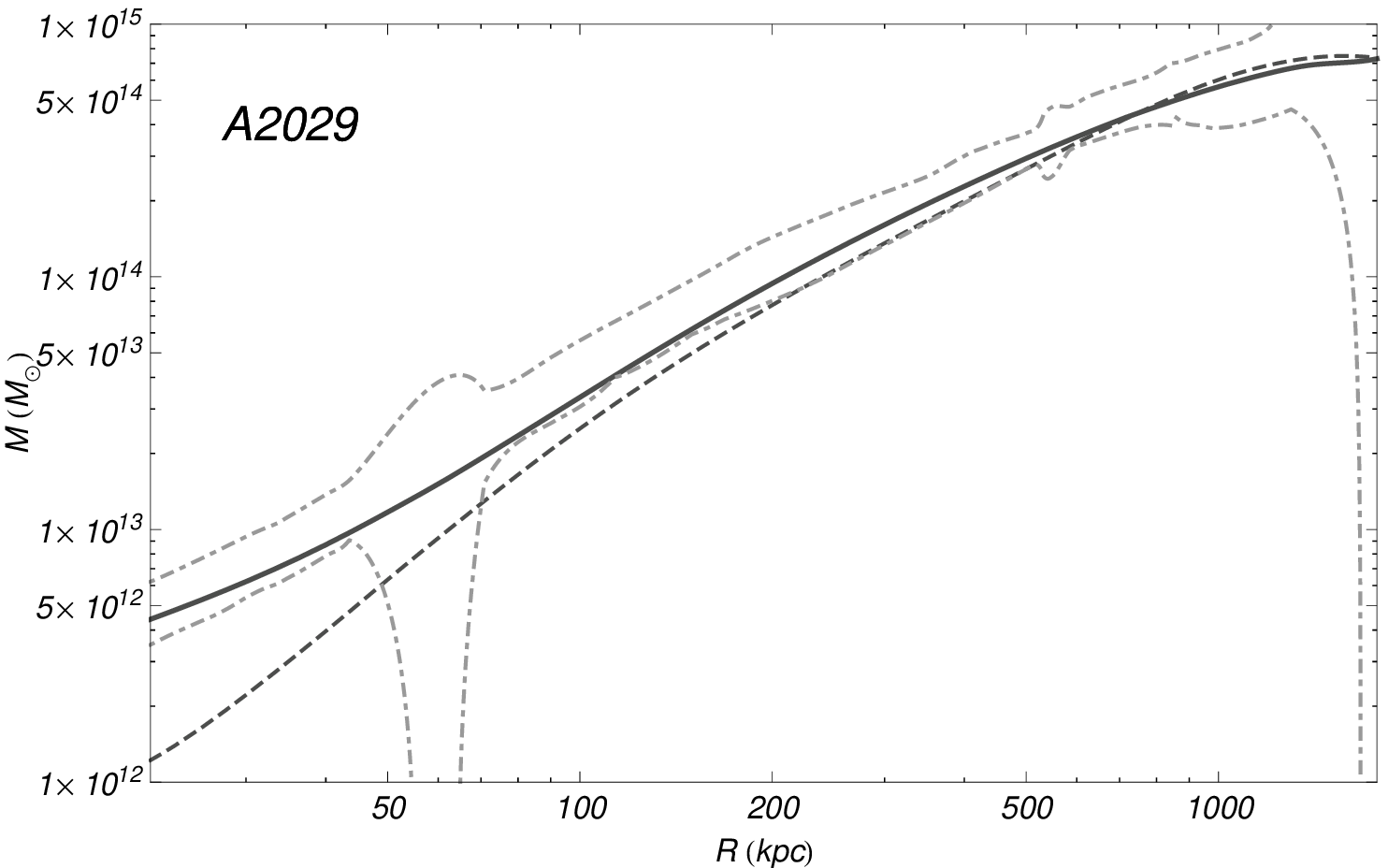}
  \includegraphics[width=80mm]{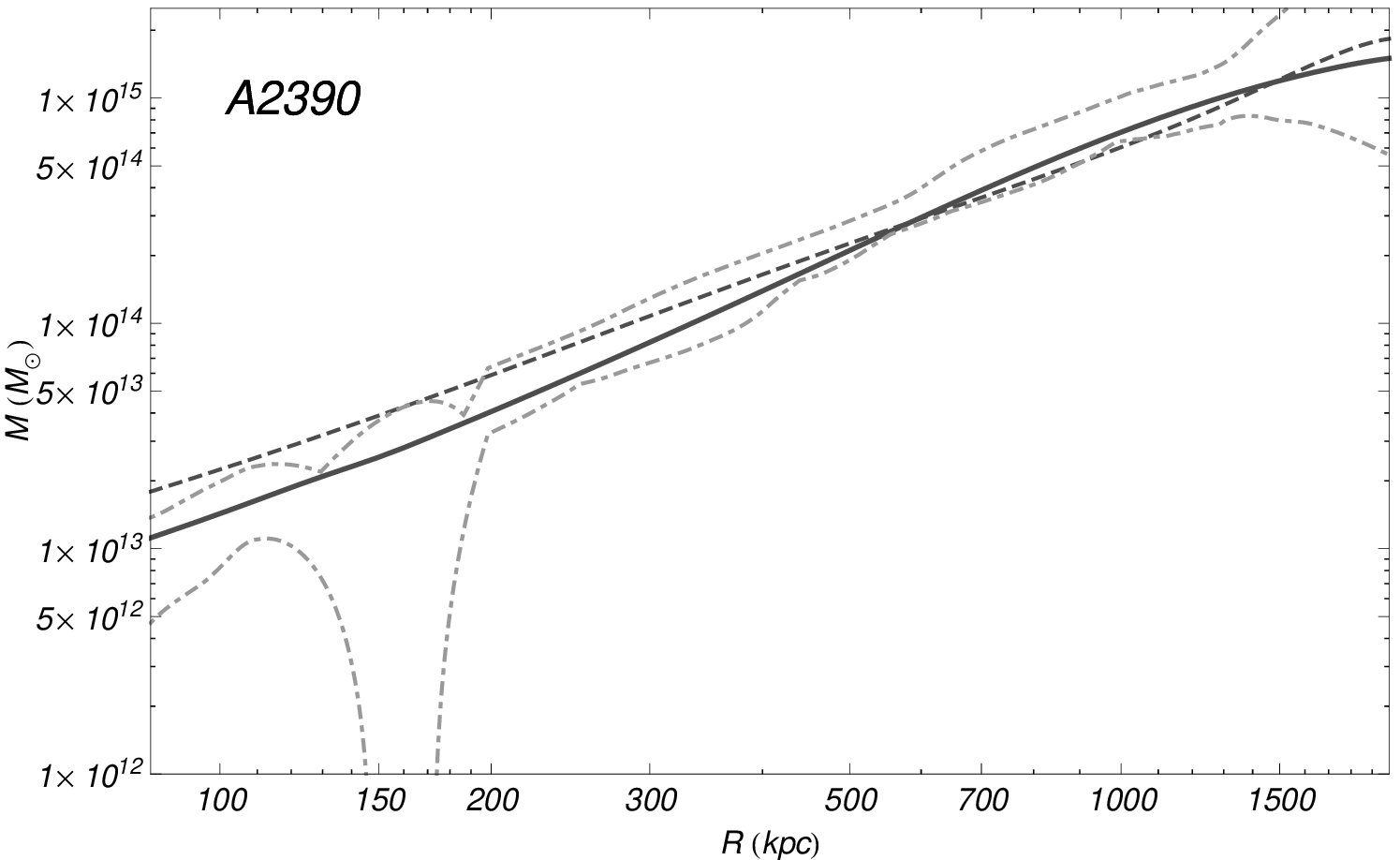}
  \includegraphics[width=80mm]{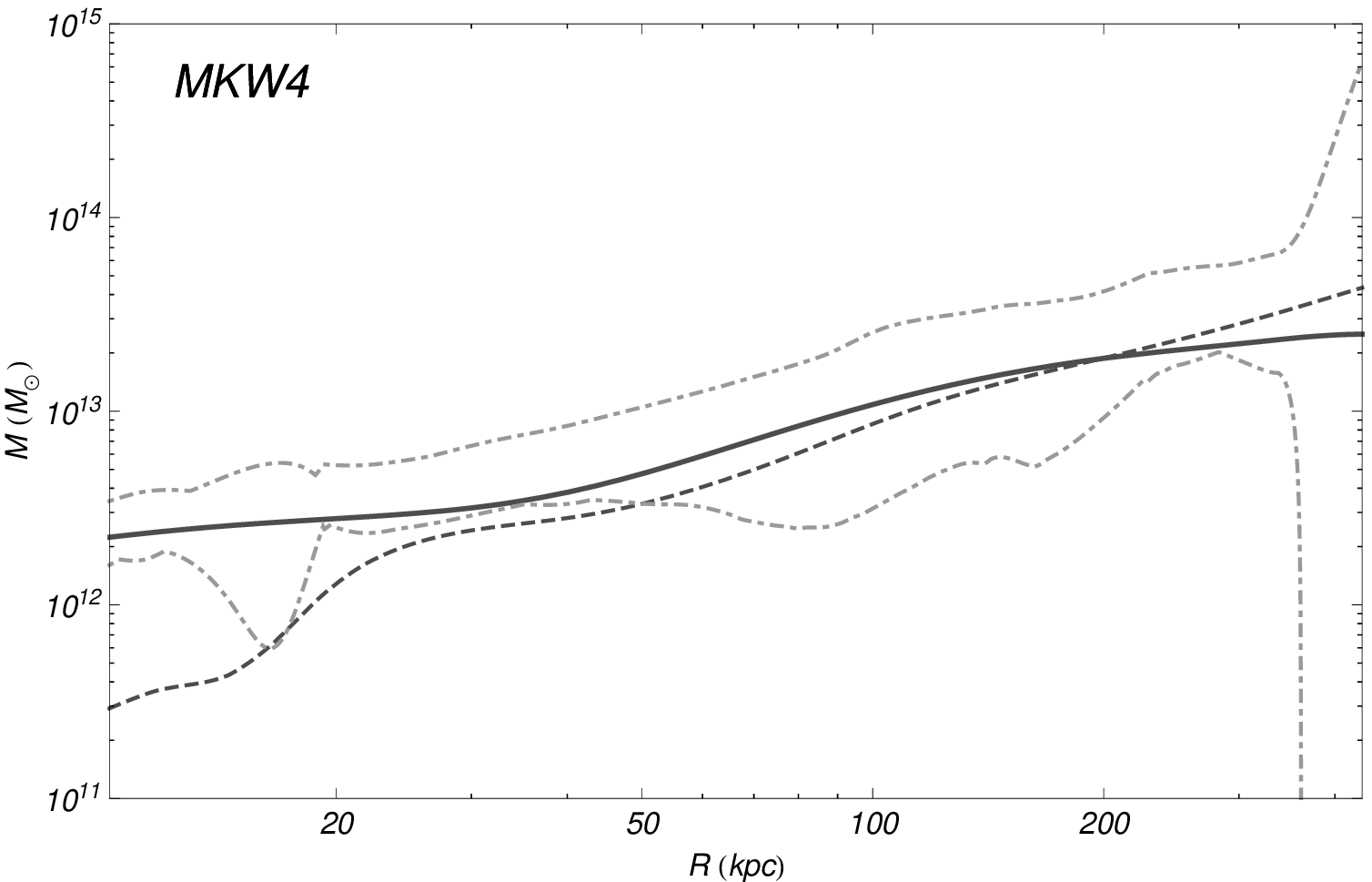}
  \includegraphics[width=80mm]{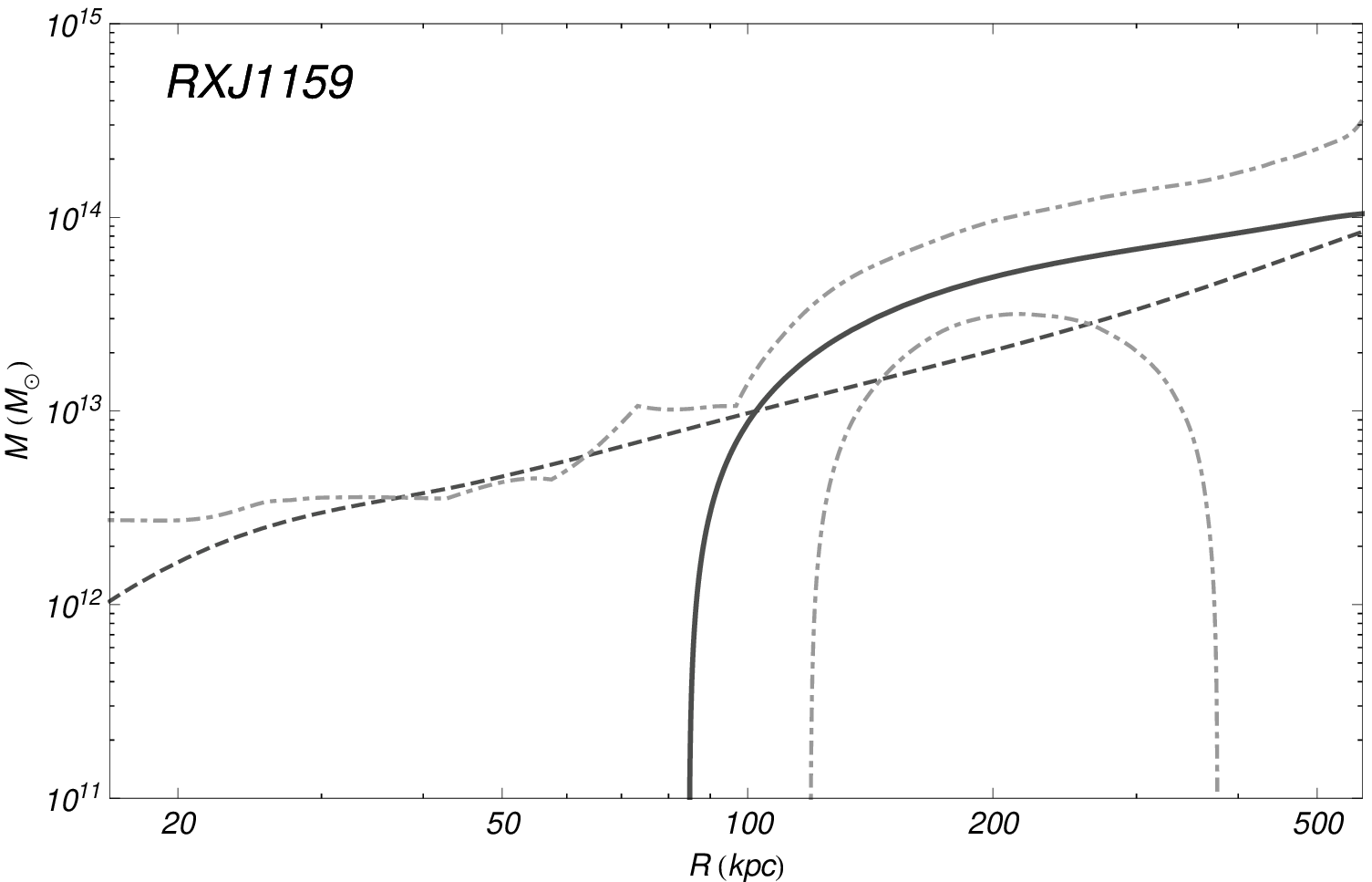}
  \caption{Same of Fig.~(\ref{fig:cham_cl1}).\label{fig:cham_cl2}}
\end{figure*}

As it is possible to see by simple visual inspection, the only bad
fit corresponds to the cluster RXJ1159: using the modelled matter
densities described in
\S~(\ref{sec:gas_model})~-~(\ref{sec:gal_model}), we obtain a too
fast decreasing mass profile in the inner region reaching
unphysical negative values. For this reason we will not consider
it anymore in our considerations. \\
On the contrary, for all the other clusters we have good results,
with mass estimations corresponding in the $1\sigma$ confidence
level. Errors contours reported in
Fig.~(\ref{fig:cham_cl1})~-~(\ref{fig:cham_cl2}) have two
contributions: the main one comes from the statistically derived
errors on mass observations, as described in
\S~(\ref{sec:uncertainties}) and which produce the larger and
irregular borders of the $1\sigma$ confidence level; the smallest
one comes from errors on fitting parameters.

{\renewcommand{\arraystretch}{1.5}
\begin{table*}
\begin{center}
 \caption{\textit{Clusters of galaxies.} Column 1: Cluster name. Column 2: cluster total mass.
  Column 3: gas mass. Gas and total mass values are estimated at $r=r_{max}$. Columns 4: gas mass weighted
  average temperature. Column 5: virial radius. Column 6: minimum observational radius. Column 7: maximum radius.
  Column 8: coupling parameter $\beta$ from scalar field ($1\sigma$ confidence interval). Columns 9: gravitational
  length $L$ from scalar field ($1\sigma$ confidence interval).\label{tabcluster}}
  \begin{tabular}{ccccccccc}
  \tableline
  name    &          $M_{cl,N}$   &       $M_{gas}$       & $<T>$ & $r_{vir}$ & $r_{min}$ & $r_{max}$ & $\beta$ &  $L$  \\
          & $(10^{14} M_{\odot})$ & ($10^{13} M_{\odot}$) & (keV) &   (kpc)   &   (kpc)   &   (kpc)   &         & (kpc) \\
  \tableline
  \tableline
  A133    & $4.35874$  & $2.73866$  & $3.68$ & $1694.26$ &  $86$ & $1060$ & $2.524^{+0.259}_{-0.228}$ & $888.557^{+463.221}_{-287.415} $ \\
  A262    & $0.445081$ & $0.276659$ & $1.92$ & $1199.46$ &  $61$ & $ 316$ & $2.786^{+0.397}_{-0.356}$ & $147.977^{+21.798}_{-28.704}$ \\
  A383    & $2.79785$  & $2.82467$  & $4.36$ & $1822.12$ &  $52$ & $ 751$ & $2.189^{+0.206}_{-0.187}$ & $728.246^{+443.580}_{-194.675}$ \\
  A478    & $8.51832$  & $10.5583$  & $7.34$ & $2344.98$ &  $59$ & $1580$ & $2.106^{+0.149}_{-0.120}$ & $820.874^{+259.014}_{-194.548}$ \\
  A907    & $4.87657$  & $6.38070$  & $5.44$ & $2030.39$ & $563$ & $1226$ & $2.364^{+0.521}_{-0.290}$ & $594.207^{+339.605}_{-183.460}$ \\
  A1413   & $10.9598$  & $9.32466$  & $6.76$ & $2259.35$ &  $57$ & $1506$ & $2.210^{+0.108}_{-0.105}$ & $1323.890^{+158.186}_{-216.063}$ \\
  A1795   & $5.44761$  & $5.56245$  & $5.52$ & $2054.1$ &  $79$ & $1151$ & $2.224^{+0.080}_{-0.072}$ & $869.098^{+297.243}_{-145.295}$ \\
  A1991   & $1.24313$  & $1.00530$  & $2.23$ & $1338.46$ &  $55$ & $ 618$ & $2.439^{+0.693}_{-0.388}$ & $534.918^{+483.400}_{-344.322}$ \\
  A2029   & $8.92392$  & $12.4129$  & $7.59$ & $2419.03$ &  $62$ & $1771$ & $2.047^{+0.121}_{-0.112}$ & $1073.050^{+237.912}_{-267.762}$ \\
  A2390   & $20.9710$  & $21.5726$  & $9.35$ & $2481.14$ &  $83$ & $1984$ & $1.888^{+0.067}_{-0.065}$ & $1487.800^{+90.565}_{-107.860}$ \\
  MKW4    & $0.469503$ & $0.283207$ & $1.58$ & $1068.31$ &  $60$ & $ 434$ & $3.259^{+20.876}_{-0.737}$ & $148.931^{+621.309}_{-141.849}$ \\
  RXJ1159 & $0.897997$ & $0.433256$ & $1.40$ & $1115.81$ &  $64$ & $ 568$ & $3.412^{+1.702}_{-0.722}$ & $387.568^{+601.661}_{-251.839}$ \\
  \tableline
 \end{tabular}
 \end{center}
\end{table*}}

What can be first noted is that for lower scales there is a large
deviation between our theoretical estimation and the observed one.
Typically one can locate this break-point in the range $[100,150]$
kpc. This is not a new thing when describing clusters of galaxies
with modified gravities and it is not an intrinsic failure of our
theoretical model. Similar issues are present in
\citep{Salzano09}, where $f(R)$- gravity models  are applied to
clusters of galaxies. The same situation is in \cite
{Brownstein06}: they use the the Metric - Skew - Tensor - Gravity
(MSTG) as a generalization of the Einstein General Relativity and
derive the gas mass profile of a sample of clusters with gas being
the only baryonic component of the clusters. They consider some
clusters included in our sample (in particular, A133, A262, A478,
A1413, A1795, A2029, MKW4) and  the same failing trend is found
for $r \leq 200$ kpc: they overestimate gas mass in the inner
regions with respect of the expected estimation from X-ray
observations. In the same work there is also an interesting note
about MOND theory applied to clusters of galaxies: even if it is
not possible to assess it really fails, surely we can see that
MOND in clusters does not solve the dark matter problem because it
again requires including a mass contribution other than the
observed one.

The reason for this different behavior in the inner regions is in
the break of the hypothesis of hydrostatic equilibrium. If the
hypothesis of hydrostatic equilibrium is not correct, then we are
in a regime where the fundamental relations Eqs.~(\ref{Boltzmann
equation})~-~(\ref{eq:Boltzmann potential}) are not working. As
discussed in \citep{Vik05},  the central - 70 kpc - region of most
clusters is strongly affected by radiative cooling and thus its
physical properties cannot directly be related to the depth of the
cluster potential well. This means that, in this region, the gas
is not in hydrostatic equilibrium but in a  multi-phase state. In
this case, the gas temperature cannot be used as a good standard
tracer. Among the main phenomena which causes this we have cooling
flows, merger effects and asymmetric shapes. In particular,
cooling flows produce a decrease in the temperature profile and
consequently local higher gas densities which cannot be related
directly to gravitational effects.\\
A coherent behavior is shown in our plots,
Fig.~(\ref{fig:cham_cl1})~-~(\ref{fig:cham_cl2}). We remind that
there the distribution of dark matter is represented;  higher
densities from cooling flows produce higher
non-gravitational-produced gas mass profiles, which result in a
decrease of the dark matter one. In our case, we  have that our
theoretical dark matter profile (i.e. the \textit{effective} dark
matter mimicked by the different coupling of scalar field with
baryonic mass) is higher than the observationally derived one (see
for example Abell 262, Abell 383, Abell 478, Abell 1413, Abell
1991, Abell 2029; while for Abell A133 and Abell 1795 one can
perceive the same trend but unfortunately the data do not
extend too small enough radii).\\
However, a more detailed modelling for inner regions is out the
purpose of this work, while we are here interested to show that
the scalar field mechanism can be a valid alternative to dark matter
in order to explain cluster dynamics. In this sense it is very
illuminating that on the most part of objects in our sample there
is a very good agreement between the scalar field model and the
observationally derived dark matter profiles on a wide range of
distances from the centers for any cluster, approximately in the
interval $[100; 1000]$ kpc, or at least up to the maximum radii
coming from observations.

Giving a more detailed glance to the absolute values of scalar field
parameters for clusters, we can see that the coupling constant
$\beta$ looks very well constrained in a really narrow range,
$[1.888; 3.259]$; while the gravitational length $L$ can vary in the interval
$[148.931; 1487.800]$ kpc, which seems consistent with the range
depicted by other used characteristic clusters scale, as the virial radius in a $\Lambda$CDM context.\\
To be more precise, if we consider (and exclude) three peculiar
cases, these intervals can even be constrained in narrower
windows: for $\beta$, the interval may be $[2.047; 2.786]$, while
for $L$ it may be $[147.977; 1323.890]$. \\
In \citep{Vik05}, MKW4 and RXJ1159  are recognized as the
strongest outliers in temperature profile, exhibiting very compact
cooling regions and having a temperature peak smaller than other
sample objects and located in inner region at $r \approx 50$ kpc.
RXJ1159 is better classified as an X-ray Over-Luminous Elliptical
Galaxy; optically, this object appears as a nearly isolated
elliptical galaxy but its X-ray luminosity and extent is typical
of poor clusters. MKW4 is considered a group of galaxies or a poor
cluster \citep{OSullivan03} too, so we have to consider them as
different object from the rest of the sample. Moreover we cannot
give any conclusion about this gravitational class (group of
galaxies)
because only one object has no statistical weight for any analysis.\\
Always in \citep{Vik05}, the cluster Abell 2390 seems to elude
typical clusters scaling relations; this is due to its unusual
central cool region, which extends up to $r \sim 400$ kpc,
probably because the cold gas is pushed out from the center by
radio lobes.

From these considerations, it is  interesting to note that just
these three objects result having the most different values for
the coupling constant $\beta$ with respect of the other clusters
and even seem to exhibit a peculiar trend. MKW4 (and RXJ1159) have
more compact inner cool regions and this is associated with: $1.$
higher values for $\beta$, which means a larger coupling of
scalar field with ordinary matter ($\beta$ acts like a
concentration parameter) and $2.$ smaller value for the
gravitational length. On the contrary, Abell 2390 shows a larger
cooling region, which corresponds to: $1.$ smaller values for
$\beta$ and $2.$ larger for $L$.

We underline that in our approach it is very important to find
these kind of correlations: if we want to make the scalar field
mechanism the most general and basic possible, we need to find any
possible link among its parameters and the physical properties of
our analyzed gravitational systems, so being able to perform
forecasts for other scaling properties and to recognize that
a scalar field is acting even when it is not directly possible to
derive easy quantities to be compared with data.

\begin{figure*}
\centering
  \includegraphics[width=80mm]{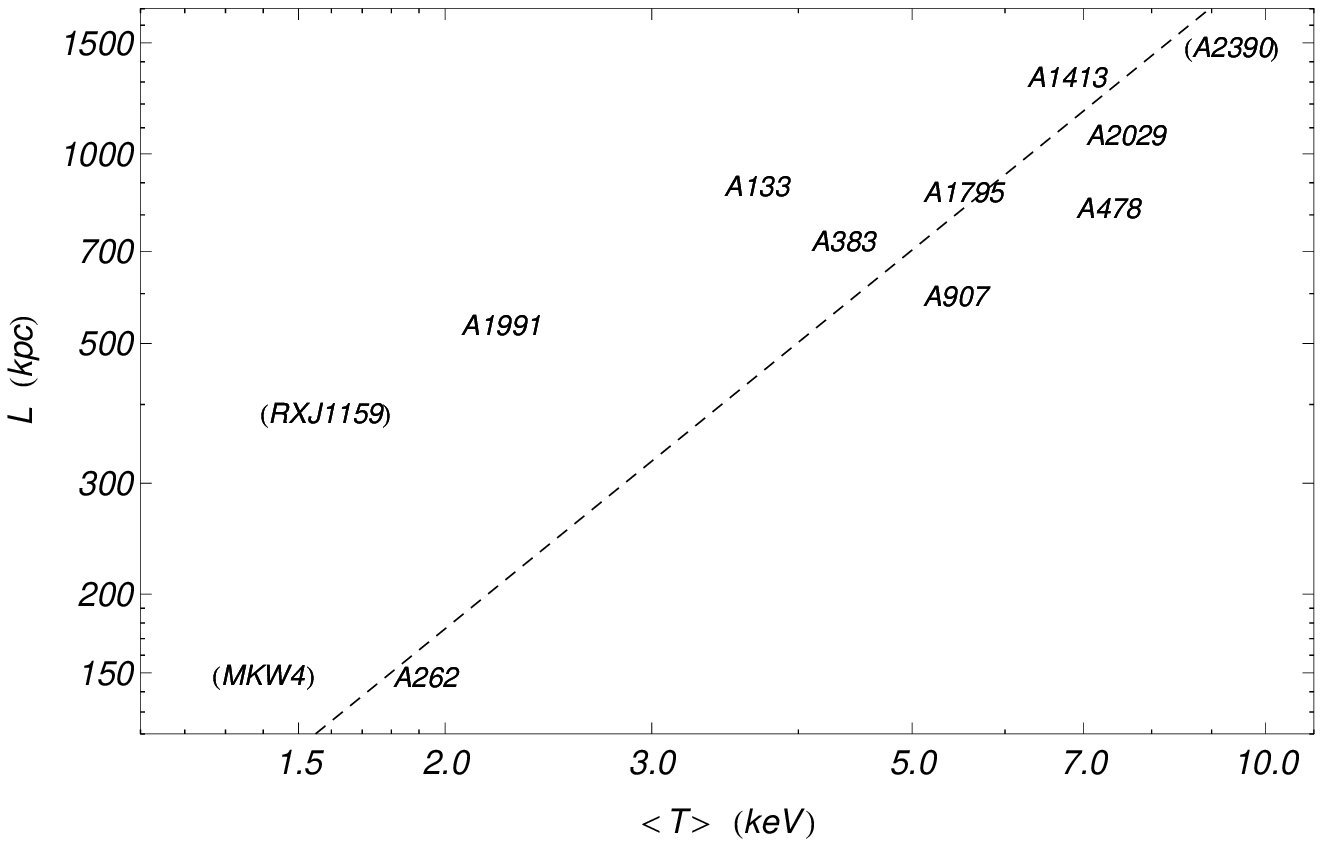}
  \includegraphics[width=80mm]{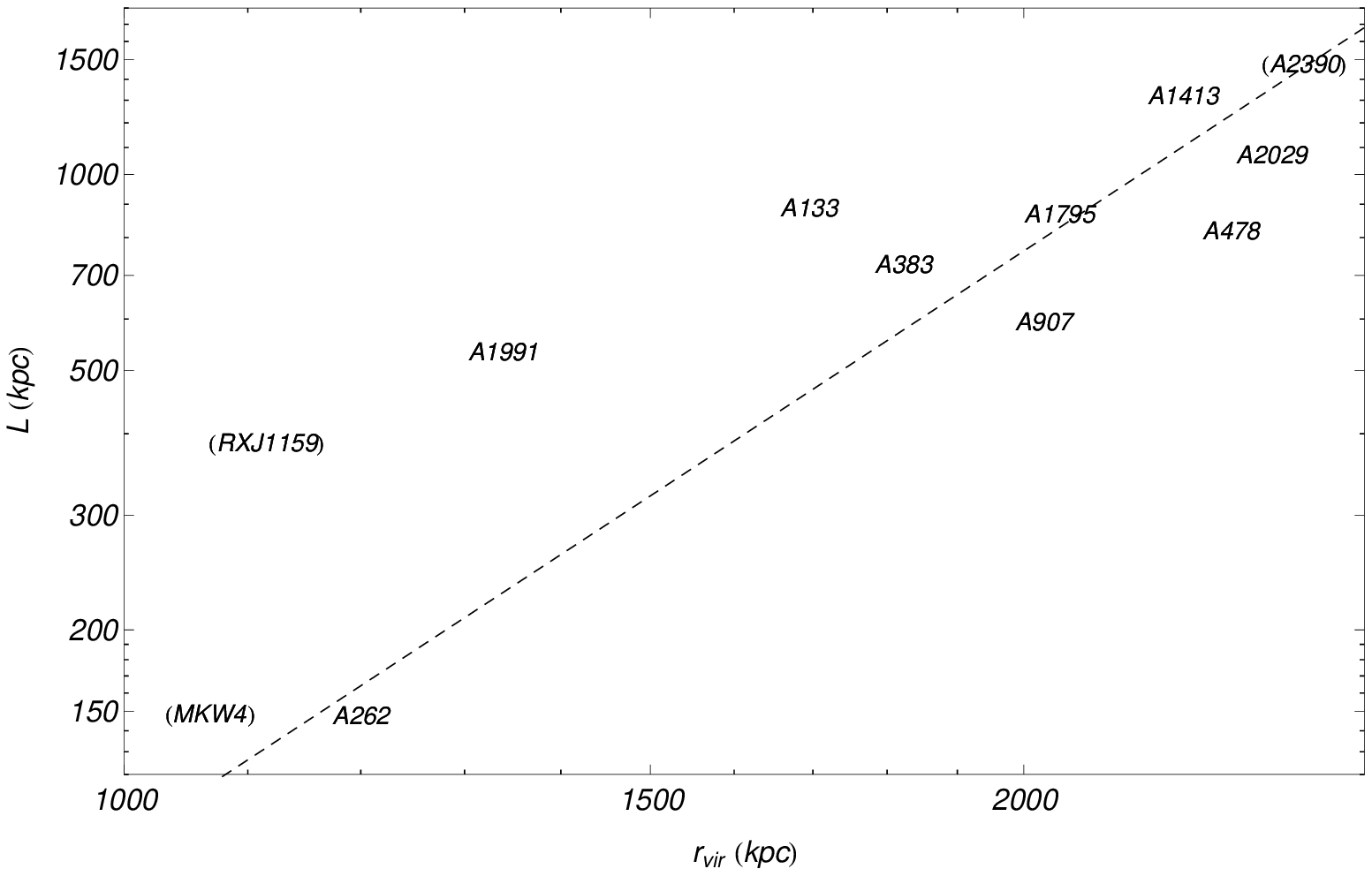}
  \caption{Scalar field length plotted versus mean (gas density weighted) cluster temperature (\textit{top panel})
  and the virial radius (\textit{bottom panel}). Objects in brackets are excluded from fits as described
  in \S~(\ref{sec:Cl results}).\label{fig:cluster_LTvirR}}
\end{figure*}

\begin{figure*}
\centering
  \includegraphics[width=80mm]{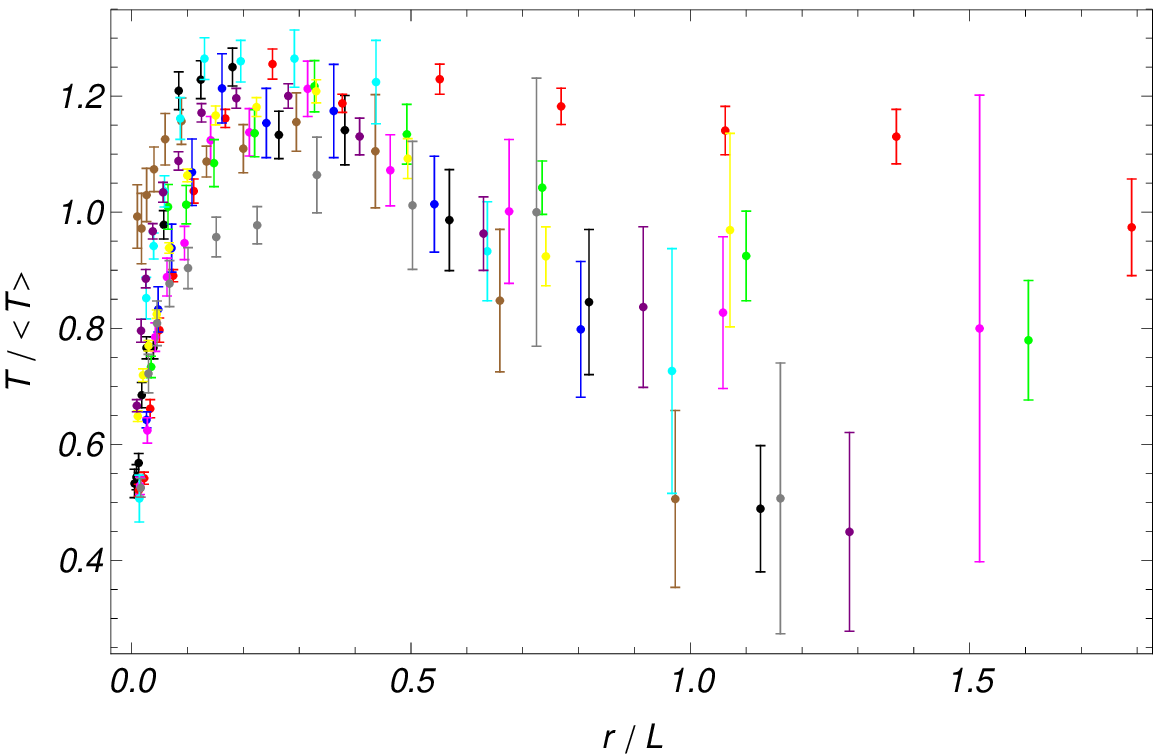}
  \includegraphics[width=80mm]{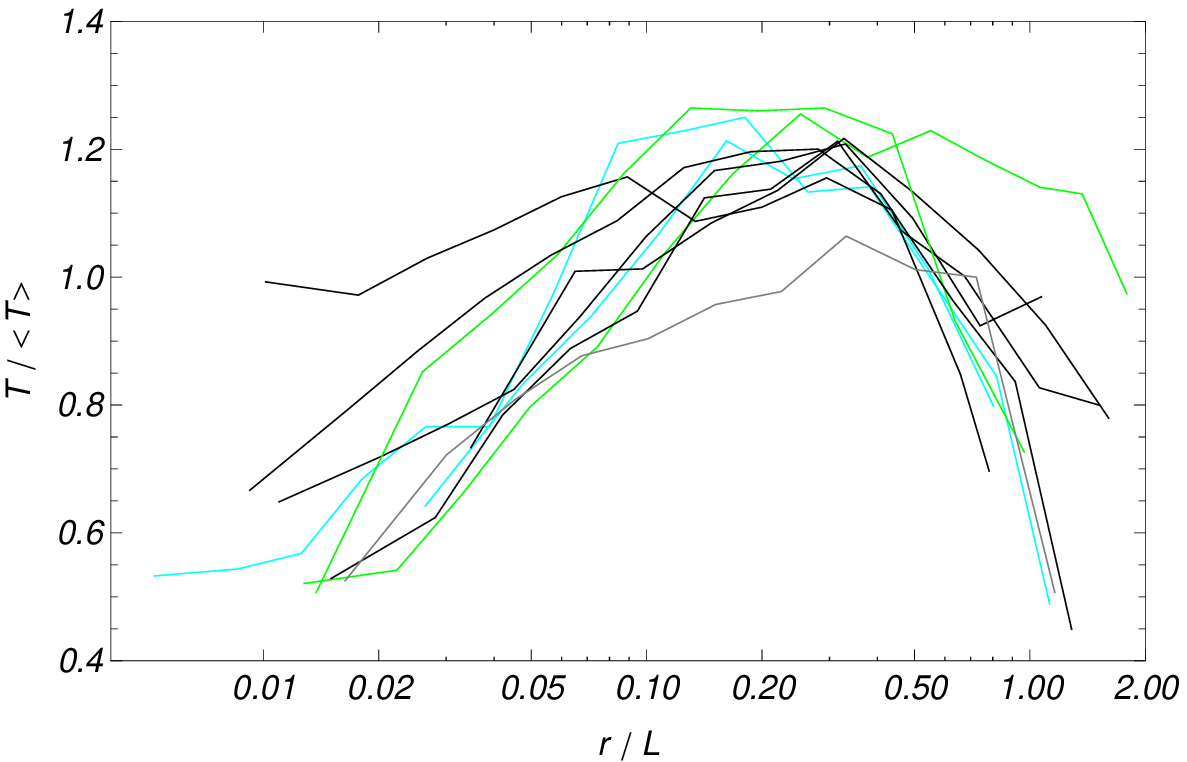}
  \caption{Temperature profiles for all clusters plotted as a function of distance from the center
  and in units of the scalar field length $L$. The temperatures are scaled to the mean (gas density weighted)
  cluster temperature. Only the two group galaxies, MKW4 and RXJ1159, have not been plotted. Colors in the bottom
  panel mean: black for clusters with $<T>$ greater than $5$ Kev; cyan for $<T>$ between $2.5$ and $5$ keV;
  green for $<T>$ less than $2.5$ KeV; grey for Abell 2390. \label{fig:cluster_T_profiles}}
\end{figure*}

Going on in this search, we have plotted the scalar field length $L$
versus the virial radius of any cluster (bottom panel in
Fig.~(\ref{fig:cluster_LTvirR})), calculated using the relation
\cite
{Bryan98,Evrard96},
\begin{equation}
r_{vir} = 1.95 \, {\mathrm{h}}^{-1} \, {\mathrm{Mpc}} \left(
\frac{<T>}{10 {\mathrm{keV}}} \right)^{1/2} \; ,
\end{equation}
where $h = 0.742$ \citep{Komatsu10} and $<T>$ is the average
temperature for any cluster, derived from gas-weighted fit to the
total cluster spectrum excluding the central region, and resulting
more directly correlated to clusters mass than the X-ray emission
weighted temperatures. Even if the two lengths result very
different, with the scalar field parameter $L$ always smaller than
the virial radius, we have a phenomenological relation between
them (we have performed a weighted linear fit which results better
than any higher order polynomial),
\begin{equation}
\log L = 3.0082 \cdot \log r_{vir} - 7.0476 \; .
\end{equation}
From it one can see that the scalar field length scales as $L \propto
r_{vir}^3 \sim M_{vir}$, so that it may possible to find a
relation between scalar field parameters and the
virial (Navarro - Frenk - White (NFW) model based) ones.\\
We have also checked a possible relation between $L$ and the mean
temperature $<T>$ (top panel in Fig.~(\ref{fig:cluster_LTvirR})),
with the final result:
\begin{equation}
\log L = 1.51085 \cdot \log <T> + 1.79122 \; .
\end{equation}
This last results is again very intriguing, because for clusters
of galaxies we have a mass-temperature relation \citep{Bryan98},
\begin{equation}
M_{\Delta} / T^{3/2} \propto H_{0}/H(z) \; ,
\end{equation}
where $\Delta$ is the overdensity level relative to the critical
density at the cluster redshift, so that $M_{180} = M_{vir}$. Then
we have to consider that for most of our clusters the redshift is
very small so that the ratio $H_{0}/H(z)$ is almost equal to $1$.
We remind that:
\begin{equation}
L \propto <T>^{3/2} \quad {\mathrm{and}} \quad L \propto
r_{vir}^{3} \sim M_{vir} \; ,
\end{equation}
so we can deduce that our scalar field length follows perfectly the
mass-temperature relation, being $L \propto <T>^{3/2} \propto
M_{vir}$ and that any scatter in the previous relation can be
attributed to differences in clusters redshifts.

Finally, we plotted in Fig.~(\ref{fig:cluster_T_profiles}) the
scaled temperature profiles versus the distance from the center of
any clusters scaled with respect of the scalar field length $L$
obtained by fit. It is extremely interesting and not obvious that
the profiles are all re-scaled and self-similar as it happens in
the usual approach. We can even say something more: in the
scalar field approach we can see that some properties which are
depicted in Fig.~(16) of \citep{Vik05}, as for example the
different profiles among subgroups of clusters with mean
temperature in different ranges (less than $2.5$ keV, between
$2.5$ and $5.$ keV and greater than $5$ keV) with distances
rescaled with respect of $r_{500}$, here disappear. In fact, all
the clusters form an homogeneous gravitational family (the only
exception being Abell 2390, as discussed before). We could say
that the scalar field length $L$ contains more fundamental
information than the virial radius.

\subsection{LSB galaxies: results}
\label{sec:LSB_results}

Taking into account  LSB galaxies we have one parameter more than
the clusters case: together with the intrinsic scalar field
parameters we have the stellar mass-to-light ratio, $Y_{\ast}$,
which we need for the conversion from the stellar surface
photometry to the stellar mass density. In principle this
parameter could be left free; but we have decided to put a prior
on it derived from literature. In \citep{vandenHoek00} they
investigate star formation history and chemical evolution of LSB
galaxies by modeling their observed spectro-photometric and
chemical properties using a galactic chemical and photometric
evolution model incorporating a detailed metallicity dependent set
of stellar input data. Results show that $Y_{\ast}$ for this class
of galaxies has usually values between $\sim 0.5$ and $\sim 2$.
For this reason we have constrained the mass-to-light ratio to the
interval $[0; 5]$ as a conservative hypothesis, and only in one
case (UGC3851) we
had to enlarge the interval up to $10$.\\
An important assumption about this parameter is that we assume it
being constant over the whole range of data; this is an
unavoidable and the most general assumption one can do without
having details of stellar population distribution inside LSB
galaxies.

A comment is in order  for a correct evaluation of the results:
during the analysis we have compared the theoretical rotation
curves coming out from our scalar field model using only the
observable-matter densities described in \citep{SwatersPhD} with
data firstly published in \cite {deBlok02}. These data, available
in the SIMBAD data base, consist of the contributions to the total
rotation curve coming from the stellar component (with an assumed
mass-to-light ratio $Y_{\ast} = 1$) and the gas one (with gas
density normalized with a factor $1.4$ for taking into account
helium contribution), together with the total observed rotation
curve (which can be considered an effect of dark matter in a CDM
scenario or coming out from the interaction with a scalar field as
in our approach). Both the total and the gas rotation curve have
been submitted to a smoothing procedure for deleting
irregularities mainly coming from two different elements. First,
the  assumption that has been made when deriving mass models from
rotation curves is that there is symmetry, that all mass in on
circular orbits and that there is continuity with radius. Raw data
show scatter and non-circular motions which can produce virtual or
ambiguous rotation velocities. Second, gas densities often show
small-scale structures, irregularities and look clumpy at any
distance from the center, and this can give possible large
fluctuations in the rotation curve. In \cite {deBlok02} the
smoothing procedure is tested of course, and the smooth rotation
curve is a very good approximation to  the raw rotation curve. But
some discrepancies with our theoretical estimations can be find
because of we have used
raw gas density profiles from \citep{SwatersPhD}. \\
If we consider that all  gas contributions to rotation curve are
multiplied to $\beta$ only, while the star ones to the combination
$\beta \cdot Y_{\ast}$, in some cases, depending on the fit values
and on their contribution, irregularities in the gas distribution
are emphasized and may affect the total rotation curve profile.

{\renewcommand{\arraystretch}{1.5}
\begin{table*}
 \begin{center}
 \caption{\textit{LSB galaxies.} Column 1: UGC number. Column 2: Distance from the source literature.
 Column 3: disk central surface brightness in R-band, corrected for galactic extinction and inclination. Column 4:
 disk scale length. Column 5: total HI gas mass. Column 6: Maximum rotation velocity. Column 7: best fit stellar mass to light ratio ($1\sigma$ confidence interval). Column 8: coupling parameter $\beta$ from scalar field ($1\sigma$ confidence interval). Columns 9: gravitational length $L$ from scalar field ($1\sigma$ confidence interval). \label{tabspiral}}
 \begin{tabular}{ccccccccc}
  \tableline
  UGC    &   D   &    $\mu_{0,R}$      & $R_{d}$ &        $M_{HI}$      & $V_{max}$ & $Y_{\ast}$   & $\beta$ &  $L$  \\
         & (Mpc) & (mag/arcsec$^{-2}$) &  (kpc)  & ($10^{8} M_{\odot}$) &   (km/s)  &$(Y_{\odot})$ &         & (kpc) \\
  \tableline
  \tableline
  U1230  & $51$   & $22.6$ & $4.5$ & $58.0$ & $103$ &$2.09^{+0.96}_{-0.66}$ & $1.328^{+0.281}_{-0.237}$ & $31.011^{+42.784}_{-14.836}$ \\
  U1281  & $5.5$  & $22.7$ & $1.7$ & $3.2$  &  $57$ &$0.77^{+0.16}_{-0.12}$ & $1.381^{+0.195}_{-0.155}$ & $4.006^{+7.362}_{-1.938}$ \\
  U3137  & $18.4$ & $23.2$ & $2.0$ & $43.6$ & $100$ &$1.97^{+0.15}_{-0.15}$ & $1.837^{+0.030}_{-0.028}$ & $75.810^{+117.703}_{-18.384}$ \\
  U3371  & $12.8$ & $23.3$ & $3.1$ & $12.2$ &  $86$ &$1.70^{+0.58}_{-0.41}$ & $1.444^{+0.206}_{-0.173}$ & $9.864^{+26.568}_{-5.450}$ \\
  U3851  & $3.4$  & $22.6$ & $1.5$ & $7.3$  &  $55$ &$6.19^{+0.44}_{-0.77}$ & $0.238^{+0.398}_{-0.189}$ & $0.348^{+0.577}_{-0.181}$ \\
  U4173  & $16.8$ & $24.3$ & $4.5$ & $21.2$ &  $57$ &$1.67^{+0.69}_{-0.64}$ & $0.957^{+0.232}_{-0.375}$ & $4.534^{+12.419}_{-3.061}$ \\
  U4278  & $10.5$ & $22.5$ & $2.3$ & $13.6$ &  $93$ &$1.23^{+0.16}_{-0.13}$ & $1.299^{+0.074}_{-0.078}$ & $71.714^{+112.411}_{-45.008}$ \\
  U4325  & $10.1$ & $21.6$ & $1.6$ & $7.5$  & $123$ &$0.18^{+0.06}_{-0.03}$ & $4.339^{+3.051}_{-1.376}$ & $1.068^{+2.315}_{-0.664}$ \\
  U5721  & $6.7$  & $20.2$ & $0.5$ & $6.6$  &  $79$ &$0.28^{+0.05}_{-0.04}$ & $2.203^{+0.148}_{-0.114}$ & $7.210^{+12.776}_{-2.429}$ \\
  U7524  & $3.5$  & $22.2$ & $2.3$ & $9.7$  &  $83$ &$1.52^{+1.09}_{-0.38}$ & $1.579^{+0.444}_{-0.539}$ & $2.212^{+0.964}_{-0.699}$ \\
  U7603  & $6.8$  & $20.8$ & $0.7$ & $5.4$  &  $64$ &$0.014^{+0.017}_{-0.010}$ & $1.770^{+0.066}_{-0.058}$ & $23.767^{+105.159}_{-15.367}$ \\
  U8286  & $4.8$  & $20.9$ & $0.8$ & $3.5$  &  $84$ &$0.243^{+0.023}_{-0.021}$ & $2.416^{+0.078}_{-0.075}$ & $29.784^{+77.252}_{-18.995}$ \\
  U8837  & $5.1$  & $23.2$ & $1.2$ & $1.6$  &  $50$ &$0.888^{+0.409}_{-0.256}$ & $1.940^{+0.689}_{-0.485}$ & $0.349^{+0.436}_{-0.154}$ \\
  U9211  & $12.6$ & $22.6$ & $1.2$ & $10.5$ &  $64$ &$1.274^{+0.677}_{-0.458}$ & $1.525^{+0.527}_{-0.281}$ & $3.484^{+8.479}_{-1.898}$ \\
  U10310 & $15.6$ & $22.0$ & $1.9$ & $12.6$ &  $75$ &$0.489^{+0.191}_{-0.123}$ & $1.570^{+0.516}_{-0.294}$ & $3.684^{+6.726}_{-1.930}$ \\
  \tableline
 \end{tabular}
 \end{center}
\end{table*}}

We can now discuss the obtained values for the
stellar mass-to-light ratio. It is easy to see that 10 galaxies
have values compatible with the prescribed range by
\citep{vandenHoek00}; one (UGC3851) has a higher value which is
hard to explain in term of reasonable population synthesis; and
four (UGC4325, UGC5721, UGC7603, UGC8286) have values smaller than
$0.5$. Among these, one is particularly problematic, i.e. UGC7603,
because it has a mass-to-light ratio too much small, $Y_{\ast} = 0.017$,
while the others can be again considered acceptable.\\
UGC3851 is a real challenge for both our approach and the more
traditional one \citep{deBlok02}. The very linear rise in the
inner region rapidly changes in a flat part at larger radii so
that it is very hard to reproduce this sharp change in the slope.
There are two possibilities for this behavior: the HI curve in
outer points is underestimated, or there are non-circular motions
related to a bar-like structure and a star forming region in the
center which affect the H$\alpha$ observations. It is interesting
to note the many similarities between our analysis and that one in
\cite {deBlok02}: both their NFW and isothermal halo model
overestimate velocity in the inner region, with a larger deviation
in the NFW model than the isothermal one, and larger in the inner
region than in the outer one, as in our case. Then their derived
maximum value for the stellar mass-to-light ratio is $Y_{\ast} =
5.4$, very close to our value. Anyway, all the following relations
will be derived without considering this galaxy because no
definitive conclusion can be done about it.

Among the four galaxies with a low mass to light ratio, one of the
most problematic case is UGC5721, showing a very narrow
bi-modality in the parameters distribution that cannot be avoided
and resolved even changing parameters of the trial distribution in
the MCMC. The convergence test inevitably fails in this case, but
we have assumed as best fit results the values associated to the
highest peak in the likelihood distribution shown in
Fig.~(\ref{fig:cham_gal4}), which produce a very good fit to the
observational rotation curve. A discrepancy can be found around
$2$ kpc but as the large error bar shows, in this region there are
some observational difficulties related to gas observations which
produce the detected faster rise in velocity.
\begin{figure*}
\centering
  \includegraphics[width=80mm]{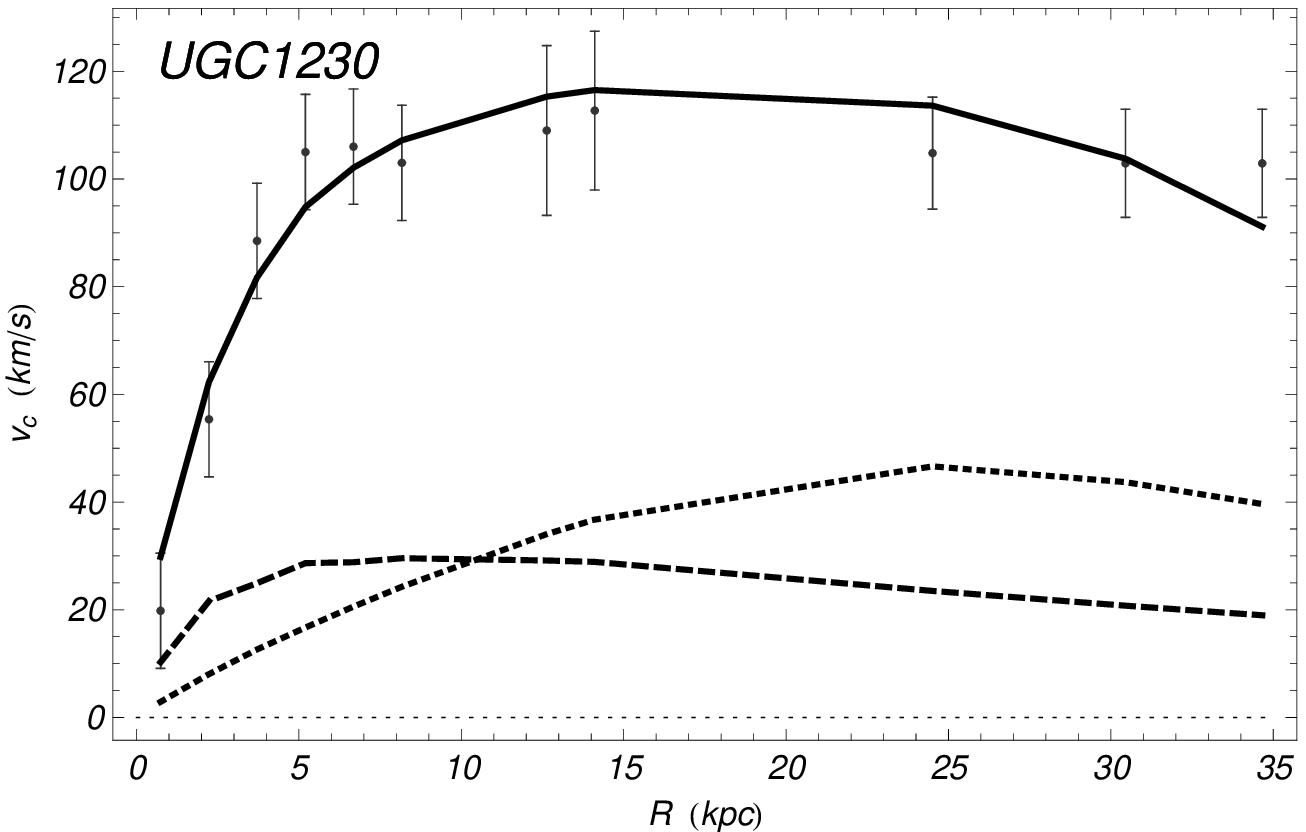}
  \includegraphics[width=80mm]{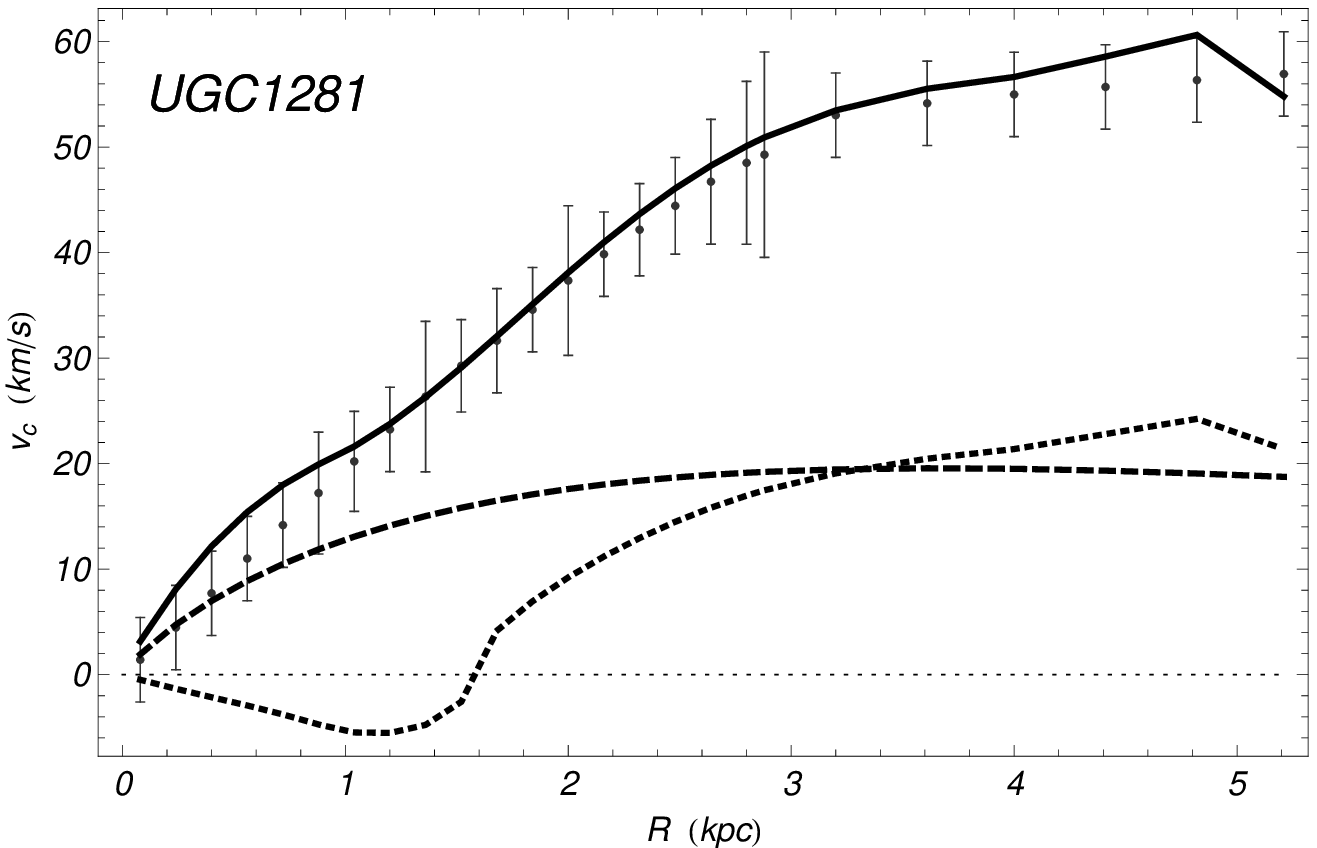}
  \includegraphics[width=80mm]{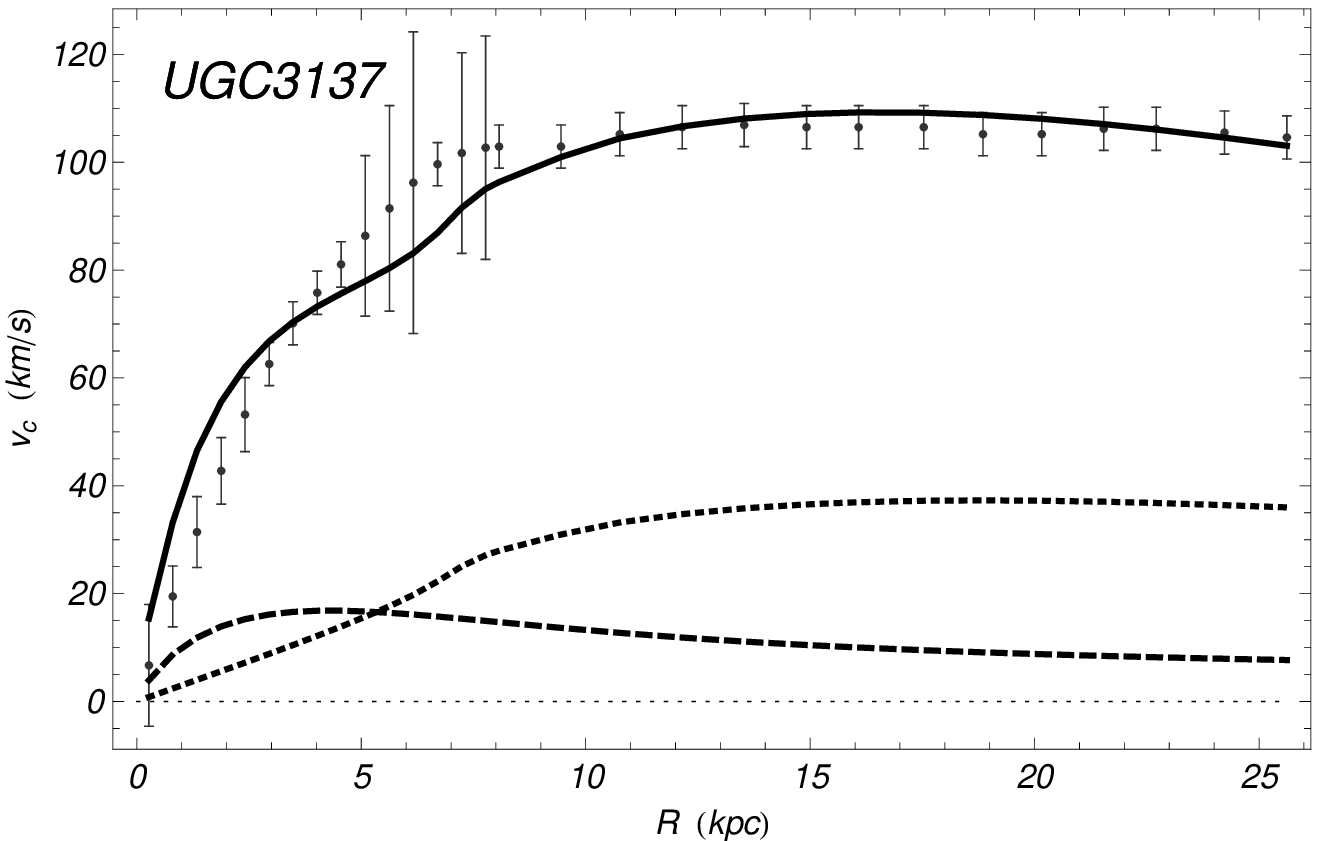}
  \includegraphics[width=80mm]{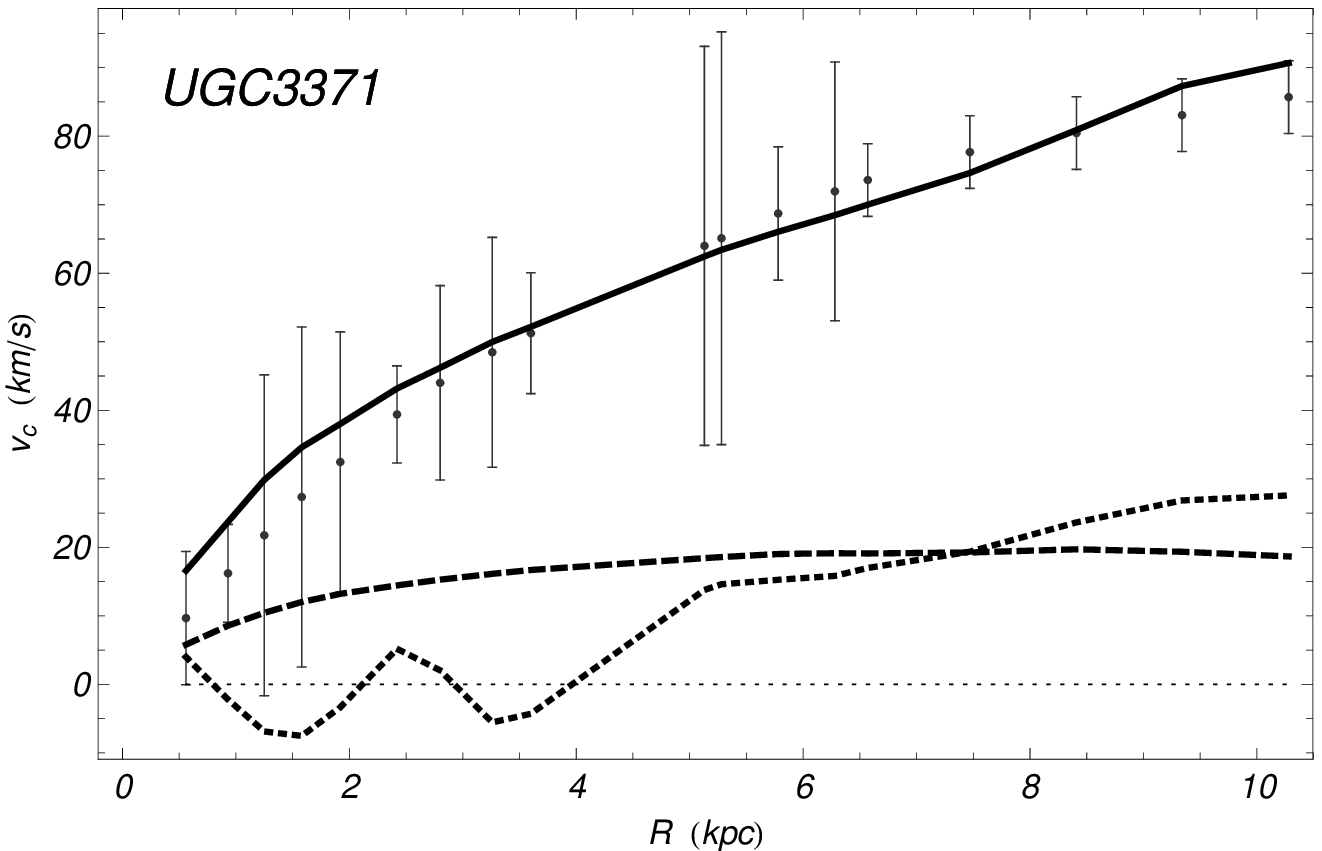}
  \includegraphics[width=80mm]{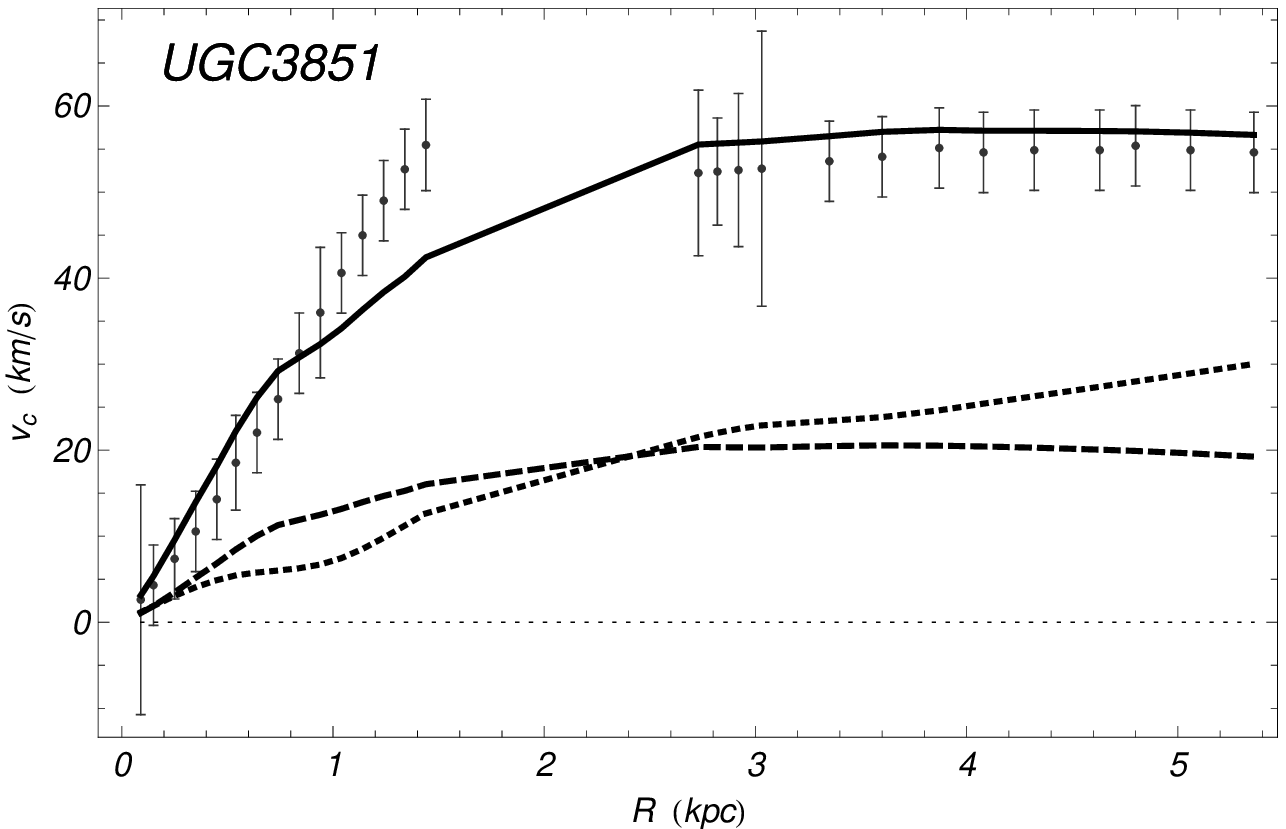}
  \includegraphics[width=80mm]{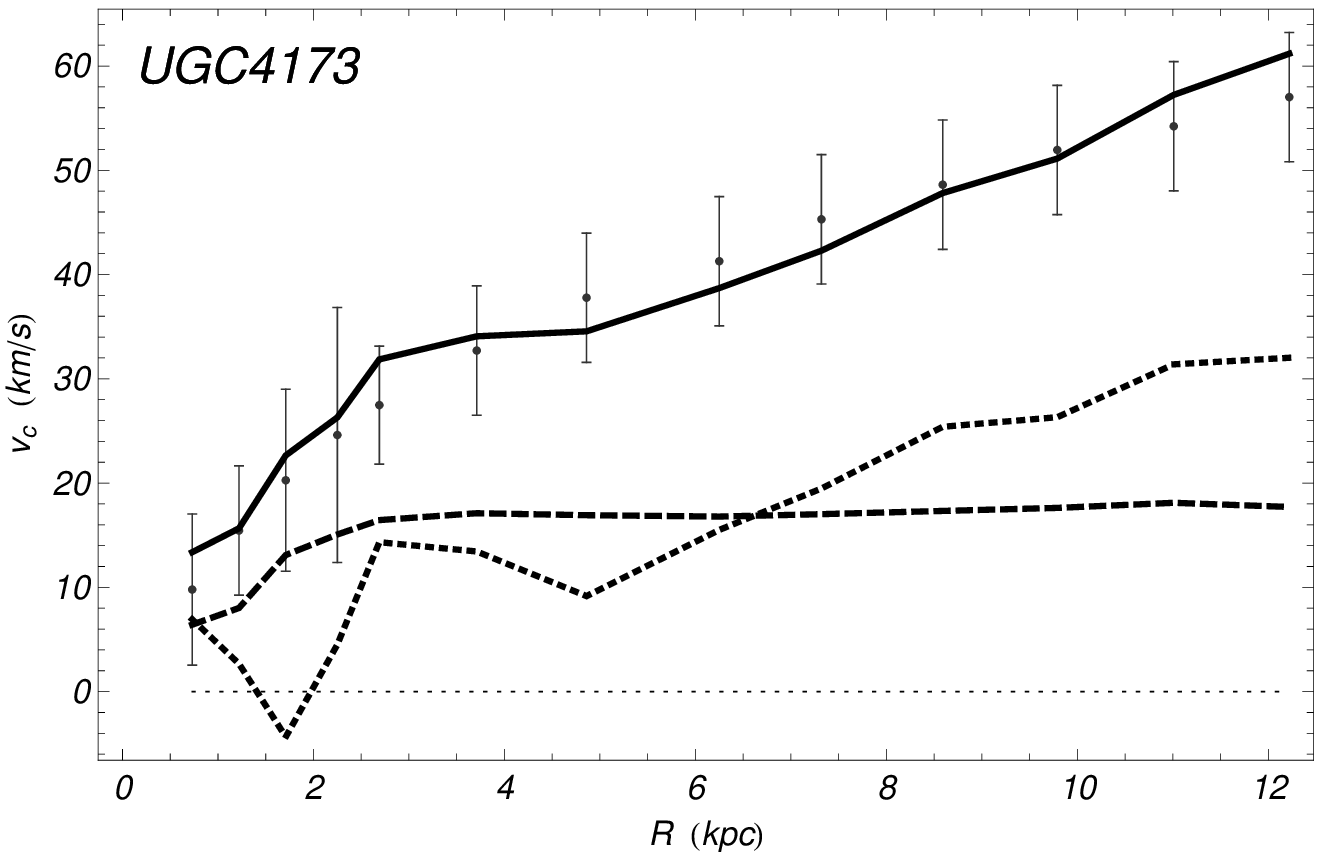}
  \caption{Rotation curves of LSB galaxies. Dots are velocities from data; solid line is the
  theoretical model, Eq.~(\ref{eq:rot_vel_final}); dashed line is the stellar contribution to
  the rotation curve assuming $Y_{\ast} = 1$; dotted line is the gas contribution. \label{fig:cham_gal1}}
\end{figure*}
\begin{figure*}
\centering
  \includegraphics[width=80mm]{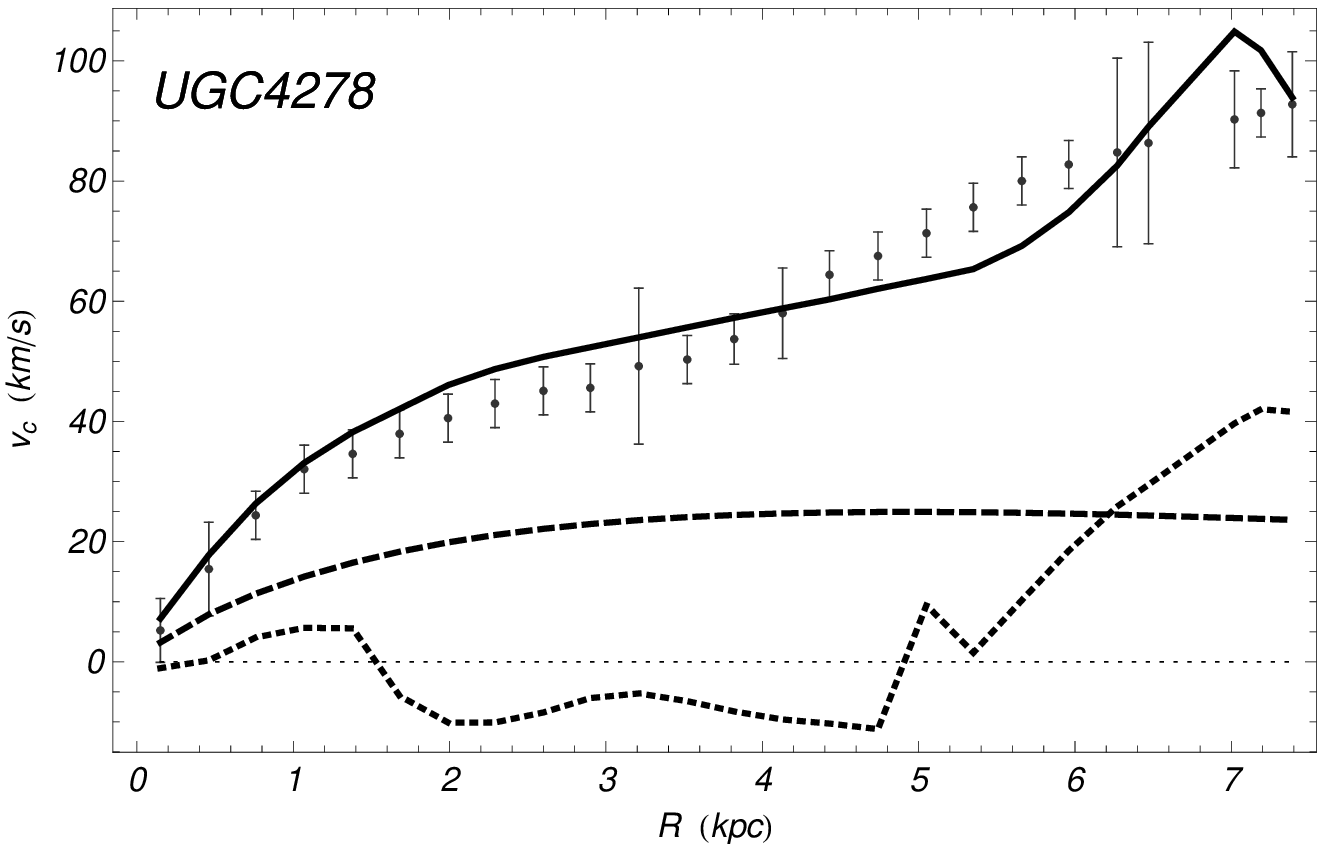}
  \includegraphics[width=80mm]{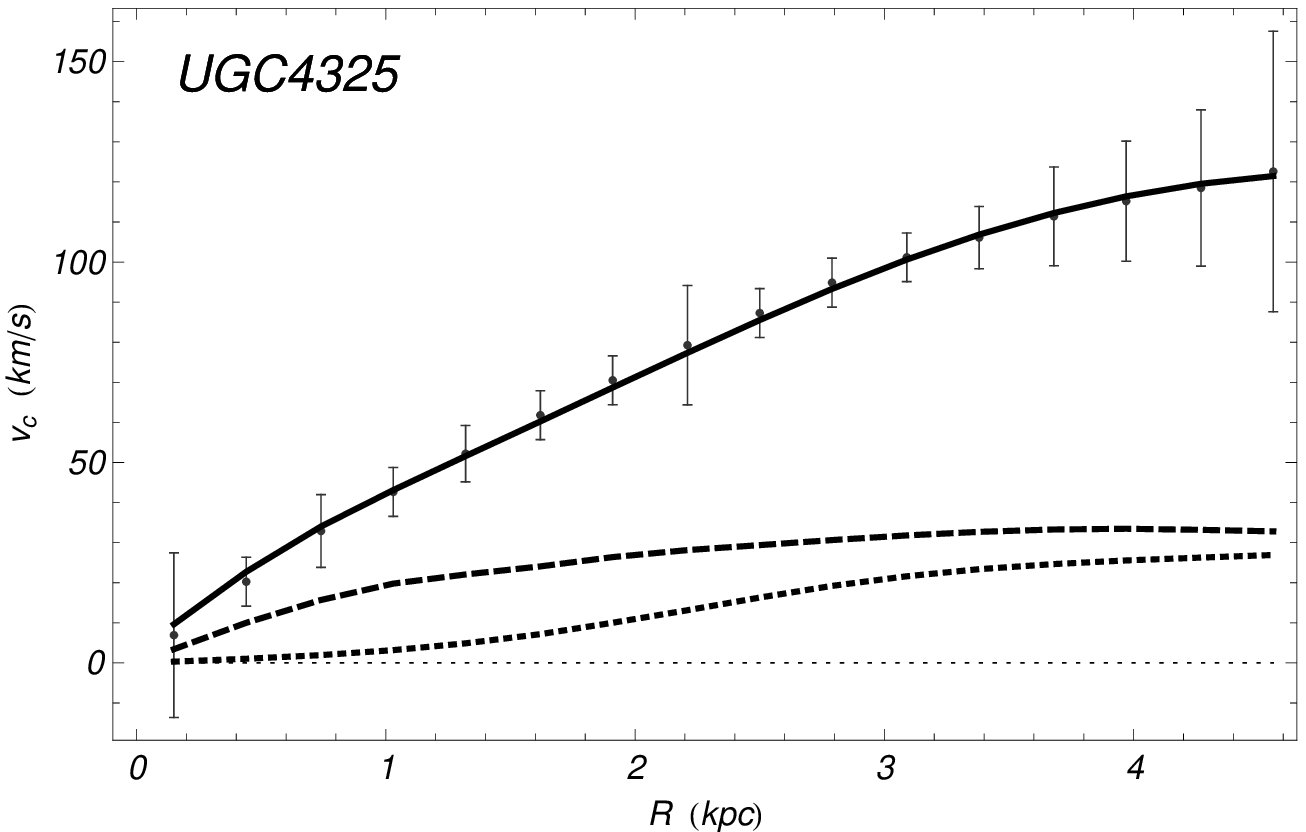}
  \includegraphics[width=80mm]{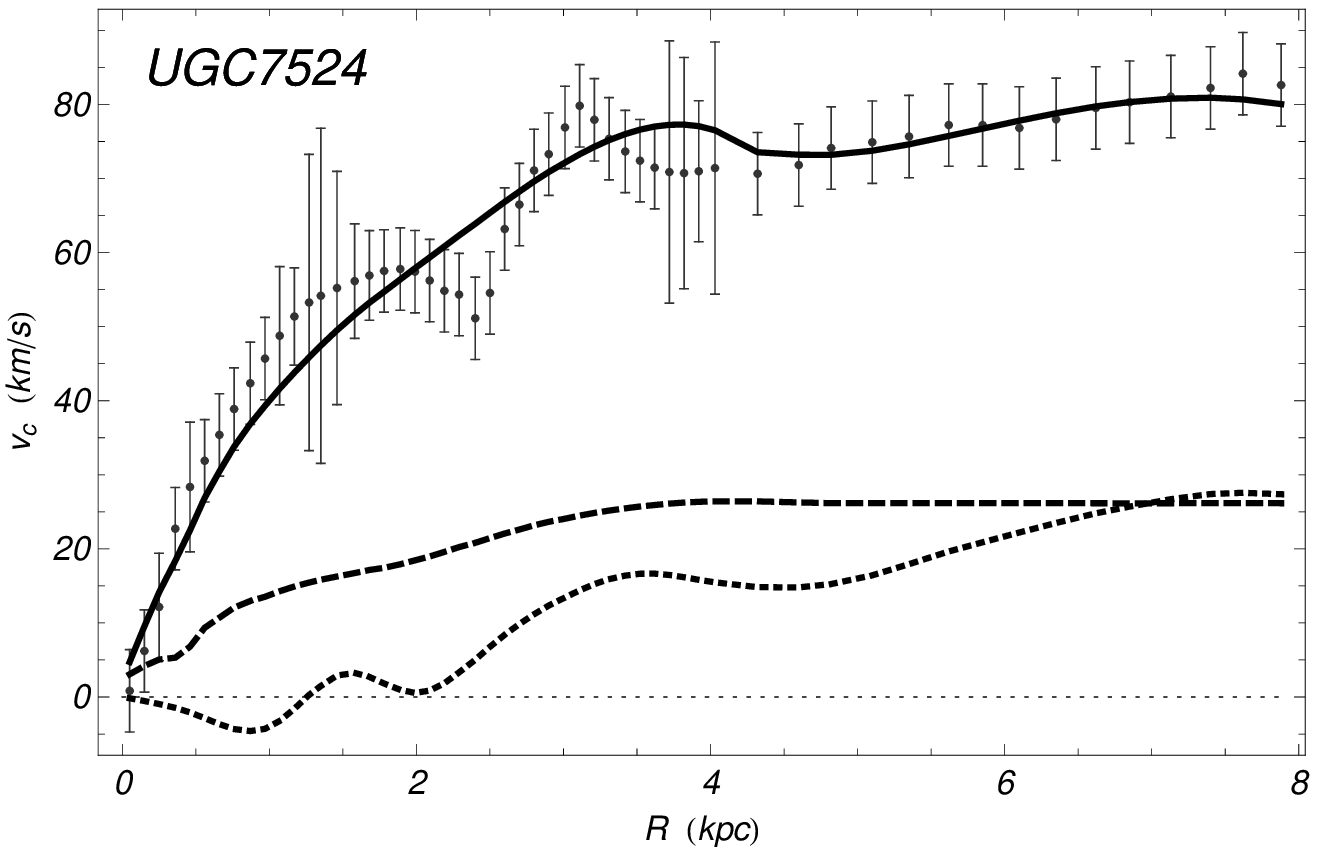}
  \includegraphics[width=80mm]{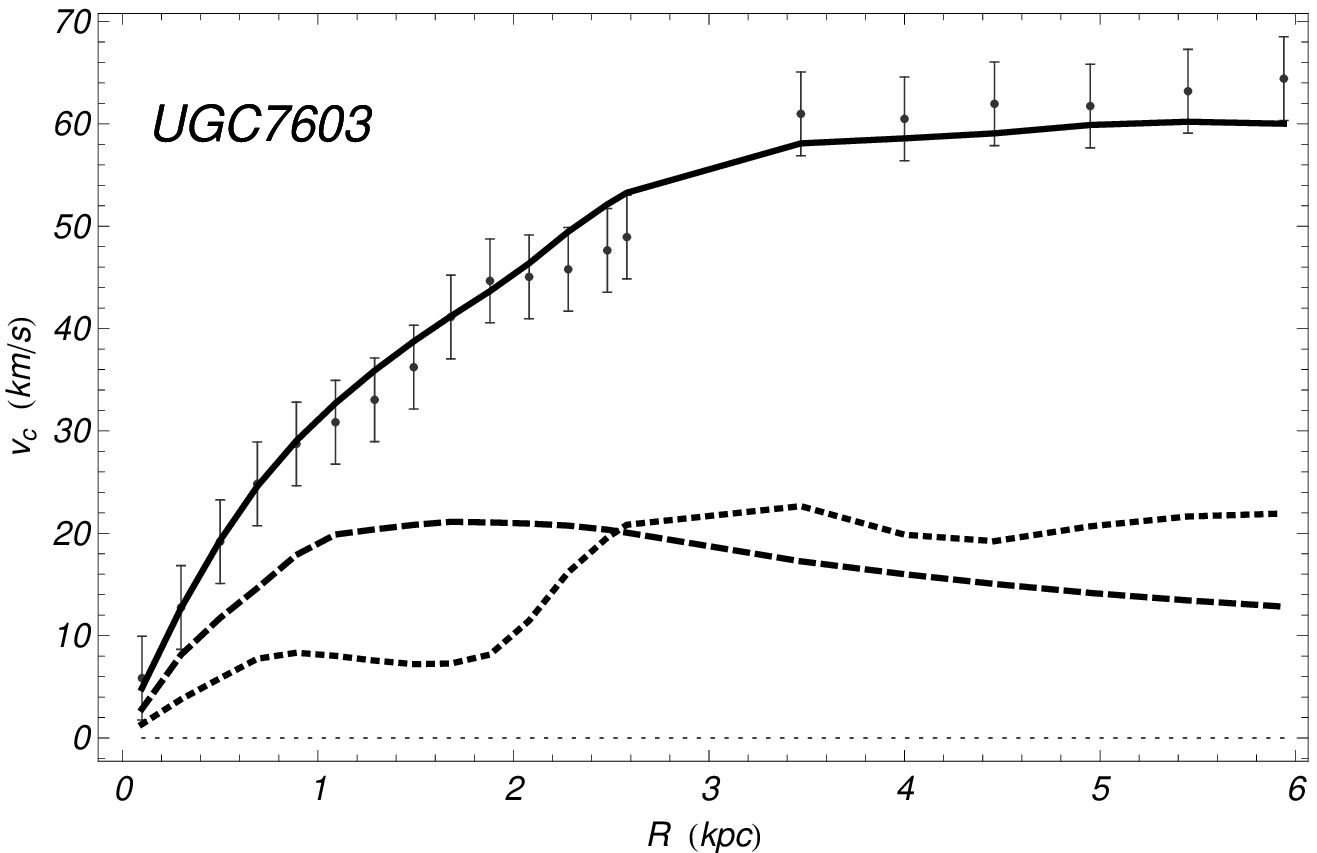}
  \includegraphics[width=80mm]{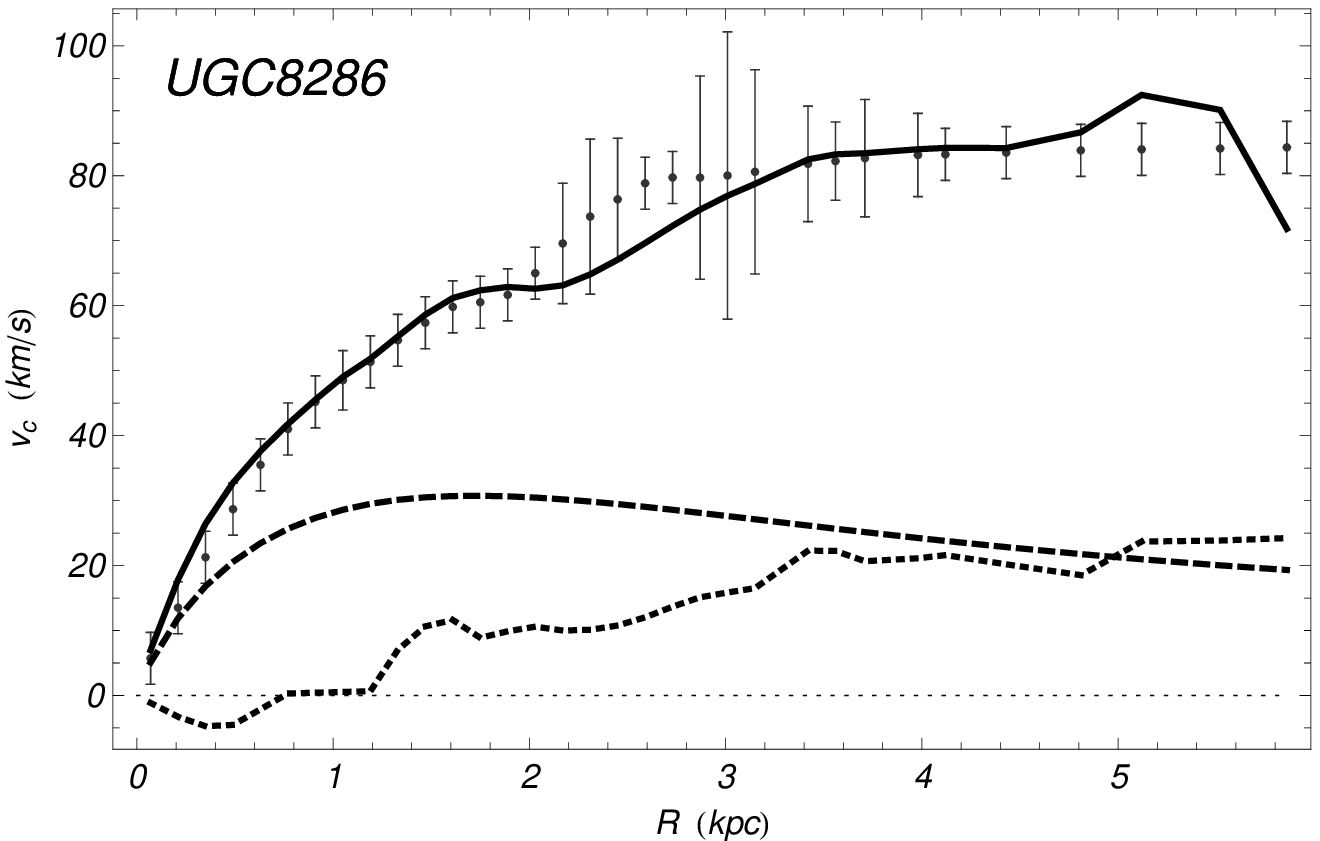}
  \includegraphics[width=80mm]{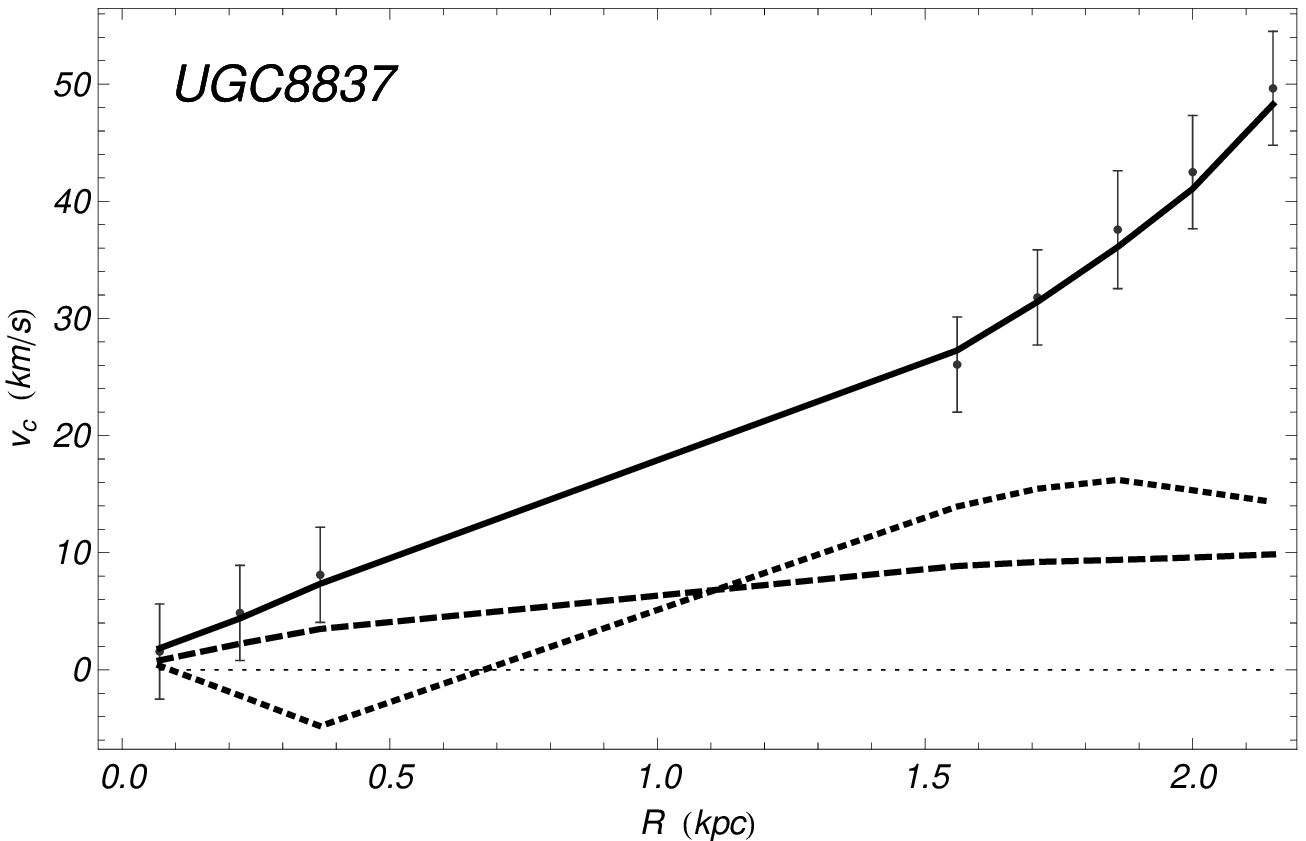}
  \includegraphics[width=80mm]{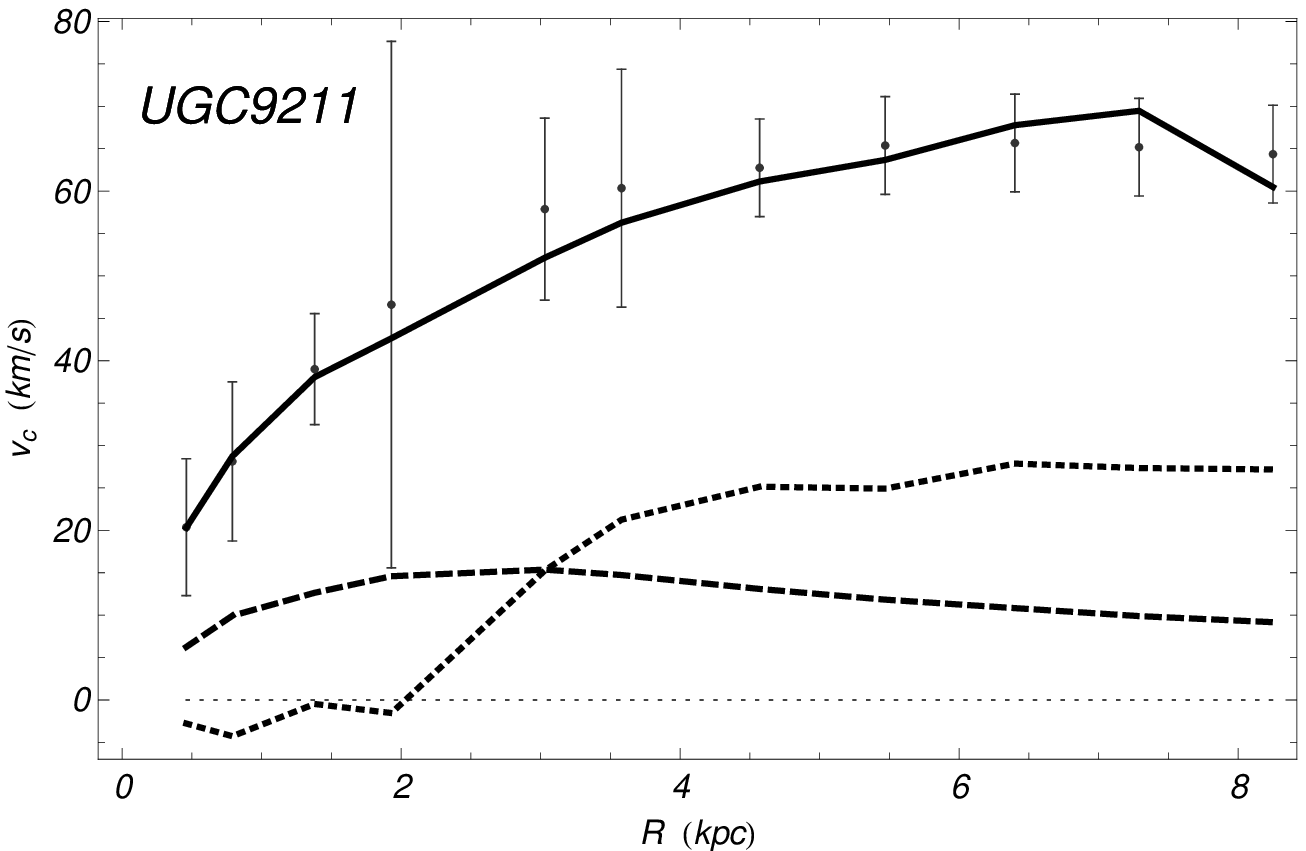}
  \includegraphics[width=80mm]{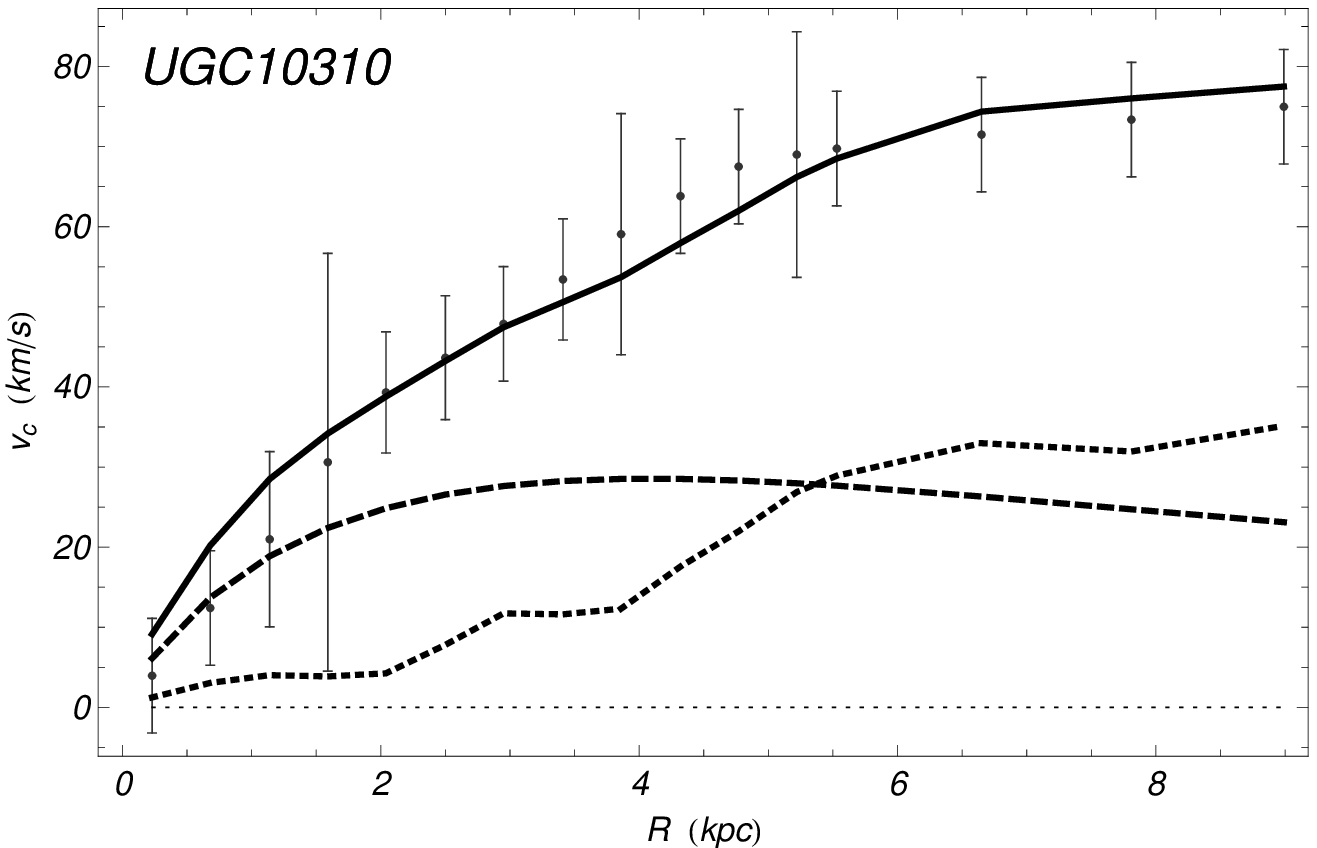}
  \caption{Same of Fig.~(\ref{fig:cham_gal1}). \label{fig:cham_gal2}}
\end{figure*}
%
%
\begin{figure*}
\centering
  \includegraphics[width=80mm]{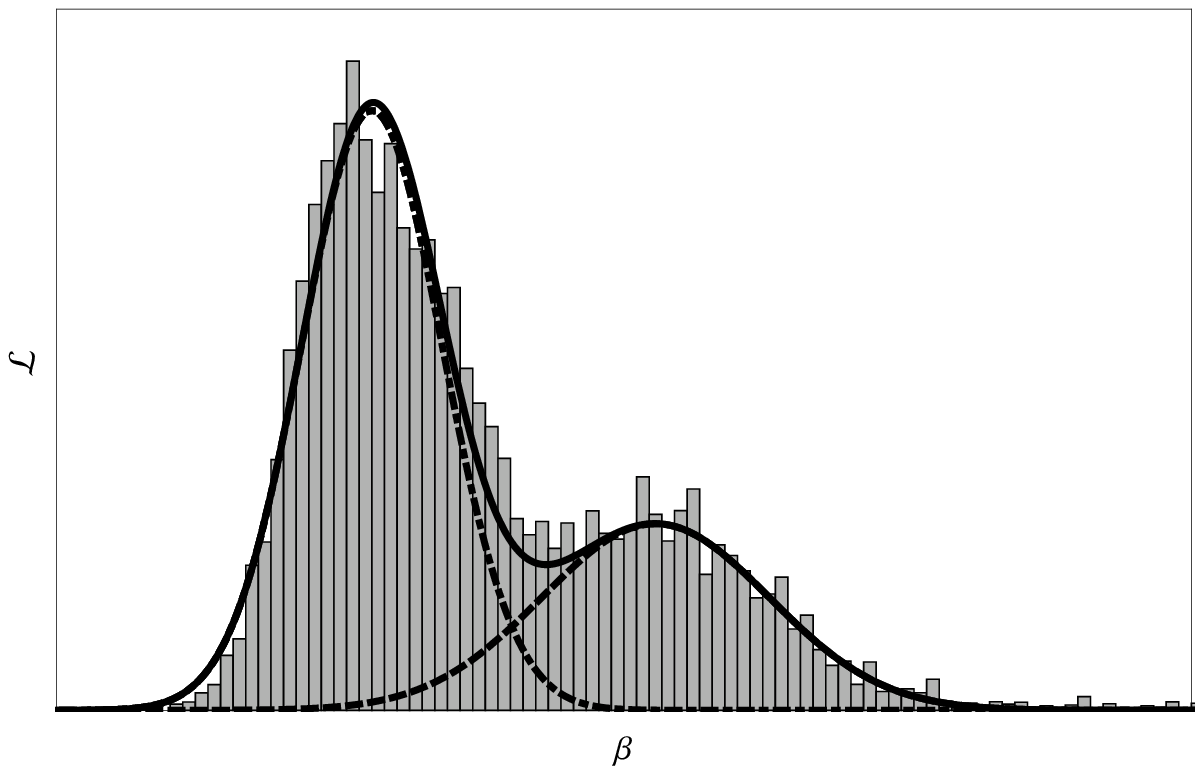}
  \includegraphics[width=80mm]{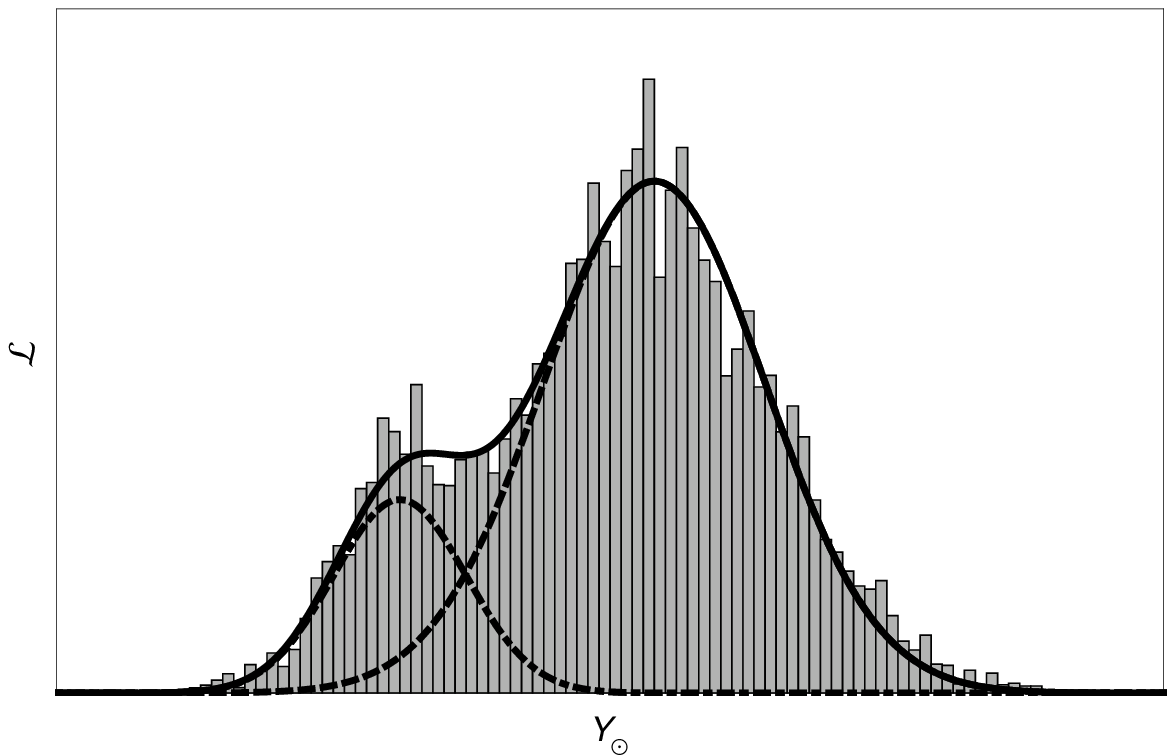}
  \includegraphics[width=80mm]{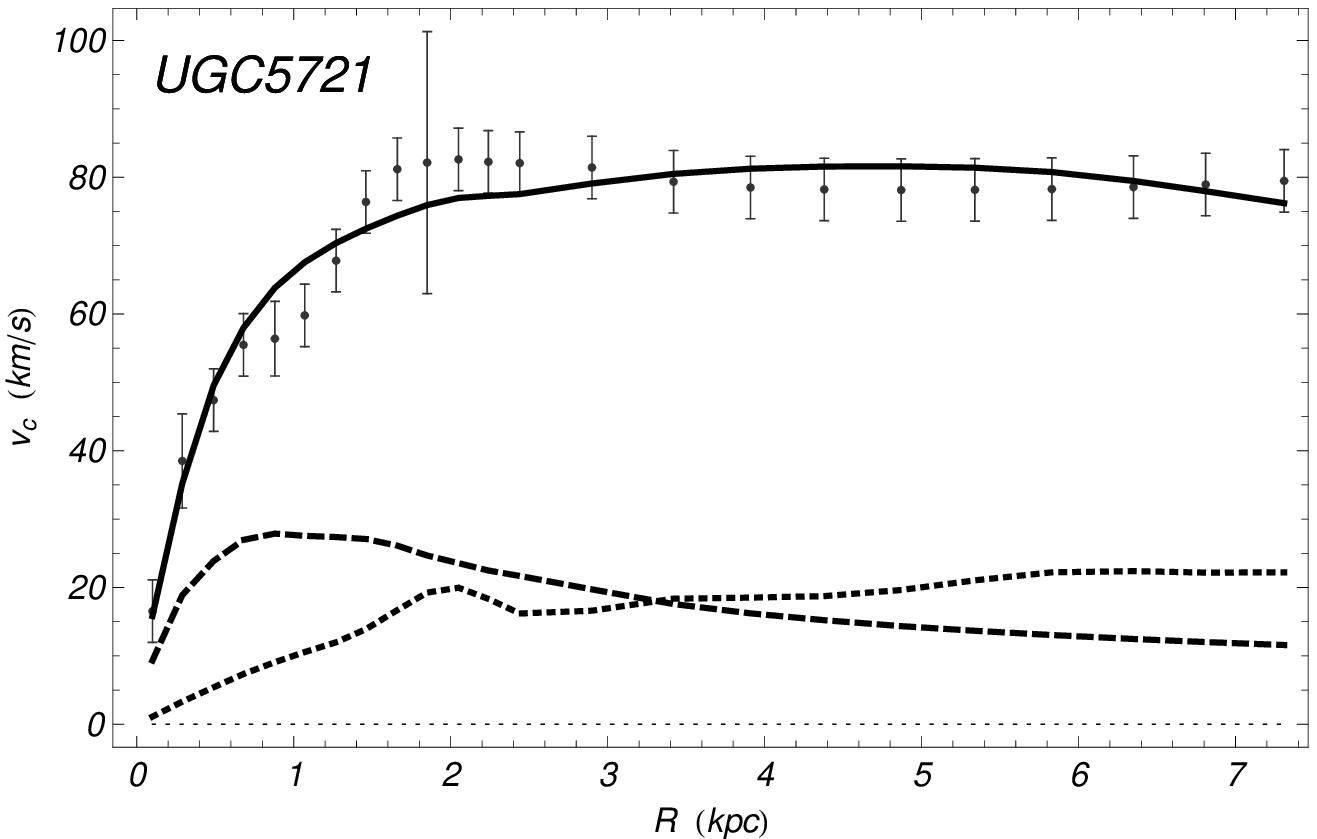}
  \caption{Same of Fig.~(\ref{fig:cham_gal1}) for UGC5721. In upper plots are shown the marginalized probability
  distributions for $\beta$ and $Y_{\ast}$ obtained from MCMCs and the unresolved bi-modality discussed in
\S~(\ref{sec:LSB_results}).\label{fig:cham_gal4}}
\end{figure*}

For what concerning the scalar field parameters, the coupling constant
$\beta$ seems to be very well constrained in the range $[0.957; 2.416]$, if
we exclude the previously discussed case of UGC3851, whose value is very low,
$\beta = 0.238$, probably because of the degeneracy with $Y_{\ast}$,
and the opposite situation of UGC4325, which exhibits a large value, $\beta = 4.339$,
and also larger errors on this parameter than other galaxies. It is not
a causality that this is also one of the four galaxies with a stellar mass-to-light
ratio lower than the simulation limit. \\
As in  clusters case, the scalar field length shows a larger spread, mainly
scaling with the dimension of any LSB galaxy. One problematic situation is UGC8837,
which clearly deviates from the general trend
for what concerning the value of scalar field length $L$, resulting it very
small. Even if the fit of our model to data is very good, we do not
consider anymore this galaxy in the following discussion because its results
are strongly related to the low quality of data. As discussed in
\cite
{deBlok02}, the H$\alpha$ are of good quality, but do not correspond
very well to the HI profile, with a large difference in their
systemic velocity probably due to the interference of inner
star formation region and there located non-circular motions. A further
error source can be also the inclination, being this galaxy
almost edge-on, the most problematic configuration in this case.

Just by visual inspection it is possible to see that the general
shapes of the rotation curves are quite well predicted by our model inside the error
confidence region, even considering some singular cases where are
present irregularities coming from gas distribution (clumpy
peaks, change in the profile convexity). In evaluating the
goodness of our fitted parameters we have to note that all the
rotation curves are limited to a certain distance from the center
of any galaxy. It is well known that spiral galaxies show a great
variety of rotation curves profiles, with flat velocity plateaux,
different slopes (both increasing and decreasing) in outer regions;
but in our case the necessity to match stellar photometry with
more extended gas observations, results in a limited distance range
which may affect our results. As pointed out in
\citep{Capozziello07} this could not solve the previously described
degeneracies between the fitting parameters, or to induce their wrong estimations, so
requiring more extended data for a more exact analysis.\\

\begin{figure*}
\centering
  \includegraphics[width=120mm]{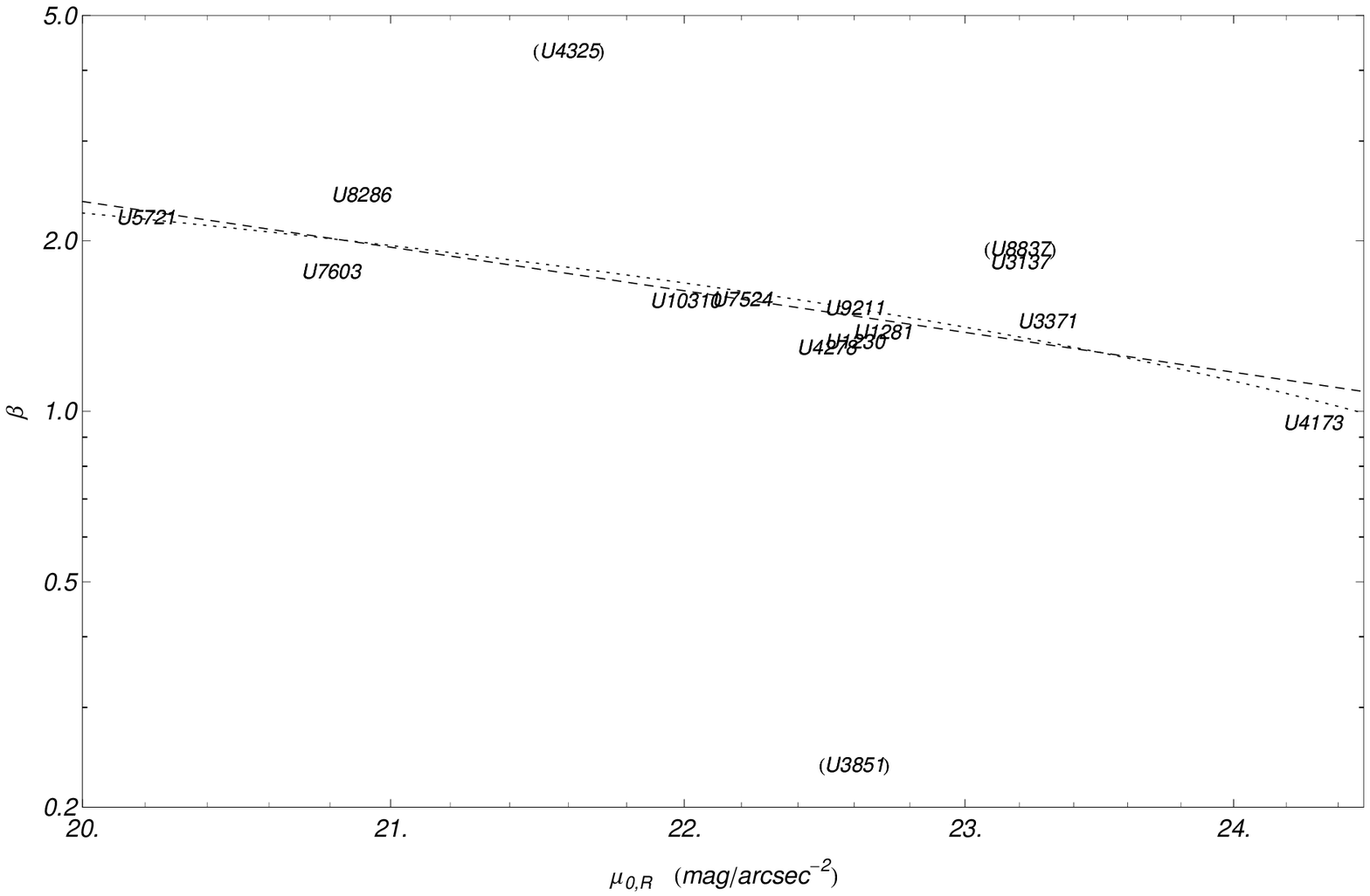}
  \includegraphics[width=120mm]{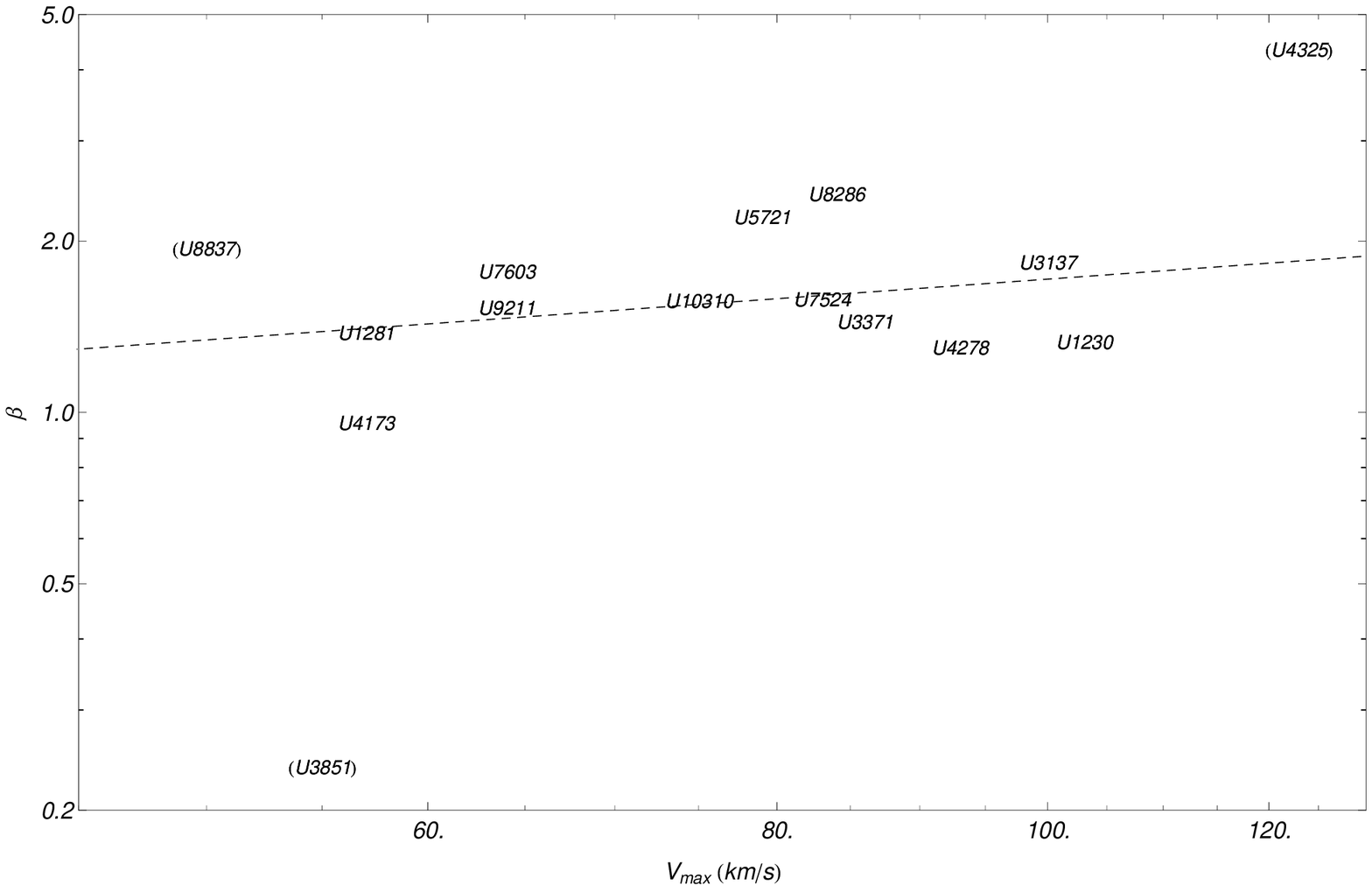}
  \includegraphics[width=120mm]{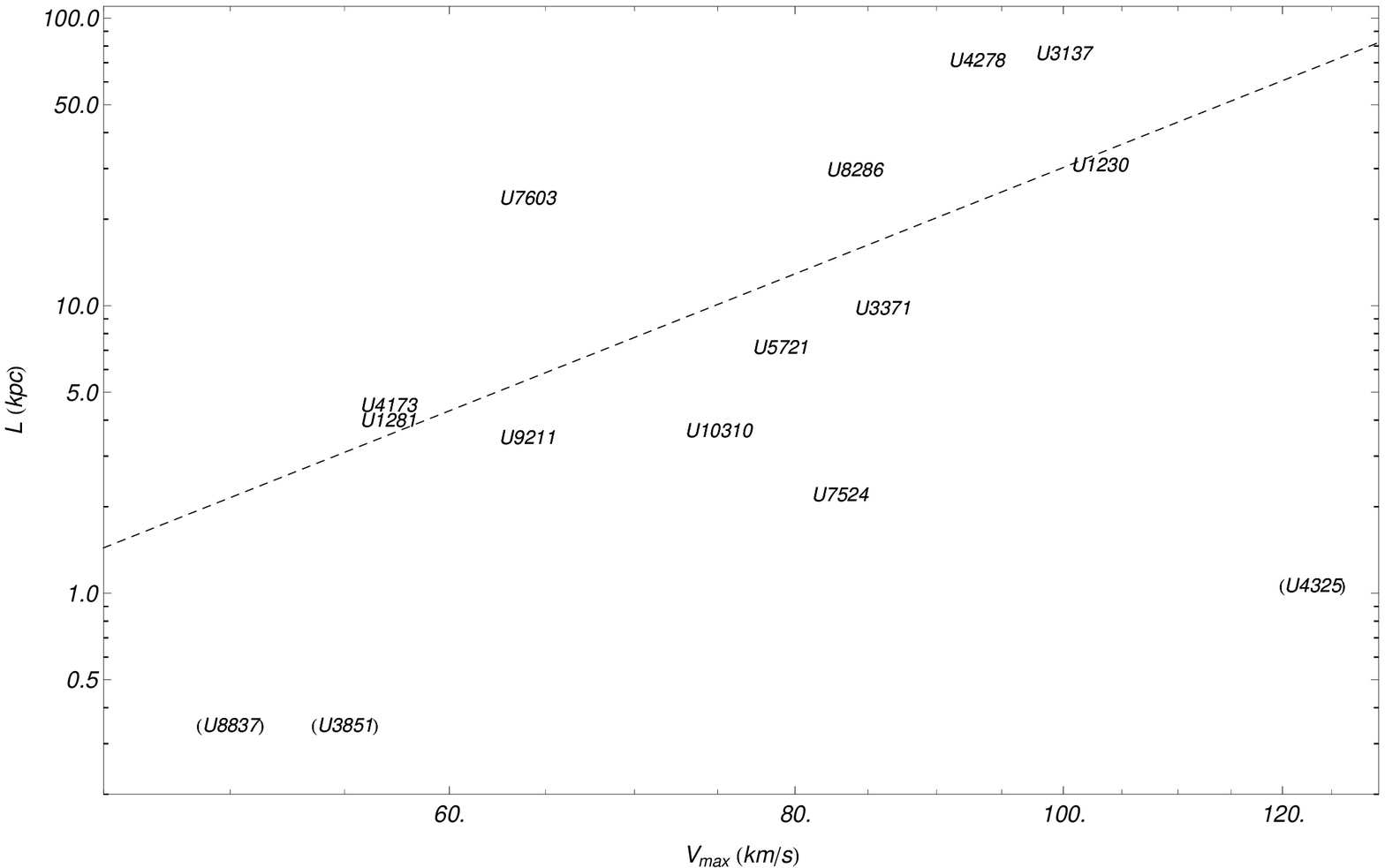}
  \caption{Correlations among scalar field parameters and structural properties of LSB galaxies. (\textit{Top panel.})
  Coupling constant $\beta$ versus the central disk surface brightness. The dotted line is the linear
  fit; the dashed line is the log-log fit. (\textit{Middle panel.}) Coupling constant $\beta$ versus
  the maximum rotation velocity. (\textit{Bottom panel.}) Scalar field length $L$ versus the maximum rotation velocity.
  Objects in brackets are excluded from fits as described in \S~(\ref{sec:results}).\label{fig:LSB_bLMuVmax}}
\end{figure*}

Then, for best appreciating the statistical validity of our analysis,
we can compare them with a recent work \citep{Swaters10}
on the same class of galaxies but based on MOND as alternative
theoretical model to dark matter. MOND is well-know to be the main
and most successful alternative scenario in explaining rotation curves
of spiral galaxies, even if not being satisfactory (when even completely
unable) in describing mass profiles in clusters of galaxies.
In \citep{Swaters10} MOND results to predict quite well LSB rotation curves;
a look to their figures show that even MOND is unable to explain
all the features that appear in such profiles, and even some open questions remain,
that seem to be intrinsic in the model while only weakly depending
on possible observational sources of uncertainties. \\
A positive step of our model, is that we are able to obtain
sufficiently good fits for two very different gravitational
systems as clusters of galaxies and LSB and opening at the same
time the possibility of a unique theoretical background
underlying both them. \\
Moreover, some of the correlations they find between MOND
acceleration and physical properties of LSB galaxies can be
found in our approach too. For example, we find a correlation
among the parameter $\beta$ and the extrapolated central disk
surface brightness $\mu_{0,R}$, even more general than their
on, because in our case it well fit all the LSB galaxies,
without any cut-of dependence (their correlation strongly depends
on galaxies with higher surface densities). Two possible fits are:
\begin{equation}
\beta = -0.277167 \cdot \mu_{0,R} + 7.78297
\end{equation}
and
\begin{equation}
\log \beta = -3.80827 \cdot \log \mu_{0,R} + 5.32519 \; ,
\end{equation}
and, as shown in the top panel of Fig.~(\ref{fig:LSB_bLMuVmax}),
they are almost indistinguishable.
Then, we also have a correlation among both scalar field parameters
and the maximum rotation velocity, while in MOND no correlation
has been found. In the middle and bottom panels of Fig.~(\ref{fig:LSB_bLMuVmax})
we have:
\begin{equation}
\log \beta = 0.354674 \cdot \log V_{max} -0.475189 \quad \mathrm{and}
\end{equation}
\begin{equation}
\log L = 3.81376 \cdot \log V_{max} -6.14695 \; .
\end{equation}
Still some questions have to be better
studied, because for LSB galaxies we could not have performed a detailed analysis of scaling
or structural parameters as we did with clusters of galaxies.

Anyway, the selected sample has given important details and
perspectives about the possibility that scalar field works well even
at these scales. But the large dispersion in scalar field parameters
show that something more accurate should be done. The used sample
of LSB galaxies is limited by the fact that many rotation curves
are not smooth, not extended to large radii; and then they
constitute a limited sample of a more complex and extended class
of gravitational systems as galaxies. In a forthcoming paper we
are going to revisit all these questions, enlarging the galaxy
sample from dwarf and irregular galaxies, to high surface
brightness (HSB) spiral galaxies, and to elliptical galaxies.

\subsection{Unified picture}
\label{sec:unified}

Even with all the previous caveats in mind, it seems that results
are consistent when we compare clusters and LSB analysis. We can
say more: even if we will only derive phenomenological and
\textit{visual} relations, whose physical meaning has to be
studied in more detail, they can help us in finding problematic
cases and thus verifying \textit{a posteriori} if they are such
for intrinsic problems of our model or for something different.

\begin{figure*}
\centering
  \includegraphics[width=150mm]{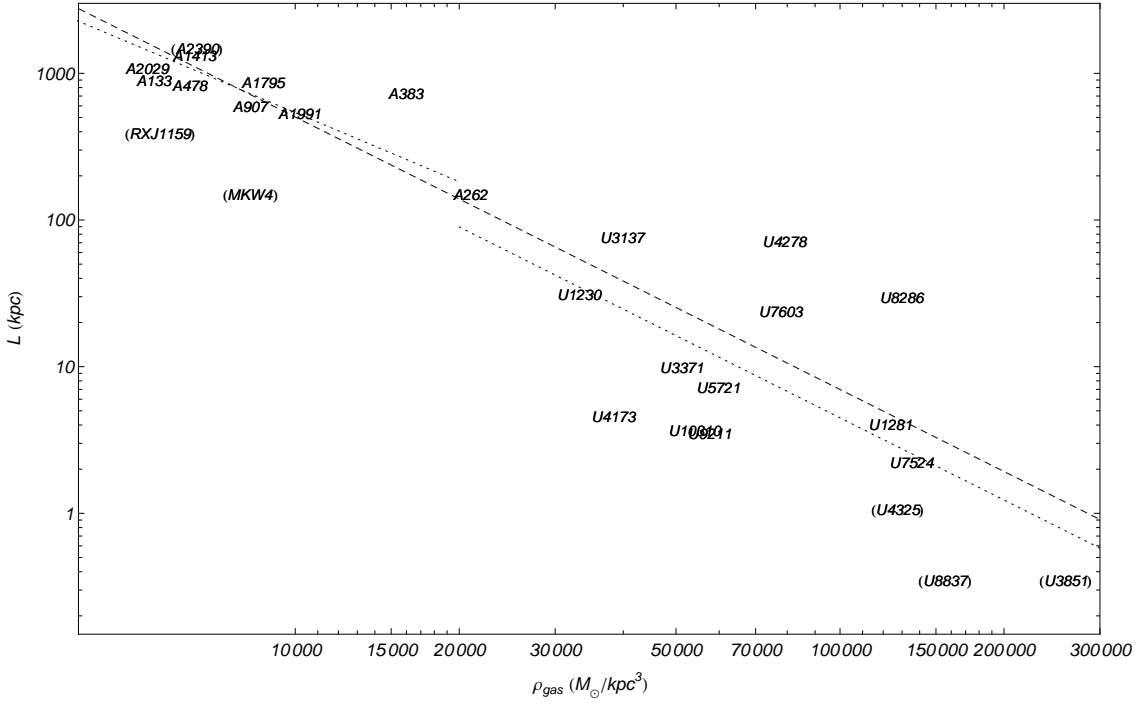}
  \caption{$L$ vs $\rho_{gas}$. The dotted lines are the singular fits to clusters and LSB samples.
  The dashed line is the fit to the total (clusters + LSB) sample. The fits are weighted with errors
  on parameters derived from MCMCs.
  Objects in brackets are excluded from fits as described in \S~(\ref{sec:results}).}
\label{fig:cham_csLd}
\end{figure*}

\begin{figure*}
\centering
  \includegraphics[width=150mm]{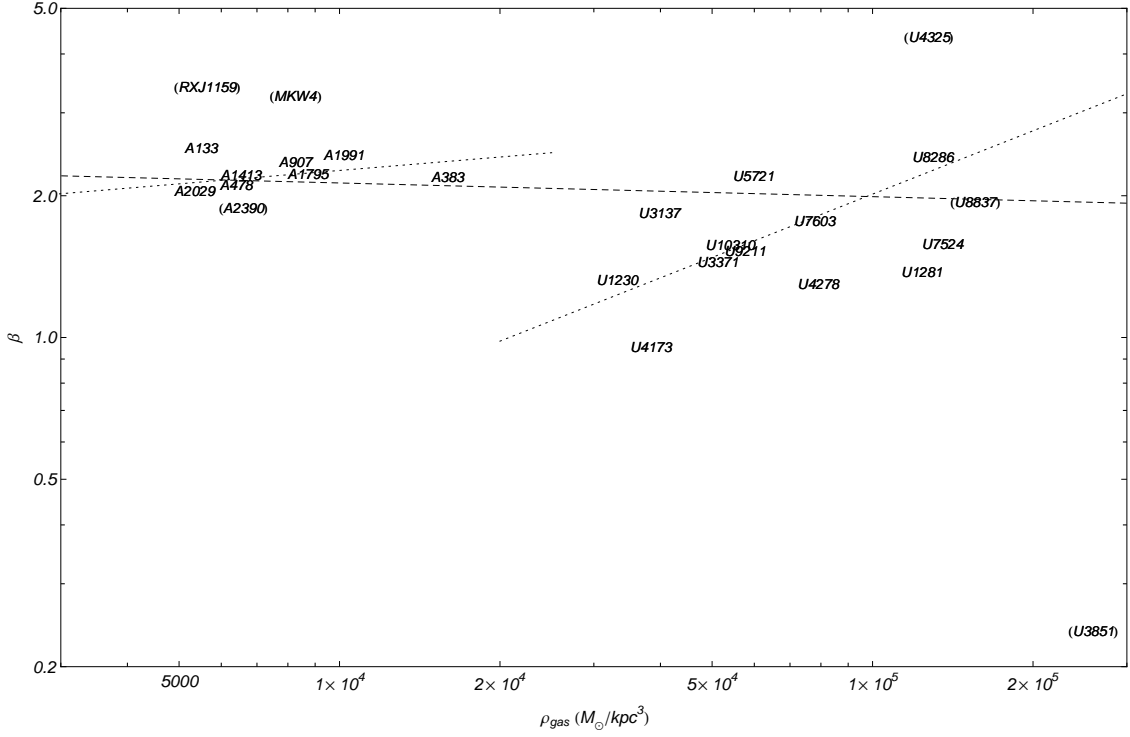}
  \caption{$\beta$ vs $\rho_{gas}$. The dotted lines are the singular fits to clusters and LSB samples.
  The dashed line is the fit to the total (clusters + LSB) sample. The fits are weighted with errors
  on parameters derived from MCMCs. Objects in brackets are excluded from fits as described in \S~(\ref{sec:results}).}
\label{fig:cham_csbd}
\end{figure*}

\begin{figure*}
\centering
  \includegraphics[width=150mm]{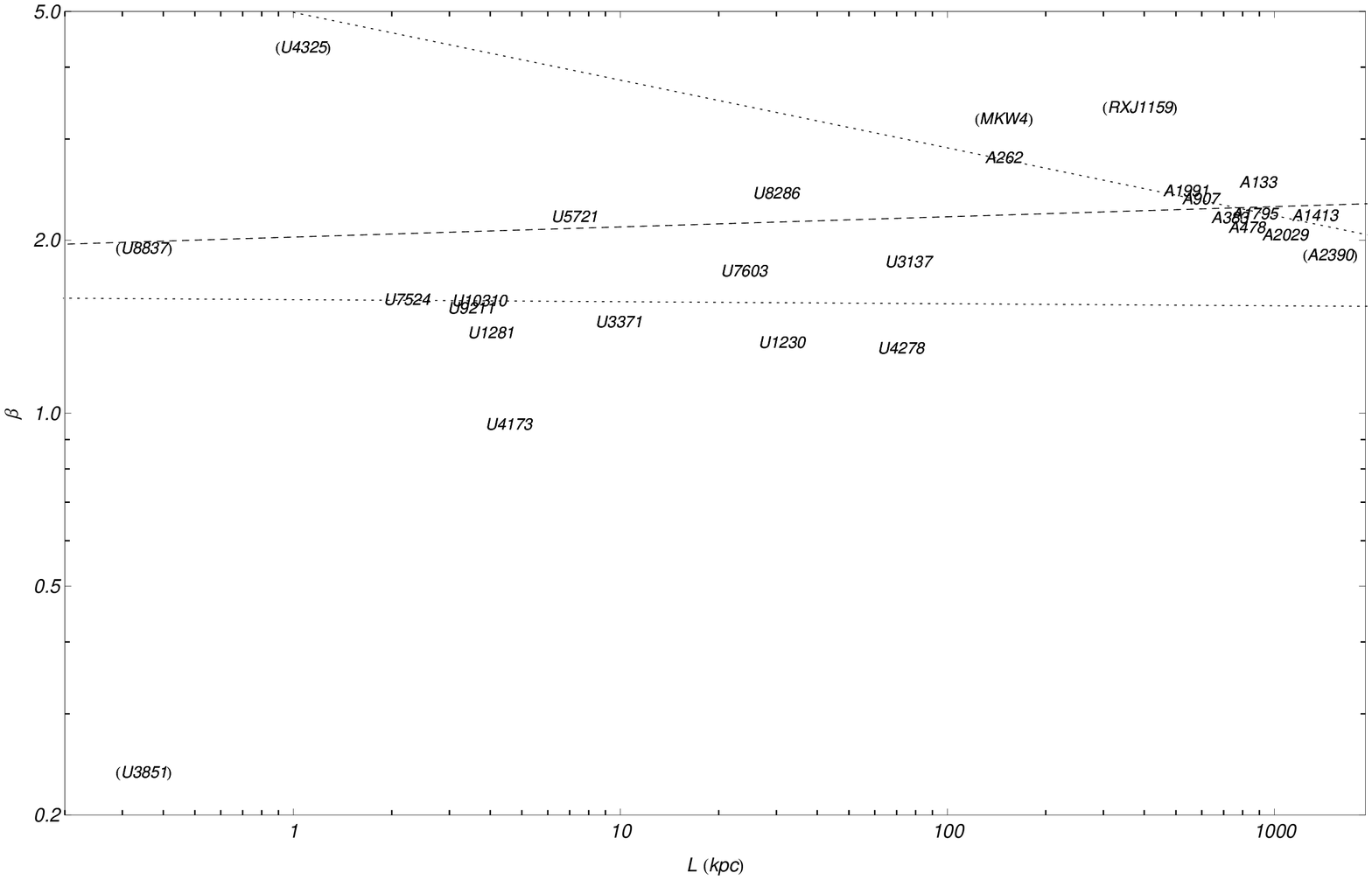}
  \caption{$\beta$ vs $L$. The dotted lines are the singular fits to clusters and LSB samples.
  The dashed line is the fit to the total (clusters + LSB) sample. The fits are weighted with errors
  on parameters derived from MCMCs. Objects in brackets are excluded from fits as described in \S~(\ref{sec:results}).}
\label{fig:cham_csbL2}
\end{figure*}

\begin{figure*}
\centering
  \includegraphics[width=150mm]{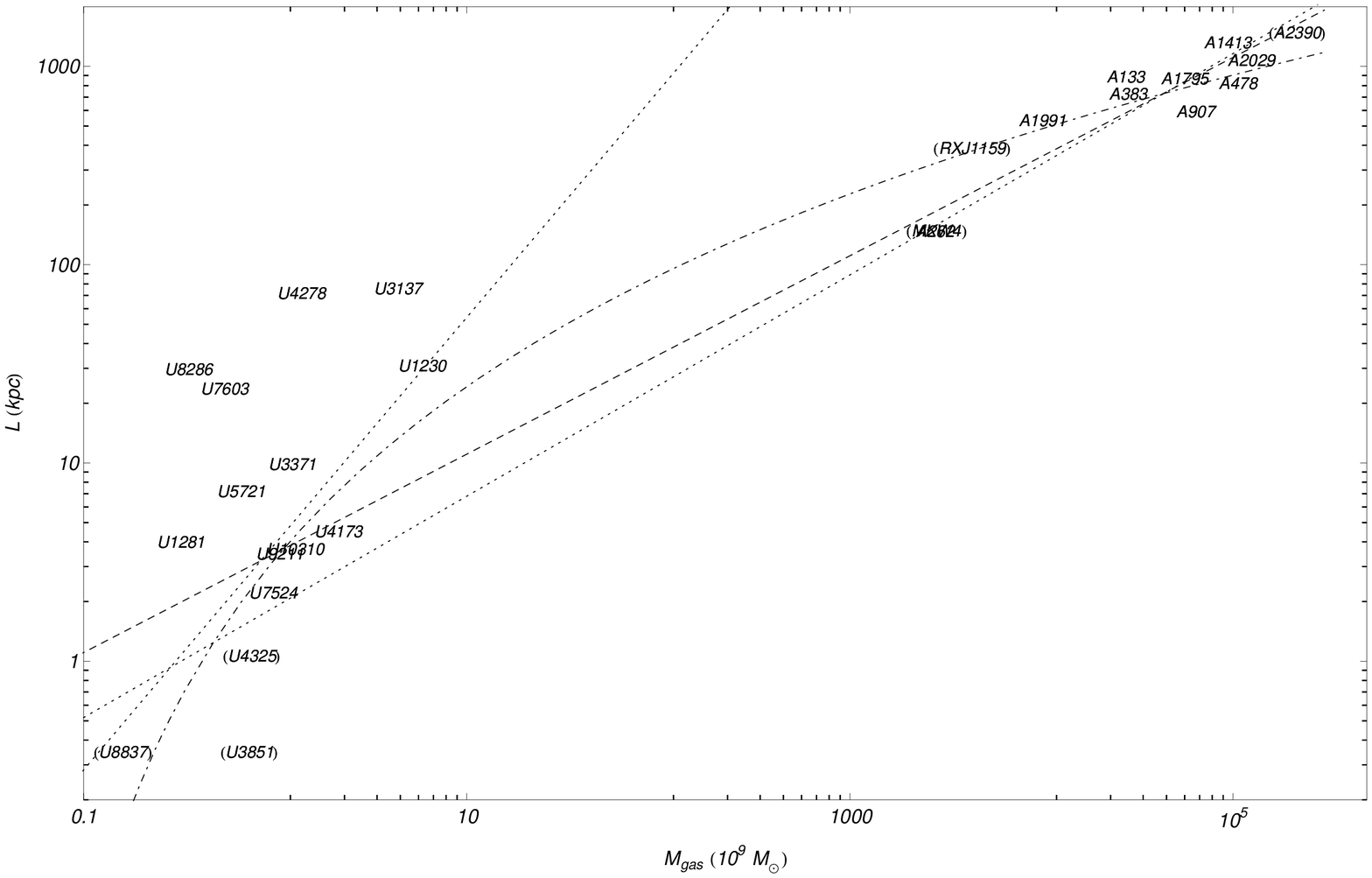}
  \caption{$L$ vs $M_{gas}$. The dotted lines are the singular fit to clusters and LSB samples.
  The dashed line is the linear fit to the total (clusters + LSB) sample. The dot-dashed line is
  the logarithmic fit to the total (clusters + LSB) sample. The fits are weighted with errors
  on parameters derived from MCMCs. Objects in brackets are excluded from fits as described in \S~(\ref{sec:results}).}
\label{fig:cham_csLM}
\end{figure*}

\begin{figure*}
\centering
  \includegraphics[width=150mm]{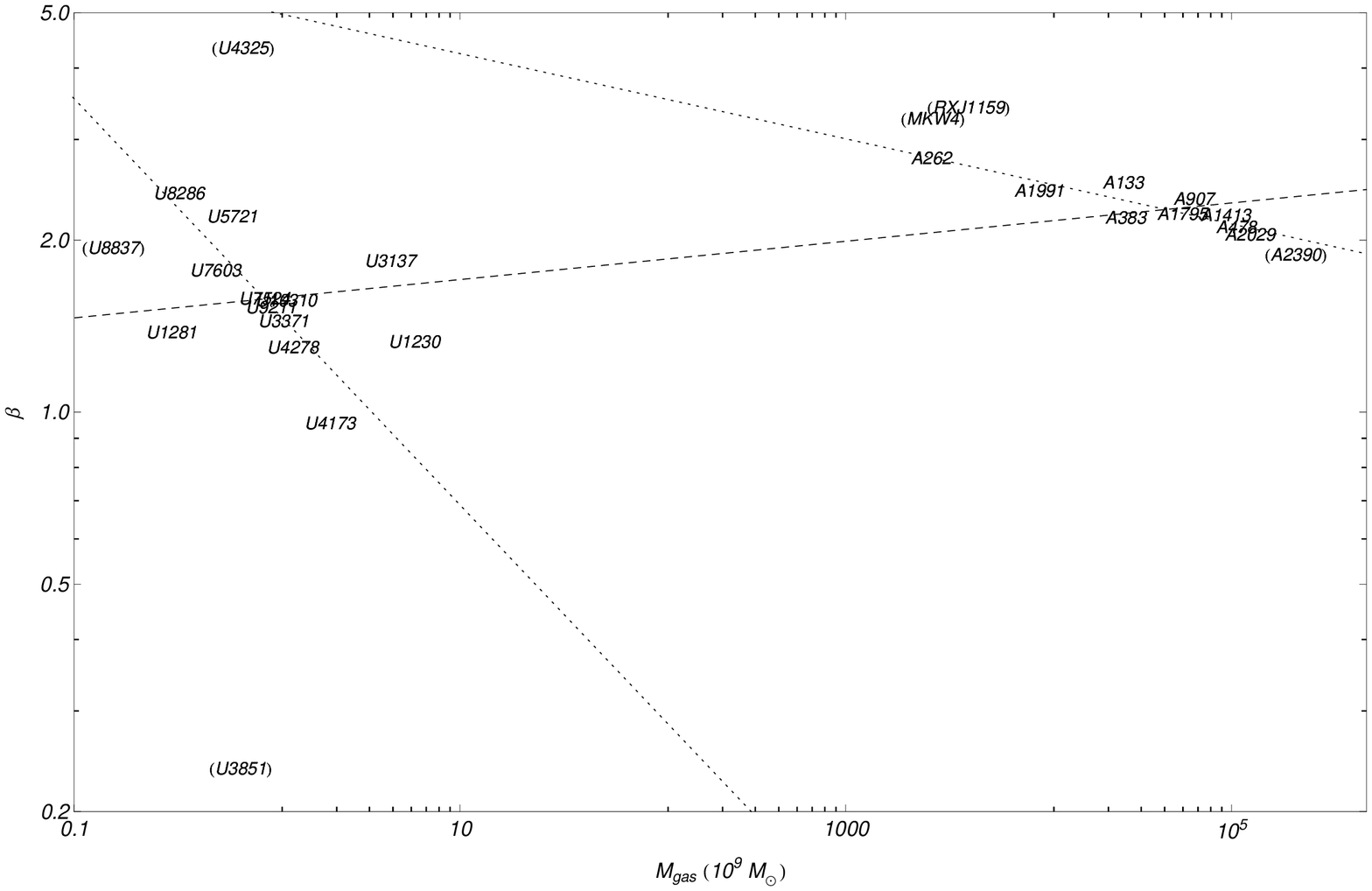}
  \caption{$\beta$ vs $M_{gas}$. The dotted lines are the singular fit to clusters and LSB samples.
  The dashed line is the fit to the total (clusters + LSB) sample. The fits are weighted with errors
  on parameters derived from MCMCs. Objects in brackets are excluded from fits as described in \S~(\ref{sec:results}).}
\label{fig:cham_csbM}
\end{figure*}

Figs.~(\ref{fig:cham_csLd})~-~(\ref{fig:cham_csbM}) show that a possible general trend including both
clusters and LSB galaxies is feasible.\\
In Fig.~(\ref{fig:cham_csLd}) a correlation between the scalar field
length $L$ and the gas density of any object is possible. We have
considered only gas contribution, because we do not want to take
into account   dark matter and use only visible matter: in
clusters of galaxies, gas is largely the main contribution to the
visible mass; and for a self-consistent discussion, we have
considered an analogous quantity for LSB galaxies too, also
considering that for calculating the stellar mass contribution, we
have to use the mass-to-luminosity ratio $Y_{\ast}$, being this a
parameter fit.

For clusters (without the problematic cases: Abell 2390, MKW4, RXJ1159) we have:
\begin{equation}
\log L = -1.56605 \cdot \log \rho_{gas} + 8.99653 \; ;
\end{equation}
for LSB galaxies (without UGC3851, UGC4325, UGC8837):
\begin{equation}
\log L = -1.8617 \cdot \log \rho_{gas} + 9.95973 \; ;
\end{equation}
while for the total sample (without the previous exceptions):
\begin{equation}
\log L = -1.85764 \cdot \log \rho_{gas} + 10.1326
\end{equation}
We can note that a small difference in the slope between the total
and the cluster-only sample, while the one from spiral is
practically equivalent. This may lead us to say that a possible
general trend is present, and that, likely, it can be made more
clear when adding further intermediate data, as group of galaxies
and elliptical galaxies, or smaller scales ones.

The same, but for the coupling constant $\beta$, is shown in Fig.~(\ref{fig:cham_csbd}).
For clusters (without Abell 2390, MKW4, RXJ1159) it is:
\begin{equation}
\log \beta = 0.0946259 \cdot \log \rho_{gas} - 0.0237813 \; ;
\end{equation}
for LSB galaxies (without UGC3851, UGC4325, UGC8837):
\begin{equation}
\log \beta = 0.446629 \cdot \log \rho_{gas} -1.92872 \; ;
\end{equation}
and for the total sample (without the previous exceptions):
\begin{equation}
\log \beta = -0.0290609 \cdot \log \rho_{gas} + 0.444376
\end{equation}
In this case we see that the total-sample slope is different
from the other two cases, even if the one from the cluster-only sample is very
small, and also by visual inspection clusters seem to be less spread
around the main general relation than LSB galaxies. These last ones,
on the contrary, seem to show a proper intrinsic slope, even if it can
depend on the previously described problems we are facing with when
working with a restricted galaxy sample, or with not enough extended
rotation curves.

We note that it is much important to verify a very low value
($\approx 0$) for $\beta$ because we remind that in
\S.\ref{sec:modified_potential} we assumed that $\beta$ is
constant or at least has a weak dependence on scale. This
hypothesis is partially confirmed by clusters and by the total
sample fit.

The same conclusion can be derived from Fig.~(\ref{fig:cham_csbL2}) where
we plot $\beta$ versus $L$: it is evident that there is quite no
dependence for $\beta$ on the gravitational scale.
In this case we also note
that the global fit feels more clusters than LSB, having performed a
weighted fit and clusters showing best constraints on scalar field parameters.
For clusters (without Abell 2390, MKW4, RXJ1159) we have:
\begin{equation}
\log \beta = -0.11781 \cdot \log L + 0.697316 \; ;
\end{equation}
for LSB galaxies (without UGC3851, UGC4325, UGC8837):
\begin{equation}
\log \beta = -0.00355508 \cdot \log L + 0.197588 \; ;
\end{equation}
and for the total sample (without the previous exceptions):
\begin{equation}
\log \beta = 0.0176545 \cdot \log L + 0.306458
\end{equation}

When plotting the scalar field parameters versus the total gas mass enclosed
in the considered gravitational structures it is more evident the need
of more objects for giving more detailed and best constrained results. In
Figs.~(\ref{fig:cham_csLM})~-~(\ref{fig:cham_csbM}) it is evident the big
void lying among clusters and LSB regions.

We have given possible fits even
in this case; for clusters (without Abell 2390, MKW4, RXJ1159) we have:
\begin{equation}
\log L = 0.558047 \cdot \log \frac{M_{gas}}{10^9 M_{\odot}} + 0.273985 \; ,
\end{equation}
\begin{equation}
\log \beta = -0.0745605 \cdot \log \frac{M_{gas}}{10^9 M_{\odot}} + 0.702031 \; ;
\end{equation}
while for LSB galaxies (without UGC3851, UGC4325, UGC8837):
\begin{equation}
\log L = 1.141 \cdot \log \frac{M_{gas}}{10^9 M_{\odot}} +0.592915 \; ,
\end{equation}
\begin{equation}
\log \beta = -0.355447 \cdot \log \frac{M_{gas}}{10^9 M_{\odot}} + 0.193349 \; ;
\end{equation}
For the total sample (without the previous exceptions) we have tried two different
fits for the length $L$:
\begin{equation}
\log L = 1.141 \cdot \log \frac{M_{gas}}{10^9 M_{\odot}} +0.592915 \; ,
\end{equation}
\begin{equation}
\log L = 3.67182 \cdot \log \left( 1.38315 \log \frac{M_{gas}}{10^9 M_{\odot}} \right) + 1.59465 \; ,
\end{equation}
but without intermediate data we cannot infer any conclusion; while for $\beta$
we have:
\begin{equation}
\log \beta = 0.0335265 \cdot \log \frac{M_{gas}}{10^9 M_{\odot}} + 0.198459 \; .
\end{equation}

\section{Conclusions}
\label{sec:conclusions}

In this work we studied  the dynamical properties  of several
astrophysical systems within the theoretical framework of
scalar theories. 
We investigate  whether there are evidences for a scalar field in the considered
astrophysical systems and if it is possible to observationally
detect it. We have taken into account three different classes of
objects: supernovae, low surface brightness spiral galaxies and
clusters of galaxies. Results show that: \textit{$i)$} there is an
intrinsic difficulty in extracting information about scalar field
mechanism (or more generally about a varying gravitational
constant) from supernovae; \textit{$ii)$} a scalar field can
fairly well reproduce the matter profile in clusters of galaxies,
estimated by X-ray observations and without the need of any
additional dark matter; \textit{$iii)$}  good fits to the rotation
curves of low surface brightness galaxies, using visible stellar
and gas mass components, are obtained.

These results show that different astrophysical  system can be
used as different tracers of the same physical mechanism.
Moreover, they point towards the possibility of a unifying view
of dark matter and dark energy via a scalar field, at least at
galactic and cluster scales \citep{Cardone}. The main criticism of
the approach is related to the fact that the very different
physical properties and evolution of the considered astrophysical
systems could insert unwanted biases and priors leading to a wrong
overall picture of the underlying cosmological model. This
shortcoming could be partially avoided if homogeneous and well
calibrated samples of data at low, medium and high redshifts will
be achieved in future.

\section{Acknowledgments}
DFM thanks the Research Council of Norway FRINAT grant 197251/V30
and the Abel extraordinary chair UCM-EEA-ABEL-03-2010. DFM is also
partially supported by the projects CERN/FP/109381/2009 and
PTDC/FIS/102742/2008. VS has been partially funded by the Research
Council of Norway with a fellow under the YGGDRASIL programme
$2009$-$2010$ and is now working at UPV/EHU under the project
``Convocatoria para la concesi$\acute{\mathrm{o}}$n de Ayuda a la
Especializaci$\acute{\mathrm{o}}$n para Investigadores Doctores en
la UPV/EHU-2009''. SC acknowledges the support of INFN (Sez. di
Napoli) and the ERASMUS/SOCRATES European program. VS acknowledges
V. F. Cardone for helpful comments and suggestions.

\label{lastpage}

\end{document}